\documentclass[aps,prl,10pt,twocolumn,superscriptaddress]{revtex4-2}
\usepackage[ascii]{inputenc}
\usepackage{amsmath,amssymb,amsfonts,amsthm}
\usepackage{mathtools}
\usepackage{tikz}
\usetikzlibrary{3d,patterns,decorations.pathmorphing,decorations.shapes, decorations.pathreplacing, shapes.geometric}
\usepackage{braket}
\usepackage{bbm}
\usepackage{todonotes}
\usepackage[caption=false]{subfig}
\usepackage[pdftex,bookmarks=false,colorlinks=true,linkcolor=blue,
citecolor=blue,filecolor=black,urlcolor=blue]{hyperref}

\graphicspath{{figures/}}

\newtheorem{theorem}{Theorem}

\newcommand{\N}{\mathbb{N}}
\newcommand{\Z}{\mathbb{Z}}
\newcommand{\R}{\mathbb{R}}
\newcommand{\C}{\mathbb{C}}

\DeclarePairedDelimiter\norm{\lVert}{\rVert}
\DeclarePairedDelimiter\abs{\lvert}{\rvert}

\DeclareMathOperator{\Tr}{Tr}
\DeclareMathOperator*{\comp}{\text{\Large $\bigcirc$}}
\DeclareMathOperator{\diag}{diag}
\DeclareMathOperator{\End}{End}

\newcommand{\drawcube}[2]{%
\begin{scope}[canvas is xz plane at y=-#1/2]
\draw[fill=#2, fill opacity=0.5] (-#1/2,-#1/2) rectangle ( #1/2, #1/2);
\end{scope}
\begin{scope}[canvas is xy plane at z=#1/2]
\draw[fill=#2, fill opacity=0.5] (-#1/2,-#1/2) rectangle ( #1/2, #1/2);
\end{scope}
\begin{scope}[canvas is yz plane at x=#1/2]
\draw[fill=#2, fill opacity=0.5] (-#1/2,-#1/2) rectangle ( #1/2, #1/2);
\end{scope}
\draw[dashed] (-#1/2,-#1/2,-#1/2) -- (-#1/2, #1/2,-#1/2) -- ( #1/2, #1/2,-#1/2);
\draw[dashed] (-#1/2, #1/2,-#1/2) -- (-#1/2, #1/2, #1/2);
}

\newcommand{\drawcubeoutline}[2]{%
\begin{scope}[canvas is xz plane at y=-#1/2]
\draw[#2] (-#1/2,-#1/2) rectangle ( #1/2, #1/2);
\end{scope}
\begin{scope}[canvas is xy plane at z=#1/2]
\draw[#2] (-#1/2,-#1/2) rectangle ( #1/2, #1/2);
\end{scope}
\begin{scope}[canvas is yz plane at x=#1/2]
\draw[#2] (-#1/2,-#1/2) rectangle ( #1/2, #1/2);
\end{scope}
\draw[#2, dashed] (-#1/2,-#1/2,-#1/2) -- (-#1/2, #1/2,-#1/2) -- ( #1/2, #1/2,-#1/2);
\draw[#2, dashed] (-#1/2, #1/2,-#1/2) -- (-#1/2, #1/2, #1/2);
}

\makeatletter
\tikzoption{canvas is plane}[]{\@setOxy#1}
\def\@setOxy O(#1,#2,#3)x(#4,#5,#6)y(#7,#8,#9)%
    {\def\tikz@plane@origin{\pgfpointxyz{#1}{#2}{#3}}%
     \def\tikz@plane@x{\pgfpointxyz{#1+#4}{#2+#5}{#3+#6}}%
     \def\tikz@plane@y{\pgfpointxyz{#1+#7}{#2+#8}{#3+#9}}%
     \tikz@canvas@is@plane
    }
\makeatother

\definecolor{mblue}  {rgb}{0.368417, 0.506779, 0.709798}
\definecolor{morange}{rgb}{0.880722, 0.611041, 0.142051}
\definecolor{mgreen} {rgb}{0.560181, 0.691569, 0.194885}
\definecolor{mred}   {rgb}{0.922526, 0.385626, 0.209179}
\definecolor{mpurple}{rgb}{0.647624, 0.37816,  0.614037}
\definecolor{mcyan}  {rgb}{0.363898, 0.618501, 0.782349}

\colorlet{lightred}  {mred!33!white}
\colorlet{lightblue} {mblue!33!white}
\colorlet{lightgreen}{mgreen!33!white}

\begin{document}

\title{Ternary unitary quantum lattice models and circuits in $2 + 1$ dimensions}

\author{Richard M. Milbradt}
\email{r.milbradt@tum.de}
\affiliation{Technical University of Munich, Department of Informatics, Boltzmannstra{\ss}e 3, 85748 Garching, Germany}

\author{Lisa Scheller}
\email{lisa.scheller@tum.de}
\affiliation{Technical University of Munich, Department of Informatics, Boltzmannstra{\ss}e 3, 85748 Garching, Germany}

\author{Christopher A{\ss}mus}
\email{chris.assmus@tum.de}
\affiliation{Technical University of Munich, Department of Informatics, Boltzmannstra{\ss}e 3, 85748 Garching, Germany}

\author{Christian B.~Mendl}
\email{christian.mendl@tum.de}
\affiliation{Technical University of Munich, Department of Informatics, Boltzmannstra{\ss}e 3, 85748 Garching, Germany}
\affiliation{Technical University of Munich, Institute for Advanced Study, Lichtenbergstra{\ss}e 2a, 85748 Garching, Germany}

\date{\today}

\begin{abstract}
We extend the concept of dual unitary quantum gates [Phys.~Rev.~Lett.~123, 210601 (2019)] to quantum lattice models in $2 + 1$ dimensions, by introducing and studying \emph{ternary unitary} four-particle gates, which are unitary in time and both spatial dimensions. When used as building blocks of lattice models with periodic boundary conditions in time and space (corresponding to infinite temperature states), dynamical correlation functions exhibit a light-ray structure. We also generalize solvable MPS [Phys.~Rev.~B 101, 094304 (2020)] to two spatial dimensions with cylindrical boundary conditions, by showing that the analogous \emph{solvable PEPS} can be identified with matrix product unitaries. In the resulting tensor network for evaluating equal-time correlation functions, the bulk ternary unitary gates cancel out. We delineate and implement a numerical algorithm for computing such correlations by contracting the remaining tensors.
\end{abstract}

\maketitle

\paragraph{Introduction.}

Dual unitary gates are special two-particle gates which are unitary both in time and space direction, forming the building blocks of a class of quantum lattice models in $1 + 1$ dimensions for which an exact evaluation of correlation functions is feasible \cite{Bertini2019}. Here we generalize this framework to two spatial dimensions, which are known for phenomena that do not exist in one dimension, like anyons \cite{Kitaev2003, Kitaev2006}. Specifically, we construct and analyze four-particle gates which are unitary in time and along both spatial dimensions, denoted ``ternary unitary'' gates. We will show that corresponding quantum lattice models exhibit light-ray correlation functions, pictorially along the edges of a pyramid. We also generalize corresponding ``solvable'' quantum states \cite{Piroli2020} to two spatial dimensions, assuming cylindrical boundary conditions. In analogy to matrix product states (MPS) in Ref.~\cite{Piroli2020}, we employ projected entangled pair states (PEPS) \cite{Verstraete2004, Hastings2006} as Ansatz for the generalisation. We will show that these ``solvable'' PEPS can be identified with concatenations of matrix product unitaries \cite{Cirac2017, Sahinoglu2018}.

\paragraph{Ternary unitary gates.}

We examine a subset of four-particle gates $U \in \End(\mathcal{H}^{\otimes 4})$, where $\mathcal{H}$ is the $d$-dimensional local Hilbert space, usually chosen as $\mathcal{H} = \C^d$, with $d = 2$ for qubits (or spin-$\frac{1}{2}$ particles). The particles are geometrically arranged as a $2 \times 2$ plaquette. In common tensor network language, $U$ is a $8$-tensor, drawn as a cube:
\begin{equation}\label{eq:IndexConvention}
\begin{tikzpicture}[>=stealth, scale=0.75, baseline=(current bounding box.center)]
\pgfsetxvec{\pgfpoint{1cm}{0cm}}
\pgfsetyvec{\pgfpoint{0.4cm}{0.3cm}}
\pgfsetzvec{\pgfpoint{0cm}{1cm}}
\node at (-1.5, 0, 0) {$U = $};
\drawcube{1}{lightred};
\draw (-0.5,-0.5,-0.5) node[below] {\tiny $5$} -- (-0.7,-0.7,-0.7);
\draw ( 0.5,-0.5,-0.5) node[below] {\tiny $6$} -- ( 0.7,-0.7,-0.7);
\draw (-0.5, 0.5,-0.5) node[below] {\tiny $7$} -- (-0.7, 0.7,-0.7);
\draw ( 0.5, 0.5,-0.5) node[below] {\tiny $8$} -- ( 0.7, 0.7,-0.7);
\draw (-0.5,-0.5, 0.5) node[above] {\tiny $1$} -- (-0.7,-0.7, 0.7);
\draw ( 0.5,-0.5, 0.5) node[above] {\tiny $2$} -- ( 0.7,-0.7, 0.7);
\draw (-0.5, 0.5, 0.5) node[above] {\tiny $3$} -- (-0.7, 0.7, 0.7);
\draw ( 0.5, 0.5, 0.5) node[above] {\tiny $4$} -- ( 0.7, 0.7, 0.7);
\node at (1, 0, 0) {$,$};
\begin{scope}[shift={(4,0,0)}]
\node at (-1.5, 0, 0) {$U^\dagger = $};
\drawcube{1}{lightblue};
\draw (-0.5,-0.5, 0.5) node[above] {\tiny $1$} -- (-0.7,-0.7, 0.7);
\draw ( 0.5,-0.5, 0.5) node[above] {\tiny $2$} -- ( 0.7,-0.7, 0.7);
\draw (-0.5, 0.5, 0.5) node[above] {\tiny $3$} -- (-0.7, 0.7, 0.7);
\draw ( 0.5, 0.5, 0.5) node[above] {\tiny $4$} -- ( 0.7, 0.7, 0.7);
\draw (-0.5,-0.5,-0.5) node[below] {\tiny $5$} -- (-0.7,-0.7,-0.7);
\draw ( 0.5,-0.5,-0.5) node[below] {\tiny $6$} -- ( 0.7,-0.7,-0.7);
\draw (-0.5, 0.5,-0.5) node[below] {\tiny $7$} -- (-0.7, 0.7,-0.7);
\draw ( 0.5, 0.5,-0.5) node[below] {\tiny $8$} -- ( 0.7, 0.7,-0.7);
\end{scope}
\begin{scope}[shift={(-2.75,-1,-1)}]
\draw[->] ( 0, 0, 0) -- ( 0.75, 0, 0) node[below] {\small $x_1$};
\draw[->] ( 0, 0, 0) -- ( 0, 0.75, 0) node[right] {\small $x_2$};
\draw[->] ( 0, 0, 0) -- ( 0, 0, 0.75) node[above] {\small $t$};
\end{scope}
\end{tikzpicture}
\end{equation}
The legs labelled $5, 6, 7, 8$ are the input dimensions, and $1, 2, 3, 4$ the output dimensions (in conformance with the convention that the leading index of a matrix corresponds to its output). In this picture, the adjoint $U^\dagger$ is a complex-conjugated copy of $U$ mirrored at the spatial $x_1$-$x_2$-plane.

We can now define new matrix products $\circ_1$ and $\circ_2$, interpreted as multiplication along the $x_1$- and $x_2$-direction rather than along the $t$-direction. This means contracting different legs compared to the usual matrix product. We define a \emph{ternary unitary} operator $U$ as an operator that is unitary with respect to the usual matrix product and the two products $\circ_1$ and $\circ_2$. Visually, we represent the usual unitary property as
\begin{equation}
\begin{tikzpicture}[>=stealth, scale=0.75]
\pgfsetxvec{\pgfpoint{1cm}{0cm}}
\pgfsetyvec{\pgfpoint{0.4cm}{0.3cm}}
\pgfsetzvec{\pgfpoint{0cm}{1cm}}
\node at (-0.5, 0, 2.5) {$U^\dagger U \qquad = \qquad U U^\dagger \quad = \quad I$};
\begin{scope}[shift={(-3, 0, 0)}]
\begin{scope}[shift={( 0, 0,-0.85)}]
\drawcube{1}{lightred};
\draw (-0.5,-0.5,-0.5) -- (-0.7,-0.7,-0.7);
\draw ( 0.5,-0.5,-0.5) -- ( 0.7,-0.7,-0.7);
\draw (-0.5, 0.5,-0.5) -- (-0.7, 0.7,-0.7);
\draw ( 0.5, 0.5,-0.5) -- ( 0.7, 0.7,-0.7);
\end{scope}
\begin{scope}[shift={( 0, 0, 0.85)}]
\drawcube{1}{lightblue};
\draw (-0.5,-0.5, 0.5) -- (-0.7,-0.7, 0.7);
\draw ( 0.5,-0.5, 0.5) -- ( 0.7,-0.7, 0.7);
\draw (-0.5, 0.5, 0.5) -- (-0.7, 0.7, 0.7);
\draw ( 0.5, 0.5, 0.5) -- ( 0.7, 0.7, 0.7);
\end{scope}
% connecting legs
\begin{scope}[canvas is plane={O(-0.5,-0.5,-0.35)x(-0.7071,-0.7071,0)y( 0, 0, 1)}]
\draw ( 0, 0) to [out=45, in=-45] ( 0, 0.7);
\end{scope}
\begin{scope}[canvas is plane={O( 0.5,-0.5,-0.35)x( 0.7071,-0.7071,0)y( 0, 0, 1)}]
\draw ( 0, 0) to [out=45, in=-45] ( 0, 0.7);
\end{scope}
\begin{scope}[canvas is plane={O(-0.5, 0.5,-0.35)x(-0.7071, 0.7071,0)y( 0, 0, 1)}]
\draw ( 0, 0) to [out=45, in=-45] ( 0, 0.7);
\end{scope}
\begin{scope}[canvas is plane={O( 0.5, 0.5,-0.35)x( 0.7071, 0.7071,0)y( 0, 0, 1)}]
\draw ( 0, 0) to [out=45, in=-45] ( 0, 0.7);
\end{scope}
\end{scope}
\node at (-1.5, 0, 0) {$=$};
\begin{scope}[shift={( 0, 0, 0)}]
\begin{scope}[shift={( 0, 0,-0.85)}]
\drawcube{1}{lightblue};
\draw (-0.5,-0.5,-0.5) -- (-0.7,-0.7,-0.7);
\draw ( 0.5,-0.5,-0.5) -- ( 0.7,-0.7,-0.7);
\draw (-0.5, 0.5,-0.5) -- (-0.7, 0.7,-0.7);
\draw ( 0.5, 0.5,-0.5) -- ( 0.7, 0.7,-0.7);
\end{scope}
\begin{scope}[shift={( 0, 0, 0.85)}]
\drawcube{1}{lightred};
\draw (-0.5,-0.5, 0.5) -- (-0.7,-0.7, 0.7);
\draw ( 0.5,-0.5, 0.5) -- ( 0.7,-0.7, 0.7);
\draw (-0.5, 0.5, 0.5) -- (-0.7, 0.7, 0.7);
\draw ( 0.5, 0.5, 0.5) -- ( 0.7, 0.7, 0.7);
\end{scope}
% connecting legs
\begin{scope}[canvas is plane={O(-0.5,-0.5,-0.35)x(-0.7071,-0.7071,0)y( 0, 0, 1)}]
\draw ( 0, 0) to [out=45, in=-45] ( 0, 0.7);
\end{scope}
\begin{scope}[canvas is plane={O( 0.5,-0.5,-0.35)x( 0.7071,-0.7071,0)y( 0, 0, 1)}]
\draw ( 0, 0) to [out=45, in=-45] ( 0, 0.7);
\end{scope}
\begin{scope}[canvas is plane={O(-0.5, 0.5,-0.35)x(-0.7071, 0.7071,0)y( 0, 0, 1)}]
\draw ( 0, 0) to [out=45, in=-45] ( 0, 0.7);
\end{scope}
\begin{scope}[canvas is plane={O( 0.5, 0.5,-0.35)x( 0.7071, 0.7071,0)y( 0, 0, 1)}]
\draw ( 0, 0) to [out=45, in=-45] ( 0, 0.7);
\end{scope}
\end{scope}
\node at (1.5, 0, 0) {$=$};
\begin{scope}[shift={(3, 0, 0)}]
\draw (-0.7,-0.7,-0.85-0.7) -- (-0.5,-0.5,-1.35) -- (-0.5,-0.5, 1.35) -- (-0.7,-0.7, 0.85+0.7);
\draw ( 0.7,-0.7,-0.85-0.7) -- ( 0.5,-0.5,-1.35) -- ( 0.5,-0.5, 1.35) -- ( 0.7,-0.7, 0.85+0.7);
\draw (-0.7, 0.7,-0.85-0.7) -- (-0.5, 0.5,-1.35) -- (-0.5, 0.5, 1.35) -- (-0.7, 0.7, 0.85+0.7);
\draw ( 0.7, 0.7,-0.85-0.7) -- ( 0.5, 0.5,-1.35) -- ( 0.5, 0.5, 1.35) -- ( 0.7, 0.7, 0.85+0.7);
\end{scope}
\end{tikzpicture} ,
\end{equation}
and the unitary condition in $x_1$-direction as
\begin{equation}\label{eq:unitary_x1}
\begin{tikzpicture}[>=stealth, scale=0.75, baseline=(current bounding box.center)]
\pgfsetxvec{\pgfpoint{1cm}{0cm}}
\pgfsetyvec{\pgfpoint{0.4cm}{0.3cm}}
\pgfsetzvec{\pgfpoint{0cm}{1cm}}
\node at (-0.75, 0, 2.5) {$U^{\dagger_1} \circ_1 U \quad = \quad I$};
\begin{scope}[shift={(-2, 0, 0)}]
\begin{scope}[shift={( 0, 0,-0.85)}]
\drawcube{1}{lightred};
\draw (-0.5,-0.5,-0.5) -- (-0.7,-0.7,-0.7);
\draw (-0.5, 0.5,-0.5) -- (-0.7, 0.7,-0.7);
\draw (-0.5,-0.5, 0.5) -- (-0.7,-0.7, 0.7);
\draw (-0.5, 0.5, 0.5) -- (-0.7, 0.7, 0.7);
\end{scope}
\begin{scope}[shift={( 0, 0, 0.85)}]
\drawcube{1}{lightblue};
\draw (-0.5,-0.5,-0.5) -- (-0.7,-0.7,-0.7);
\draw (-0.5, 0.5,-0.5) -- (-0.7, 0.7,-0.7);
\draw (-0.5,-0.5, 0.5) -- (-0.7,-0.7, 0.7);
\draw (-0.5, 0.5, 0.5) -- (-0.7, 0.7, 0.7);
\end{scope}
% connecting legs
\begin{scope}[canvas is plane={O( 0.5,-0.5,-0.35)x( 0.7071,-0.7071,0)y( 0, 0, 1)}]
\draw ( 0, 0) to [out=45, in=-45] ( 0, 0.7);
\end{scope}
\begin{scope}[canvas is plane={O( 0.5,-0.5,-1.35)x( 0.7071,-0.7071,0)y( 0, 0, 1)}]
\draw ( 0, 0) to [out=-45, in=45, looseness=1.5] ( 0, 2.7);
\end{scope}
\begin{scope}[canvas is plane={O( 0.5, 0.5,-0.35)x( 0.7071, 0.7071,0)y( 0, 0, 1)}]
\draw ( 0, 0) to [out=45, in=-45] ( 0, 0.7);
\end{scope}
\begin{scope}[canvas is plane={O( 0.5, 0.5,-1.35)x( 0.7071, 0.7071,0)y( 0, 0, 1)}]
\draw ( 0, 0) to [out=-45, in=45, looseness=1.5] ( 0, 2.7);
\end{scope}
\end{scope}
\node at ( 0, 0, 0) {$=$};
\begin{scope}[shift={( 1.5, 0, 0)}]
\begin{scope}[shift={( 0, 0,-0.85)}]
\draw (-0.7,-0.7,-0.7) -- (-0.5,-0.5,-0.5);
\draw (-0.7, 0.7,-0.7) -- (-0.5, 0.5,-0.5);
\draw (-0.7,-0.7, 0.7) -- (-0.5,-0.5, 0.5);
\draw (-0.7, 0.7, 0.7) -- (-0.5, 0.5, 0.5);
\end{scope}
\begin{scope}[shift={( 0, 0, 0.85)}]
\draw (-0.7,-0.7,-0.7) -- (-0.5,-0.5,-0.5);
\draw (-0.7, 0.7,-0.7) -- (-0.5, 0.5,-0.5);
\draw (-0.7,-0.7, 0.7) -- (-0.5,-0.5, 0.5);
\draw (-0.7, 0.7, 0.7) -- (-0.5, 0.5, 0.5);
\end{scope}
\begin{scope}[canvas is xz plane at y=-0.5]
\draw (-0.5,-0.35) to [out=0, in=0] (-0.5, 0.35);
\end{scope}
\begin{scope}[canvas is xz plane at y=-0.5]
\draw (-0.5,-1.35) to [out=0, in=0, looseness=1.2] (-0.5, 1.35);
\end{scope}
\begin{scope}[canvas is xz plane at y=0.5]
\draw (-0.5,-0.35) to [out=0, in=0] (-0.5, 0.35);
\end{scope}
\begin{scope}[canvas is xz plane at y=0.5]
\draw (-0.5,-1.35) to [out=0, in=0, looseness=1.2] (-0.5, 1.35);
\end{scope}
\end{scope}
\end{tikzpicture} \, , \, %
\begin{tikzpicture}[>=stealth, scale=0.75, baseline=(current bounding box.center)]
\pgfsetxvec{\pgfpoint{1cm}{0cm}}
\pgfsetyvec{\pgfpoint{0.4cm}{0.3cm}}
\pgfsetzvec{\pgfpoint{0cm}{1cm}}
\node at (-0.75, 0, 2.5) {$U \circ_1 U^{\dagger_1} \quad = \quad I$};
\begin{scope}[shift={(-1.5, 0, 0)}]
\begin{scope}[shift={( 0, 0,-0.85)}]
\drawcube{1}{lightred};
\draw ( 0.5,-0.5,-0.5) -- ( 0.7,-0.7,-0.7);
\draw ( 0.5, 0.5,-0.5) -- ( 0.7, 0.7,-0.7);
\draw ( 0.5,-0.5, 0.5) -- ( 0.7,-0.7, 0.7);
\draw ( 0.5, 0.5, 0.5) -- ( 0.7, 0.7, 0.7);
\end{scope}
\begin{scope}[shift={( 0, 0, 0.85)}]
\drawcube{1}{lightblue};
\draw ( 0.5,-0.5,-0.5) -- ( 0.7,-0.7,-0.7);
\draw ( 0.5, 0.5,-0.5) -- ( 0.7, 0.7,-0.7);
\draw ( 0.5,-0.5, 0.5) -- ( 0.7,-0.7, 0.7);
\draw ( 0.5, 0.5, 0.5) -- ( 0.7, 0.7, 0.7);
\end{scope}
% connecting legs
\begin{scope}[canvas is plane={O(-0.5,-0.5,-0.35)x(-0.7071,-0.7071,0)y( 0, 0, 1)}]
\draw ( 0, 0) to [out=45, in=-45] ( 0, 0.7);
\end{scope}
\begin{scope}[canvas is plane={O(-0.5,-0.5,-1.35)x(-0.7071,-0.7071,0)y( 0, 0, 1)}]
\draw ( 0, 0) to [out=-45, in=45, looseness=1.5] ( 0, 2.7);
\end{scope}
\begin{scope}[canvas is plane={O(-0.5, 0.5,-0.35)x(-0.7071, 0.7071,0)y( 0, 0, 1)}]
\draw ( 0, 0) to [out=45, in=-45] ( 0, 0.7);
\end{scope}
\begin{scope}[canvas is plane={O(-0.5, 0.5,-1.35)x(-0.7071, 0.7071,0)y( 0, 0, 1)}]
\draw ( 0, 0) to [out=-45, in=45, looseness=1.5] ( 0, 2.7);
\end{scope}
\end{scope}
\node at ( 0, 0, 0) {$=$};
\begin{scope}[shift={( 1.2, 0, 0)}]
\begin{scope}[shift={( 0, 0,-0.85)}]
\draw ( 0.7,-0.7,-0.7) -- ( 0.5,-0.5,-0.5);
\draw ( 0.7, 0.7,-0.7) -- ( 0.5, 0.5,-0.5);
\draw ( 0.7,-0.7, 0.7) -- ( 0.5,-0.5, 0.5);
\draw ( 0.7, 0.7, 0.7) -- ( 0.5, 0.5, 0.5);
\end{scope}
\begin{scope}[shift={( 0, 0, 0.85)}]
\draw ( 0.7,-0.7,-0.7) -- ( 0.5,-0.5,-0.5);
\draw ( 0.7, 0.7,-0.7) -- ( 0.5, 0.5,-0.5);
\draw ( 0.7,-0.7, 0.7) -- ( 0.5,-0.5, 0.5);
\draw ( 0.7, 0.7, 0.7) -- ( 0.5, 0.5, 0.5);
\end{scope}
\begin{scope}[canvas is xz plane at y=-0.5]
\draw ( 0.5,-0.35) to [out=180, in=180] ( 0.5, 0.35);
\end{scope}
\begin{scope}[canvas is xz plane at y=-0.5]
\draw ( 0.5,-1.35) to [out=180, in=180, looseness=1.2] ( 0.5, 1.35);
\end{scope}
\begin{scope}[canvas is xz plane at y=0.5]
\draw ( 0.5,-0.35) to [out=180, in=180] ( 0.5, 0.35);
\end{scope}
\begin{scope}[canvas is xz plane at y=0.5]
\draw ( 0.5,-1.35) to [out=180, in=180, looseness=1.2] ( 0.5, 1.35);
\end{scope}
\end{scope}
\end{tikzpicture} .
\end{equation}
An analogous depiction can be found for the $x_2$-direction. For more details on the new matrix products, refer to the supplemental material.

A prototypical ternary unitary gate is the generalization of the SWAP gate, denoted ``t-SWAP'':
\begin{equation}
U_{\text{t-SWAP}} = \sum \ket{a_1, a_2, a_3, a_4} \bra{a_4, a_3, a_2, a_1} %
= \raisebox{0.15cm}{\begin{tikzpicture}[>=stealth, scale=0.75, baseline=(current bounding box.center)]
\pgfsetxvec{\pgfpoint{1cm}{0cm}}
\pgfsetyvec{\pgfpoint{0.4cm}{0.3cm}}
\pgfsetzvec{\pgfpoint{0cm}{1cm}}
\drawcubeoutline{1}{gray};
\draw plot [smooth] coordinates {(-0.5,-0.5,-0.5) ( 0, 0,-0.1) ( 0.5, 0.5, 0.5)};
\draw plot [smooth] coordinates {(-0.5, 0.5,-0.5) ( 0, 0, 0  ) ( 0.5,-0.5, 0.5)};
\draw plot [smooth] coordinates {( 0.5,-0.5,-0.5) ( 0, 0, 0  ) (-0.5, 0.5, 0.5)};
\draw plot [smooth] coordinates {( 0.5, 0.5,-0.5) ( 0, 0, 0.1) (-0.5,-0.5, 0.5)};
\draw (-0.5,-0.5,-0.5) -- (-0.7,-0.7,-0.7);
\draw ( 0.5,-0.5,-0.5) -- ( 0.7,-0.7,-0.7);
\draw (-0.5, 0.5,-0.5) -- (-0.7, 0.7,-0.7);
\draw ( 0.5, 0.5,-0.5) -- ( 0.7, 0.7,-0.7);
\draw (-0.5,-0.5, 0.5) -- (-0.7,-0.7, 0.7);
\draw ( 0.5,-0.5, 0.5) -- ( 0.7,-0.7, 0.7);
\draw (-0.5, 0.5, 0.5) -- (-0.7, 0.7, 0.7);
\draw ( 0.5, 0.5, 0.5) -- ( 0.7, 0.7, 0.7);
\end{tikzpicture}},
\end{equation}
where $\left\{ \ket{a} \right\}_{a=0}^{d-1}$ is a basis of $\mathcal{H}$, and the sum runs over the basis $\left\{ \ket{a_1, a_2, a_3, a_4} \right\}$ of $\mathcal{H}^{\otimes  4}$. This gate swaps the opposite sites of a plaquette.

\paragraph{Constructing ternary unitary gates.}

We leave a characterization and encompassing parametrization of ternary unitaries for future work. Here we note that one can construct classes of ternary unitaries from dual unitaries, by combining four (possibly different) dual unitary gates, visualized in dark red in the following diagram:
\begin{equation}
\label{eq:four_dual_construction}
U = \raisebox{0.15cm}{%
\begin{tikzpicture}[>=stealth, scale=0.75, baseline=(current bounding box.center)]
\pgfsetxvec{\pgfpoint{1cm}{0cm}}
\pgfsetyvec{\pgfpoint{0.4cm}{0.3cm}}
\pgfsetzvec{\pgfpoint{0cm}{1cm}}
\begin{scope}[canvas is xz plane at y=0.5]
\draw[thick, fill=mred] (-0.5,-0.5) rectangle ( 0.5,-0.1);
\end{scope}
\begin{scope}[canvas is yz plane at x=-0.5]
\draw[thick, fill=mred] (-0.5, 0.1) rectangle ( 0.5, 0.5);
\end{scope}
\drawcube{1}{lightred};
\begin{scope}[canvas is xz plane at y=-0.5]
\draw[thick, fill=mred] (-0.5,-0.5) rectangle ( 0.5,-0.1);
\end{scope}
\begin{scope}[canvas is yz plane at x=0.5]
\draw[thick, fill=mred] (-0.5, 0.1) rectangle ( 0.5, 0.5);
\end{scope}
\draw (-0.5,-0.5,-0.5) -- (-0.7,-0.7,-0.7);
\draw ( 0.5,-0.5,-0.5) -- ( 0.7,-0.7,-0.7);
%\draw (-0.5, 0.5,-0.5) -- (-0.7, 0.7,-0.7);
\draw ( 0.5, 0.5,-0.5) -- ( 0.7, 0.7,-0.7);
\draw (-0.5,-0.5, 0.5) -- (-0.7,-0.7, 0.7);
\draw ( 0.5,-0.5, 0.5) -- ( 0.7,-0.7, 0.7);
\draw (-0.5, 0.5, 0.5) -- (-0.7, 0.7, 0.7);
\draw ( 0.5, 0.5, 0.5) -- ( 0.7, 0.7, 0.7);
\end{tikzpicture}}
\end{equation}
For example, if we choose all four dual unitaries as SWAP gates, the above construction yields the t-SWAP gate. It is straightforward to verify that \eqref{eq:four_dual_construction} is indeed a ternary unitary. While it is a challenge to find ternary unitary gates directly, dual unitary gates can be explicitly parametrized in the qubit case $d = 2$ \cite{Bertini2019}. For higher dimensions, constructions were found as well, albeit not general ones \cite{Borsi2022}. Our construction for ternary unitaries corresponds to a nearest neighbor interaction, when used in the time evolution that we will define below. A different way to construct ternary unitaries is to connect opposite sites by a dual unitary gate:
\begin{equation}
\label{eq:cross_construction}
U = \raisebox{0.15cm}{%
\begin{tikzpicture}[>=stealth, scale=0.75, baseline=(current bounding box.center)]
\pgfsetxvec{\pgfpoint{1cm}{0cm}}
\pgfsetyvec{\pgfpoint{0.4cm}{0.3cm}}
\pgfsetzvec{\pgfpoint{0cm}{1cm}}
\drawcube{1}{lightred};
% dual unitaries
\begin{scope}[canvas is plane={O(0,0,0)x(-0.5, 0.5, 0)y( 0, 0, 1)}]
\filldraw[thick, fill=mred] (1,-0.1) rectangle (-1,-0.5);
\end{scope}
\begin{scope}[canvas is plane={O(0,0,0)x( 0.5, 0.5, 0)y( 0, 0, 1)}]
\filldraw[thick, fill=mred] (1, 0.1) rectangle (-1, 0.5);
\end{scope}
% re-draw some of the cube edges
\draw (-0.5,-0.5, 0.5) -- ( 0.5,-0.5, 0.5);
\draw ( 0.5,-0.5, 0.5) -- ( 0.5, 0.5, 0.5);
\draw ( 0.5,-0.5,-0.5) -- ( 0.5,-0.5, 0.5);
% legs
\draw (-0.5,-0.5,-0.5) -- (-0.7,-0.7,-0.7);
\draw ( 0.5,-0.5,-0.5) -- ( 0.7,-0.7,-0.7);
\draw (-0.5, 0.5,-0.5) -- (-0.7, 0.7,-0.7);
\draw ( 0.5, 0.5,-0.5) -- ( 0.7, 0.7,-0.7);
\draw (-0.5,-0.5, 0.5) -- (-0.7,-0.7, 0.7);
\draw ( 0.5,-0.5, 0.5) -- ( 0.7,-0.7, 0.7);
\draw (-0.5, 0.5, 0.5) -- (-0.7, 0.7, 0.7);
\draw ( 0.5, 0.5, 0.5) -- ( 0.7, 0.7, 0.7);
\end{tikzpicture}}
\end{equation}

\paragraph{Physical setting.}

We consider a quantum system on a two-dimensional square lattice, where each site is associated with a local Hilbert space $\mathcal{H} = \C^d$. To simplify the discussion, we assume an $L \times L$ lattice with periodic boundary conditions, were $L$ is even. Each site is indexed by coordinates $(i,j) \in (\Z_{/L})^2$.

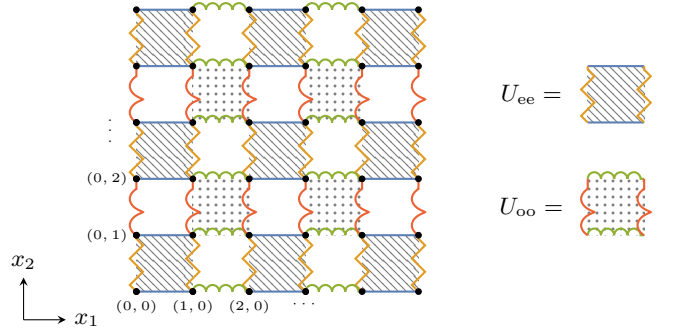
\begin{figure}[!ht]
\centering
\begin{tikzpicture}[>=stealth, scale=0.75]
\foreach \i in {0,...,2} {
    \foreach \j in {0,...,2}{
        \fill[pattern=north west lines, pattern color=gray] (2*\i, 2*\j) rectangle (2*\i+1, 2*\j+1);
    }
}
\foreach \i in {0,...,1} {
    \foreach \j in {0,...,1}{
        \fill[pattern=dots, pattern color=gray] (2*\i+1, 2*\j+1) rectangle (2*\i+2, 2*\j+2);
    }
}
\foreach \i in {0,...,2} {
    \foreach \j in {0,...,5}{
        \draw[thick, mblue] (2*\i, \j) -- (2*\i+1, \j);
    }
}
\foreach \i in {0,...,5} {
    \foreach \j in {0,...,2}{
        \draw[thick, morange, decorate, decoration=zigzag] (\i, 2*\j) -- (\i, 2*\j+1);
    }
}
\foreach \i in {0,...,1} {
    \foreach \j in {0,...,5}{
        \draw[thick, mgreen, decorate, decoration=bumps] (2*\i+1, \j) -- (2*\i+2, \j);
    }
}
\foreach \i in {0,...,5} {
    \foreach \j in {0,...,1}{
        \draw[thick, mred, decorate, decoration=coil] (\i, 2*\j+1) -- (\i, 2*\j+2);
    }
}
\foreach \i in {0,...,5} {
    \foreach \j in {0,...,5}{
        \filldraw (\i,\j) circle (0.05);
    }
}
\filldraw (0, 0) circle (0.05) node[below] {\tiny $(0,0)$};
\filldraw (1, 0) circle (0.05) node[below] {\tiny $(1,0)$};
\filldraw (2, 0) circle (0.05) node[below] {\tiny $(2,0)$};
\filldraw (3, 0) circle (0.05) node[below] {\tiny $\cdots$};
\filldraw (0, 1) circle (0.05) node[left]  {\tiny $(0,1)$};
\filldraw (0, 2) circle (0.05) node[left]  {\tiny $(0,2)$};
\filldraw (0, 3) circle (0.05) node[left=0.2] {\tiny $\vdots$};
\begin{scope}[shift={( 8, 3)}]
\node at (-1, 0.5) {$U_{\text{ee}} = $};
\fill[pattern=north west lines, pattern color=gray] ( 0, 0) rectangle ( 1, 1);
\draw[thick, mblue]                                ( 0, 0) -- ( 1, 0);
\draw[thick, mblue]                                ( 0, 1) -- ( 1, 1);
\draw[thick, morange, decorate, decoration=zigzag] ( 0, 0) -- ( 0, 1);
\draw[thick, morange, decorate, decoration=zigzag] ( 1, 0) -- ( 1, 1);
\end{scope}
\begin{scope}[shift={( 8, 1)}]
\node at (-1, 0.5) {$U_{\text{oo}} = $};
\fill[pattern=dots, pattern color=gray] ( 0, 0) rectangle ( 1, 1);
\draw[thick, mgreen, decorate, decoration=bumps] ( 0, 0) -- ( 1, 0);
\draw[thick, mgreen, decorate, decoration=bumps] ( 0, 1) -- ( 1, 1);
\draw[thick, mred,   decorate, decoration=coil]  ( 0, 0) -- ( 0, 1);
\draw[thick, mred,   decorate, decoration=coil]  ( 1, 0) -- ( 1, 1);
\end{scope}
\begin{scope}[shift={(-2,-0.5)}]
\draw[->] ( 0, 0) -- ( 0.75, 0) node[right] {\small $x_1$};
\draw[->] ( 0, 0) -- ( 0, 0.75) node[above] {\small $x_2$};
\end{scope}
\end{tikzpicture}
\caption{Pattern sequence of dual unitary gates which can be subsumed as two ``time steps'' of ternary unitary gates of the form \eqref{eq:four_dual_construction}.}
\label{fig:dual_unitary_sequence}
\end{figure}

In our model, a discrete time step can be regarded as trotterized time evolution governed by a Hamiltonian with nearest neighbor interactions:
\begin{equation}
\mathbb{U} = \mathbb{U}_{\text{oo,vert}} \mathbb{U}_{\text{oo,horz}} \mathbb{U}_{\text{ee,vert}} \mathbb{U}_{\text{ee,horz}} .
\end{equation}
Here $\mathbb{U}_{\text{ee,horz}}$ is the interaction between even-indexed sites with their right neighbors, and $\mathbb{U}_{\text{ee,vert}}$ with their upper neighbors. Analogously, $\mathbb{U}_{\text{oo,vert}}$ and $\mathbb{U}_{\text{oo,horz}}$ start from odd-indexed sites. The operators consist of non-overlapping two-particle gates, see Fig.~\ref{fig:dual_unitary_sequence}. If we choose these gates as dual unitaries, we can combine them to form ternary unitary gates as in Eq.~\eqref{eq:four_dual_construction}:
\begin{equation}
\mathbb{U}_{\text{ee}} = \mathbb{U}_{\text{ee,vert}} \mathbb{U}_{\text{ee,horz}}, \quad %
\mathbb{U}_{\text{oo}} = \mathbb{U}_{\text{oo,vert}} \mathbb{U}_{\text{oo,horz}}.
\end{equation}

Our method works for more general situations as well. For conciseness, we denote a $2 \times 2$ plaquette anchored at a site $x = (x_1,x_2)$ as $p(x) = \{(x_1,x_2),(x_1+1,x_2),(x_1,x_2+1),(x_1+1,x_2+1)\}$. Now define the two operators
\begin{eqnarray}
\mathbb{U}_{\text{ee}} &=& \bigotimes_{(i,j) \in \{ 0, \cdots , \frac{L}{2}-1 \}^2} U_{p(2i,2j)}, \\
\mathbb{U}_{\text{oo}} &=& \bigotimes_{(i,j) \in \{ 0, \cdots , \frac{L}{2}-1 \}^2} U_{p(2i+1,2j+1)},
\end{eqnarray}
where $U_S$ is the ternary unitary $U$ acting on sites $S$. We remark that the following derivations are straightforwardly generalizable for the case of differing ternary gates at each plaquette and time step. In particular, the ``light ray'' correlation structure (see below) persists. A discrete time step, motivated by trotterized time evolution, is then
\begin{equation}\label{eq:time_step}
\mathbb{U} = \mathbb{U}_{\text{oo}}\mathbb{U}_{\text{ee}}.
\end{equation}
The time dependence (in the Heisenberg picture) of a local operator $a_x$ acting on lattice site $x$ is defined as
\begin{equation}
\label{eq:local_op_dynamics}
a_x(t) = \mathbb{U}^{-t} a_x \mathbb{U}^t,
\end{equation}
where $t$ is an integer.

\paragraph{Dynamic correlations.}

Based on these definitions, we introduce dynamic correlation functions:
\begin{equation}\label{eq:corr_function}
D^{\alpha \beta}(x, y, t) = \frac{1}{d^{L^2}} \Tr\!\big[ a_x^\alpha \mathbb{U}^{-t} a_y^\beta \mathbb{U}^t\big],
\end{equation}
with $t \in \Z$ and $\{ a_x^\alpha\}_{\alpha=0}^{d^2-1}$ a basis of local operators on $\mathcal{H}_x$, the Hilbert space associated to site $x$. We can use the same trick as in the one-dimensional case \cite{Bertini2019} and choose $a^0_x = \mathbbm{1}$ and all operators to be Hilbert-Schmidt orthonormal, i.e., $\Tr[ a_x^\alpha a_x^\beta ] = d\,\delta_{\alpha \beta}$. Thus
\begin{equation}
\Tr [a_x^\alpha ] = d\,\delta_{\alpha 0}.
\end{equation}
As immediate consequence,
\begin{subequations}
\begin{eqnarray}
D^{0 \beta} (x,y,t) &=& \delta_{0 \beta}, \\
D^{\alpha 0}(x,y,t) &=& \delta_{\alpha 0}
\end{eqnarray}
\end{subequations}
for all $x$, $y$ and $t$. As derived in the supplementary, the only remaining cases of non-zero correlations require that $a_x^\alpha$ and $a_y^\beta$ are connected along a cross-diagonal ``light ray'', and the tensor diagram then simplifies to the following form (illustrated for $t = 2$):
\begin{equation}
\label{eq:corr_function_ray_diagram}
\begin{tikzpicture}[baseline=(current bounding box.center), scale=0.6]
\pgfsetxvec{\pgfpoint{1cm}{0cm}}
\pgfsetyvec{\pgfpoint{0.4cm}{0.3cm}}
\pgfsetzvec{\pgfpoint{0cm}{1cm}}
\foreach \i in {0,1,2,3}
{
    \begin{scope}[shift={(1.5*\i,1.5*\i,1.5*\i)}]
    \drawcube{1}{lightred};
    % partially shown connected legs (for visual clarity)
    \draw ( 0.5,-0.5,-0.5) -- ( 0.5,-0.5,-1.5) -- ( 0.75,-0.5,-1.25);
    \draw ( 0.5, 0.5,-0.5) -- ( 0.5, 0.5,-1.5) -- ( 0.75, 0.5,-1.25);
    \draw (-0.5, 0.5,-0.5) -- (-0.5, 0.5,-1.5) -- (-0.75, 0.5,-1.25);
    % direct connections
    \draw (-0.5, 0.5, 0.5) -- (-0.5, 0.5, 11-3*\i-0.5);
    \draw (-0.5,-0.5, 0.5) -- (-0.5,-0.5, 11-3*\i-0.5);
    \draw ( 0.5,-0.5, 0.5) -- ( 0.5,-0.5, 11-3*\i-0.5);
    \end{scope}
    \begin{scope}[shift={(1.5*\i,1.5*\i,11-1.5*\i)}]
    \drawcube{1}{lightblue}
    \draw (-0.5, 0.5, 0.5) -- (-0.5, 0.5, 1.5) -- (-0.75, 0.5, 1.25);
    \draw ( 0.5, 0.5, 0.5) -- ( 0.5, 0.5, 1.5) -- ( 0.75, 0.5, 1.25);
    \draw ( 0.5,-0.5, 0.5) -- ( 0.5,-0.5, 1.5) -- ( 0.75,-0.5, 1.25);
    \end{scope}
}
\foreach \i in {0,1,2}
{
    \begin{scope}[shift={(1.5*\i, 1.5*\i, 1.5*\i)}]
    \draw ( 0.5, 0.5, 0.5) -- ( 1, 1, 1);
    \end{scope}
    \begin{scope}[shift={(1.5*\i, 1.5*\i, 11-1.5*\i)}]
    \draw ( 0.5, 0.5,-0.5) -- ( 1, 1,-1);
    \end{scope}
}
\draw (-0.5,-0.5,-0.5) -- (-0.5,-0.5,-1.5) -- (-0.75,-0.5,-1.25) -- (-0.75,-0.5, 12.25) -- (-0.5,-0.5, 12.5) -- (-0.5,-0.5, 11.5);
\begin{scope}[canvas is xz plane at y=-0.5]
\filldraw (-0.5, 12.5) circle (0.075) node[above]{$a_x^\alpha$};
\end{scope}
\begin{scope}[shift={(4.5,4.5,4.5)}]
\draw (0.5,0.5,0.5) -- (0.5,0.5,1.5);
\begin{scope}[canvas is xz plane at y=0.5]
\filldraw (0.5,1) circle (0.075) node[right]{$a_y^\beta$};
\end{scope}
\end{scope}
\node at (-4, 0, 5.5) {$D^{\alpha \beta}(x,y,t) = \frac{1}{d^{6t+1}}$};
\end{tikzpicture} .
\end{equation}
The open hooked legs are connected as well, but are not fully drawn for visual clarity. We can condense this further by defining the operator
\begin{equation}
\label{eq:M_def}
M_z(a) = \frac{1}{d^3} \Tr_{p \setminus \bar{z}}\!\left[ U^\dagger a_z U \right],
\end{equation}
where $p = \{ (0,0), (1,0), (0,1), (1,1) \}$ enumerates the sites of a $2 \times 2$ plaquette, $z \in p$ indexes one of the sites, and $\bar{z} = (1,1) - z$ is the site opposite to $z$. For example,
\begin{equation}\label{eq:M_example}
\begin{tikzpicture}[baseline=(current bounding box.center), scale=0.75]
\pgfsetxvec{\pgfpoint{1cm}{0cm}}
\pgfsetyvec{\pgfpoint{0.4cm}{0.3cm}}
\pgfsetzvec{\pgfpoint{0cm}{1cm}}
\node at (-3, 0, 0) {$M_{(1,1)}(a) = \frac{1}{d^3}$};
% cubes
\begin{scope}[shift={(0,0,-1)}]
\drawcube{1}{lightred};
\end{scope}
\begin{scope}[shift={(0,0,1)}]
\drawcube{1}{lightblue};
\end{scope}
% middle lines
\draw (-0.5,-0.5,-0.5) -- (-0.5,-0.5, 0.5);
\draw (-0.5, 0.5,-0.5) -- (-0.5, 0.5, 0.5);
\draw ( 0.5, 0.5,-0.5) -- ( 0.5, 0.5, 0.5);
\draw ( 0.5,-0.5,-0.5) -- ( 0.5,-0.5, 0.5);
% traces
\draw ( 0.5,-0.5,-1.5) -- ( 0.5,-0.5,-2.5) -- ( 1.5,-0.5,-2.75) -- ( 1.5,-0.5, 2.75) -- ( 0.5,-0.5, 2.5) -- ( 0.5,-0.5, 1.5);
\draw ( 0.5, 0.5,-1.5) -- ( 0.5, 0.5,-2.5) -- ( 1.5, 0.5,-2.75) -- ( 1.5, 0.5, 2.75) -- ( 0.5, 0.5, 2.5) -- ( 0.5, 0.5, 1.5);
\draw (-0.5, 0.5,-1.5) -- (-0.5, 0.5,-2.5) -- (-1.5, 0.5,-2.75) -- (-1.5, 0.5, 2.75) -- (-0.5, 0.5, 2.5) -- (-0.5, 0.5, 1.5);
% remaining legs
\draw (-0.5,-0.5,-1.5) -- (-0.5,-0.5,-3.25);
\draw (-0.5,-0.5, 1.5) -- (-0.5,-0.5, 3.25);
% operator
\pgfsetxvec{\pgfpoint{1cm}{0cm}}
\pgfsetyvec{\pgfpoint{0cm}{1cm}}
\pgfsetzvec{\pgfpoint{0.4cm}{0.3cm}}
\filldraw (0.5,0,0.5) circle (0.075) node[anchor=west]{$a$};
\end{tikzpicture}\, .
\end{equation}
We remark that the maps $M_z$ are $d^2$-unistochastic quantum channels \cite{Musz2013} and contracting (by an analogous argumentation as in \cite{Bertini2019}).

Finally, we may express the correlation function (in case of $x$ connected along the light ray with $y$) as
\begin{equation}\label{eq:corr_function_M2t}
D^{\alpha \beta}(x,y,t) = \frac{1}{d}\Tr\!\left[ M_{y \text{ mod } 2}^{2t}(a^\beta) \, a^\alpha \right],
\end{equation}
where $(y \text{ mod } 2)$ is understood entry-wise. Note that the light ray condition implies $\abs{x_1 - y_1} = \abs{x_2 - y_2} = 2t$. Otherwise, $D^{\alpha \beta}(x,y,t) = 0$. Eq.~\eqref{eq:corr_function_M2t} is similar to the one-dimensional case \cite{Bertini2019}. See the supplementary material for a discussion of the computational cost of evaluating \eqref{eq:corr_function_M2t}. We can simplify \eqref{eq:corr_function_M2t} further, by writing it in the form
\begin{equation}
\label{eq:corr_function_eigenvalues}
D^{\alpha \beta}(x,y,t) = \sum_{\chi=1}^{d^2-1} c^{\alpha \beta}_{y,\chi} (t) \lambda_{y,\chi}^{2t} ,
\end{equation}
where $\left\{ \lambda_{y,\chi} \right\}_{\chi=0}^{d^2-1}$ are the eigenvalues of $M_{y \text{ mod } 2}$ and $c^{\alpha \beta}_{y,\chi} (t)$ are polynomials in $t$. The eigenvalues lie on the unit disk and have coinciding algebraic and geometric multiplicity if they are on the unit circle \cite{Bertini2019}. The eigenvalue $\lambda_{y,0} = 1$ (corresponding to the identity) is omitted from \eqref{eq:corr_function_eigenvalues}, since the operators $a_y^\beta$, $a_x^\alpha$ were assumed to be orthogonal to $\mathbbm{1}$. This result is analogous the result found in \citep{Bertini2019} and shows that the eigenvalues suffice to classify ternary unitaries by their ergodicity. A more detailed analysis of the $M$-maps for the construction \eqref{eq:four_dual_construction} can be found in the supplemental material.

\paragraph{Solvable PEPS.}

We generalize the concept of solvable states introduced in \cite{Piroli2020} to two dimensions. The goal is to construct and characterize ``solvable'' projected entangled pair states (sPEPS) as initial states, such that a semi-analytic calculation of (equal-time) correlation functions is feasible. For this purpose, we consider a scenario which renders the theoretical results in \cite{Piroli2020} applicable: an $L_1 \times L_2$ square lattice (both $L_1$ and $L_2$ even) with cylindrical geometry, i.e., arbitrary boundary conditions in $x_1$-direction and periodic boundary conditions in $x_2$-direction, and taking the thermodynamic limit $L_1, L_2 \to \infty$. For the following, a local PEPS tensor $\Lambda$ represents two neighboring sites in $x_1$-direction, so it is endowed with two physical legs. Graphically this is represented as
\begin{equation}
\begin{tikzpicture}[baseline=(current bounding box.center), >=stealth, scale=0.75]
\pgfsetxvec{\pgfpoint{1cm}{0cm}}
\pgfsetyvec{\pgfpoint{0.4cm}{0.3cm}}
\pgfsetzvec{\pgfpoint{0cm}{1cm}}
\node[anchor=east] at (-3, 0, 0){$\Lambda^{i,j}_{\mu_1 \eta_1 \mu_2 \eta_2} = $};
\draw ( 0,  -0.5, 0) -- ( 0,  -1.25, 0) node[anchor=north]{\small $\eta_1$};
\draw (-0.1, 0.5, 0) -- (-0.1, 1.25, 0);
\node[anchor=south] at (-0.3,1.2,-0.1){\small $\eta_2$};
\draw (-1, 0, 0) -- (-1.6, 0, 0) node[anchor=east]{\small $\mu_1$};
\draw ( 1, 0, 0) -- ( 1.6, 0, 0) node[anchor=west]{\small $\mu_2$};
\begin{scope}[canvas is xy plane at z=0]
\filldraw[fill=lightgray] (-1,-0.5) rectangle ( 1, 0.5);
\end{scope}
\draw (-0.5, 0, 0) -- (-0.5, 0, 0.6) node[anchor=south]{\small $i$};
\draw ( 0.5, 0, 0) -- ( 0.5, 0, 0.6) node[anchor=south]{\small $j$};
\end{tikzpicture},
\end{equation}
where the $\mu$ ($\eta$) denote the virtual bonds parallel to the $x_1$($x_2$)-direction and have bond dimension $\chi_1$ ($\chi_2$). We denote the uniform PEPS corresponding to $\Lambda$ (shift-invariant by two sites in $x_1$-direction and one site in $x_2$-direction) on an $L_1 \times L_2$ square lattice by $\ket{\Psi_{L_1 L_2}[\Lambda]}$. For brevity we will say that such a PEPS is shift-invariant.

By combining one physical leg with one $\mu$-leg each, we can reinterpret $\Lambda$ as a matrix product operator (MPO):
\begin{equation}
\label{eq:peps_grouped_legs}
\begin{tikzpicture}[baseline=(current bounding box.center), scale=0.75]
\pgfsetxvec{\pgfpoint{1cm}{0cm}}
\pgfsetyvec{\pgfpoint{0.4cm}{0.3cm}}
\pgfsetzvec{\pgfpoint{0cm}{1cm}}
\node[anchor=east] at (-3.5,0,0){$\Lambda^{(i, \mu_1), (j, \mu_2)}_{\eta_1 \eta_2} = $};

\draw ( 0,-0.5, 0) -- ( 0,-1.25, 0) node[anchor=north] {\small $\eta_1$};
\draw ( 0, 0.5, 0) -- ( 0, 1.25, 0) node[anchor=south] {\small $\eta_2$};
\draw[very thick] (-1, 0, 0) -- (-1.6, 0, 0) node[anchor=east]{\small $(i, \mu_1)$};
\draw[very thick] ( 1, 0, 0) -- ( 1.6, 0, 0) node[anchor=west]{\small $(j, \mu_2)$};

\begin{scope}[canvas is xy plane at z=0]
\filldraw[fill=lightgray] (-1,-0.5) rectangle (1,0.5);
\end{scope}
\end{tikzpicture},
\end{equation}
where the $\eta$-legs are the virtual bonds and the other legs correspond to ``physical'' dimensions. This reinterpretation will become relevant for Theorem~\ref{thm:solvable_peps} below. More notation will also be useful. We define as
\begin{equation}
\begin{tikzpicture}[baseline=(current bounding box.center), scale=0.5, >=stealth]
\useasboundingbox (-3,-3) rectangle (10, 2);
\node[anchor=east] at (-1.25, 0){$\Lambda^{L_2} = $};
\foreach \i in {0,1,3,4} {
    \begin{scope}[shift = {(2*\i,0)}]
    \filldraw[fill=lightgray] (-0.5,-0.5) rectangle (0.5,0.5);
    \draw (-1, 0) -- (-0.5, 0);
    \draw ( 0,-1) -- ( 0,  -0.5);
    \draw ( 1, 0) -- ( 0.5, 0);
    \draw ( 0, 1) -- ( 0,  0.5);
    \node[anchor=south] at (0,-0.5){$\Lambda$};
    \end{scope}
}
\node[anchor=west] at (3.25,0){$\cdots$};
\node at ( 1, 0.2) {\small $\eta$};
\node at ( 7, 0.2) {\small $\eta$};
\draw (-1, 0) to[out=130, in=50, looseness=0.75] ( 9, 0);
\draw [decorate, decoration={brace}] (8.5,-1.5) --  (-0.5,-1.5);
\node[anchor=north] at (4,-1.6){$L_2$};
\begin{scope}[shift={( 8.5,-2.5)}]
\draw[->] ( 0, 0) -- ( 1.5, 0) node[below] {\small $x_2$};
\end{scope}
\end{tikzpicture}.
\end{equation}
This kind of contraction is known as blocking of local tensors \cite{Cirac2017}. The equivalent to this in the PEPS-picture is the contraction of all $\eta$-legs of $\Lambda$-tensors with the same $x_1$-coordinate. Now we define two conditions for sPEPS that can be justified as the conditions for solvable MPS in \cite{Piroli2020}. For the latter refer to the supplementary material. Also from here on we assume that all states are normalised for all lattice sizes. A shift-invariant PEPS $\ket{\Psi_{L_1,L_2}[\tilde{\Lambda}]}$ is called solvable if
\begin{enumerate}
\item The transfer operator of $\tilde{\Lambda}^{L_2}$ defined as
\begin{equation}
\label{eq:transfer_operator}
\begin{tikzpicture}[>=stealth, scale=0.75]
\pgfsetxvec{\pgfpoint{1cm}{0cm}}
\pgfsetyvec{\pgfpoint{0.4cm}{0.3cm}}
\pgfsetzvec{\pgfpoint{0cm}{1cm}}
\pgfmathsetmacro{\ex}{0.5}
\pgfmathsetmacro{\hg}{2.5}
\begin{scope}[canvas is xy plane at z=0]
\foreach \y in {-1,...,1}
{
    \draw[fill=lightgray] ( 0-\ex, 2*\y-\ex) rectangle (1+\ex, 2*\y+\ex);
}
\foreach \y in {-0.2, 0, 0.2}
{
    \filldraw ( 0.5,-3.5+\y) circle (0.025);
    \filldraw ( 0.5, 3.5+\y) circle (0.025);
}
\end{scope}
\begin{scope}[canvas is xy plane at z=\hg]
\foreach \y in {-0.2, 0, 0.2}
{
    \filldraw ( 0.5,-3.5+\y) circle (0.025);
    \filldraw ( 0.5, 3.5+\y) circle (0.025);
}
\end{scope}
% physical legs
\foreach \y in {-1,...,1}
{
    \draw ( 0, 2*\y, 0) -- ( 0, 2*\y, \hg);
}
\foreach \y in {-1,...,1}
{
    \draw (1, 2*\y, 0) -- (1, 2*\y, \hg);
}

% virtual bonds
\foreach \y in {-1,...,0}
{
    \foreach \z in {0, \hg}
    {
        \draw (0.5, 2*\y+\ex, \z) -- (0.5, 2*\y+2-\ex, \z);
    }
}
\foreach \y in {-1,...,1}
{
    \foreach \z in {0, \hg}
    {
        \draw (-1.5+\ex, 2*\y, \z) -- (-\ex, 2*\y, \z);
        \draw (2.5-\ex, 2*\y, \z) -- (1 + \ex, 2*\y, \z);
    }
}
\begin{scope}[canvas is xy plane at z=\hg]
\foreach \y in {-1,...,1}
{
    \draw[fill=gray] ( 0-\ex, 2*\y-\ex) rectangle (1+\ex, 2*\y+\ex);
}
\end{scope}
\end{tikzpicture}
\end{equation}
where the darker local tensor are $\tilde{\Lambda}^*$, has a unique largest eigenvalue $\lambda = 1$ with an algebraic multiplicity of $1$.
\item There exists a non-zero tensor $S \in \C^{d^{L_2}} \times \C^{d^{L_2}}$ such that
\begin{equation}
\label{eq:solvable_condition}
\begin{tikzpicture}[>=stealth, scale=0.75]
\pgfsetxvec{\pgfpoint{1cm}{0cm}}
\pgfsetyvec{\pgfpoint{0.4cm}{0.3cm}}
\pgfsetzvec{\pgfpoint{0cm}{1cm}}
\pgfmathsetmacro{\ex}{0.5}
\pgfmathsetmacro{\hg}{3}
\begin{scope}[canvas is xy plane at z=0]
\foreach \y in {-1,...,1}
{
    \draw[fill=lightgray] ( 0-\ex, 2*\y-\ex) rectangle (1+\ex, 2*\y+\ex);
}
\foreach \y in {-0.2, 0, 0.2}
{
    \filldraw ( 0.5,-3.5+\y) circle (0.025);
    \filldraw ( 0.5, 3.5+\y) circle (0.025);
}
\end{scope}
\begin{scope}[canvas is xy plane at z=\hg]
\foreach \y in {-0.2, 0, 0.2}
{
    \filldraw ( 0.5,-3.5+\y) circle (0.025);
    \filldraw ( 0.5, 3.5+\y) circle (0.025);
}
\end{scope}
% physical legs
\foreach \y in {-1,...,1}
{
    \draw ( 0, 2*\y, 0) -- ( 0, 2*\y, \ex);
}
\foreach \y in {-1,...,1}
{
    \draw (1, 2*\y, 0) -- (1, 2*\y, \hg);
}
\foreach \y in {-1,...,1}
{
    \draw ( 0, 2*\y, \hg) -- ( 0, 2*\y, \hg-\ex);
}
% virtual bonds
\foreach \y in {-1,...,0}
{
    \foreach \z in {0, \hg}
    {
        \draw (0.5, 2*\y+\ex, \z) -- (0.5, 2*\y+2-\ex, \z);
    }
}
\foreach \y in {-1,...,1}
{
    \foreach \z in {0, \hg}
    {
        \draw (-1.5+\ex, 2*\y, \z) -- (-\ex, 2*\y, \z);
    }
}
\begin{scope}[canvas is xy plane at z=\hg]
\foreach \y in {-1,...,1}
{
    \draw[fill=gray] ( 0-\ex, 2*\y-\ex) rectangle (1+\ex, 2*\y+\ex);
}
\end{scope}
\foreach \y in {-1,...,1}
{
    \begin{scope}[canvas is xz plane at y=2*\y]
    \draw (1+\ex, 0) to[out=0, in=-90] (2.5, \hg/2-0.25);
    \end{scope}
}
\begin{scope}[shift={(2.5, 0, \hg/2)}]
\begin{scope}[canvas is xz plane at y=-2-\ex]
\draw[fill=lightgreen] (-0.25,-0.25) rectangle ( 0.25, 0.25);
\end{scope}
\begin{scope}[canvas is xy plane at z=0.25]
\draw[fill=lightgreen] (-0.25,-2-\ex) rectangle ( 0.25, 2+\ex);
\end{scope}
\begin{scope}[canvas is yz plane at x=0.25]
\draw[fill=lightgreen] (-2-\ex,-0.25) rectangle ( 2+\ex, 0.25);
\end{scope}
\node[anchor=west] at (0, 2+\ex, 0.5){$S$};
\end{scope}
\foreach \y in {-1,...,1}
{
    \begin{scope}[canvas is xz plane at y=2*\y]
    \draw (1+\ex, \hg) to[out=0, in= 90] (2.5, \hg/2+0.25);
    \end{scope}
}

\node at (4, 0, \hg/2) {$=$};

\begin{scope}[shift={(5.5, 0, 0)}]
\begin{scope}[canvas is xy plane at z=0]
\foreach \y in {-0.2, 0, 0.2}
{
    \filldraw ( 0.5,-3.5+\y) circle (0.025);
    \filldraw ( 0.5, 3.5+\y) circle (0.025);
}
\end{scope}
\foreach \y in {-1,...,1}
{
    \begin{scope}[canvas is xz plane at y=2*\y]
    \draw (-1+\ex, 0) -- (1+\ex, 0) to[out=0, in=-90] (2.5, \hg/2-0.25);
    \draw ( 0, \ex) to[out=-90, in=180] ( 0.25, \ex/2) to[out=0, in=-90] ( 0.5, \ex) -- ( 0.5, \hg-\ex) to[out=90, in=0] ( 0.25, \hg-\ex/2) to[out=180, in=90] ( 0, \hg-\ex) ;
    \end{scope}
}
\begin{scope}[shift={(2.5, 0, \hg/2)}]
\begin{scope}[canvas is xz plane at y=-2-\ex]
\draw[fill=lightgreen] (-0.25,-0.25) rectangle ( 0.25, 0.25);
\end{scope}
\begin{scope}[canvas is xy plane at z=0.25]
\draw[fill=lightgreen] (-0.25,-2-\ex) rectangle ( 0.25, 2+\ex);
\end{scope}
\begin{scope}[canvas is yz plane at x=0.25]
\draw[fill=lightgreen] (-2-\ex,-0.25) rectangle ( 2+\ex, 0.25);
\end{scope}
\node[anchor=west] at (0, 2+\ex, 0.5){$S$};
\end{scope}
\foreach \y in {-1,...,1}
{
    \begin{scope}[canvas is xz plane at y=2*\y]
    \draw (-1+\ex, \hg) -- (1+\ex, \hg) to[out=0, in= 90] (2.5, \hg/2+0.25);
    \end{scope}
}
\end{scope}
\end{tikzpicture}
\end{equation}
\end{enumerate}
We need two additional definitions to state the following theorem. A matrix product unitary (MPU) is an MPO that fulfills \cite{Cirac2017}
\begin{equation}\label{eq:MPU_cond}
\begin{tikzpicture}[baseline=(current bounding box.center), scale=0.5]
\foreach \i in {0,1,3}{
    \begin{scope}[shift={(2*\i,0,0)}]
    \filldraw[fill=lightgray] (-0.5,-1.5) rectangle (0.5,-0.5);
    \filldraw[fill=lightgray] (-0.5, 1.5) rectangle (0.5, 0.5);

    \draw (0,-0.5) -- (0,0.5);
    \draw (0,-1.5) -- (0,-2);
    \draw (0,1.5) -- (0,2);

    \draw (-0.5,-1) -- (-1,-1);
    \draw (0.5,-1) -- (1,-1);
    \draw (-0.5,1) -- (-1,1);
    \draw (0.5,1) -- (1,1);

    \node[anchor=south] at (0,-1.5){$\Lambda$};
    \node[anchor=south] at (0,0.5){$\Lambda^*$};

    \end{scope}
}

\node[anchor=east] at (4.5,0){\small $\cdots$};

\draw (-1,-1) -- (-1.5,-1) -- (-1.5,-1.75) -- (-1,-1.75);
\draw (-1, 1) -- (-1.5, 1) -- (-1.5, 1.75) -- (-1, 1.75);
\draw ( 7,-1) -- ( 7.5,-1) -- ( 7.5,-1.75) -- ( 7,-1.75);
\draw ( 7, 1) -- ( 7.5, 1) -- ( 7.5, 1.75) -- ( 7, 1.75);

\node[anchor=west] at (7.5,0){$=$};
\begin{scope}[shift={(-1.5,0,0)}]
\foreach \i in {0,1,3}{
    \begin{scope}[shift={(10.5+0.5*\i,0)}]
        \draw (0,-1.75) -- (0,1.75);
    \end{scope}
}
\node[anchor=west] at (11,0){\small $\cdots$};
\draw[decorate, decoration={brace}] (12,-2.5) -- (10.5,-2.5);
\node[anchor=north] at (11.25,-2.5){$L_2$};
\end{scope}

\draw[decorate, decoration={brace}] (6.25,-2.5) -- (-0.25,-2.5);
\node[anchor=north] at (3,-2.5){$L_2$};
\end{tikzpicture}
\end{equation}
for all $L_2 \in \N_1$. MPUs can be characterized in different ways \cite{Cirac2017, Sahinoglu2018}. For example, any one-dimensional quantum cellular automaton is an MPU and vice versa \cite{Cirac2017}. The second required notion is that of equivalence. We say that two sets of states $\{\ket{\Psi_{L_1 L_2}}\}_{L_1,L_2}$, $\{\ket{\Phi_{L_1 L_2}}\}_{L_1,L_2}$ defined on increasing lattice sizes $L_1  \times L_2$ are equivalent if
\begin{multline}
\lim_{L_2 \rightarrow \infty}\lim_{L_1 \rightarrow \infty} \bra{ \Psi_{L_1L_2}} \Omega_R \ket{\Psi_{L_1 L_2}} = \\ \lim_{L_2 \rightarrow \infty}\lim_{L_1 \rightarrow \infty} \bra{ \Phi_{L_1L_2}} \Omega_R \ket{\Phi_{L_1 L_2}}
\end{multline}
for all operators $\Omega_R$, where $R \subset \Z^2$ is bounded. Alternatively we just say $\ket{\Psi_{L_1 L_2}}$ is equivalent to $\ket{\Phi_{L_1 L_2}}$ in the thermodynamic limit. Now we can state the main result of this section:
\begin{theorem}[Classification of solvable PEPS states]
\label{thm:solvable_peps}
On a square lattice $L_1 \times L_2$ with cylindrical boundary conditions a solvable PEPS $\ket{\Psi_{L_1 L_2} [\tilde{\Lambda}]}$ as defined above is equivalent in the thermodynamic limit to some shift-invariant PEPS $\ket{\Psi_{L_1 L_2} [\Lambda]}$ such that the associated MPO \eqref{eq:peps_grouped_legs} is an MPU up to a scalar factor.
\end{theorem}
A detailed proof can be found in the supplementary material. Essentially, the theorem is a consequence of Theorem~1 in \cite{Piroli2020} after reducing the two-dimensional lattice geometry to a one-dimensional layout by grouping all tensors in $x_2$-direction at given $x_1$ and $t$. In particular, the local Hilbert space dimension is then $d^{L_2}$.

\begin{figure}
\begin{tikzpicture}[baseline=(current bounding box.center), scale=0.6]
\foreach \i in {-1,0,1}{
    \foreach \j in {-2,...,2}{
        \filldraw[fill=lightgray, fill opacity=0.5] (3*\i -1, 2*\j -0.5) rectangle (3*\i +1, 2*\j +0.5);
        \draw (3*\i +1, 2*\j) -- (3*\i +1.5, 2*\j);
        \draw (3*\i -1, 2*\j) -- (3*\i -1.5, 2*\j);
        \draw (3*\i, 2*\j +0.5) -- (3*\i, 2*\j +1);
        \draw (3*\i, 2*\j -0.5) -- (3*\i, 2*\j -1);
        \filldraw (3*\i +0.5, 2*\j) circle (0.05);
        \filldraw (3*\i -0.5, 2*\j) circle (0.05);
        \draw (3*\i +0.5, 2*\j) circle (0.15);
        \draw (3*\i -0.5, 2*\j) circle (0.15);
    }

    \foreach \j in {0,2}{
        \draw[color=mred] (3*\i -1.5, 2*\j +0.3) -- (3*\i -0.2, 2*\j +0.3) -- (3*\i -0.2, 2*\j -2.3) -- (3*\i -1.5, 2*\j -2.3);
        \fill[fill=lightred, fill opacity=0.5] (3*\i -1.5, 2*\j +0.3) -- (3*\i -0.2, 2*\j +0.3) -- (3*\i -0.2, 2*\j -2.3) -- (3*\i -1.5, 2*\j -2.3) -- cycle;

        \draw[color=mred] (3*\i +1.5, 2*\j +0.3) -- (3*\i +0.2, 2*\j +0.3) -- (3*\i +0.2, 2*\j -2.3) -- (3*\i +1.5, 2*\j -2.3);
        \fill[fill=lightred, fill opacity=0.5] (3*\i +1.5, 2*\j +0.3) -- (3*\i +0.2, 2*\j +0.3) -- (3*\i +0.2, 2*\j -2.3) -- (3*\i +1.5, 2*\j -2.3) -- cycle;
    }

    \draw[color=mred] (3*\i -1.5, -4 +0.3) -- (3*\i -0.2, -4 +0.3) -- (3*\i - 0.2, -4 -1);
    \fill[fill=lightred, fill opacity=0.5] (3*\i -1.5, -4 +0.3) -- (3*\i -0.2, -4 +0.3) -- (3*\i - 0.2, -4 -1) -- (3*\i -1.5, -4 -1) -- cycle;
    \draw[color=mred] (3*\i +1.5, -4 +0.3) -- (3*\i +0.2, -4 +0.3) -- (3*\i +0.2, -4 -1);
    \fill[fill=lightred, fill opacity=0.5] (3*\i +1.5, -4 +0.3) -- (3*\i +0.2, -4 +0.3) -- (3*\i +0.2, -4 -1) -- (3*\i +1.5, -4 -1) -- cycle;
}
\foreach \i in {-1,...,3}{
    \node[anchor=east] at (-4.7, 2*\i-2){\small $\i$};
}
\foreach \i in {-2,0,2}{
    \node[anchor=north] at (1.5*\i + 0.5, -5.2){\small $\i$};
}
\foreach \i in {-3,-1,1}{
    \node[anchor=north] at (1.5*\i + 1, -5.2){\small $\i$};
}

\node[anchor=south] at (-5, 4.5){\small $\vdots$};
\node[anchor=north] at (-5,-4.2){\small $\vdots$};
\node[anchor=west]  at ( 4,-5.6){\small $\cdots$};
\node[anchor=east]  at (-4,-5.6){\small $\cdots$};
\end{tikzpicture}
\caption{A top-down view of the PEPS (gray $\Lambda$ tensors) and the first layer $\mathbb{U}_{ee}$ of ternary unitary gates. Note the parity in $x_1$-direction.}
\label{fig:speps_first_layer}
\end{figure}
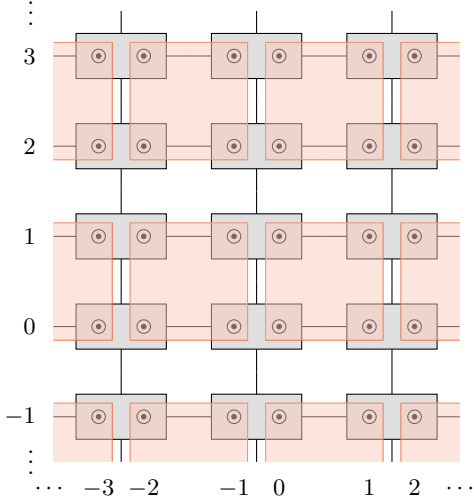
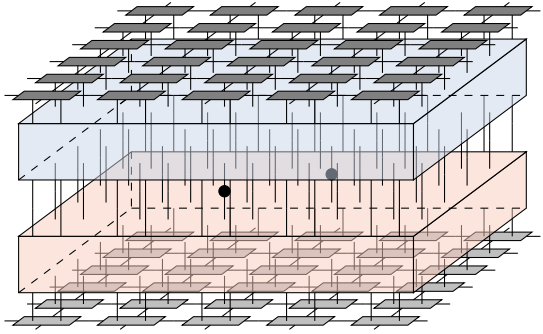
\begin{figure}
\begin{tikzpicture}[baseline=(current bounding box.center), scale=0.375]
\pgfsetxvec{\pgfpoint{1cm}{0cm}}
\pgfsetyvec{\pgfpoint{0.4cm}{0.3cm}}
\pgfsetzvec{\pgfpoint{0cm}{1cm}}

\foreach \i in {4,...,0}{
    \foreach \j in {5,...,0}{
        \begin{scope}[shift={(3*\i,2*\j,0)}]
        \begin{scope}[canvas is xy plane at z=-7]
            \filldraw[fill=lightgray] (-1,-0.5) rectangle (1,0.5);
            \draw (0,0.5) -- (0,1);
            \draw (0,-0.5) -- (0,-1);
            \draw (-1,0) -- (-1.5,0);
            \draw (1,0) -- (1.5,0);
        \end{scope}
        \draw (-0.5,0,-7) -- (-0.5,0,-6);
        \draw (0.5,0,-7) -- (0.5,0,-6);
        \end{scope}
    }
}
\begin{scope}[canvas is xz plane at y=0]
    \filldraw[fill=lightred, fill opacity=0.5] (-1,-6) rectangle (13,-4);
\end{scope}
\begin{scope}[canvas is yz plane at x=13]
    \filldraw[fill=lightred, fill opacity=0.5] (0,-6) rectangle (10,-4);
\end{scope}
\begin{scope}[canvas is xy plane at z=-4]
    \filldraw[fill=lightred, fill opacity=0.5] (-1,0) rectangle (13,10);
\end{scope}
\draw[dashed] (-1,0,-6) -- (-1,10,-6) -- (-1,10,-4);
\draw[dashed] (-1,10,-6) -- (13,10,-6);

\foreach \i in {4,...,0}{
    \foreach \j in {5,...,0}{
        \begin{scope}[shift={(3*\i,2*\j,0)}]
        \draw (-0.5, 0,-4) -- (-0.5, 0,-3);
        \draw ( 0.5, 0,-4) -- ( 0.5, 0,-3);
        \end{scope}
    }
}
\begin{scope}[canvas is xz plane at y=2]
    \filldraw (5.5,-3) circle (0.2);
\end{scope}
\begin{scope}[canvas is xz plane at y=4]
    \filldraw (8.5,-3) circle (0.2);
\end{scope}
\foreach \i in {4,...,0}{
    \foreach \j in {5,...,0}{
        \begin{scope}[shift={(3*\i,2*\j,0)}]
        \draw (-0.5, 0,-3) -- (-0.5, 0,-2);
        \draw ( 0.5, 0,-3) -- ( 0.5, 0,-2);
        \end{scope}
    }
}

\begin{scope}[canvas is xz plane at y=0]
    \filldraw[fill=lightblue,fill opacity=0.5] (-1,-2) rectangle (13,0);
\end{scope}
\begin{scope}[canvas is yz plane at x=13]
    \filldraw[fill=lightblue,fill opacity=0.5] (0,-2) rectangle (10,0);
\end{scope}
\begin{scope}[canvas is xy plane at z=0]
    \filldraw[fill=lightblue,fill opacity=0.5] (-1,0) rectangle (13,10);
\end{scope}
\draw[dashed] (-1, 0,-2) -- ( -1, 10,-2) -- (-1,10, 0);
\draw[dashed] (-1,10,-2) -- ( 13, 10,-2);

\foreach \i in {4,...,0}{
    \foreach \j in {5,...,0}{
        \begin{scope}[shift={(3*\i,2*\j,0)}]
        \draw (-0.5, 0, 0) -- (-0.5, 0, 1);
        \draw ( 0.5, 0, 0) -- ( 0.5, 0, 1);
        \end{scope}
    }
}
\foreach \i in {4,...,0}{
    \foreach \j in {5,...,0}{
        \begin{scope}[shift={(3*\i,2*\j,0)}]
        \begin{scope}[canvas is xy plane at z=1]
            \filldraw[fill=gray] (-1,-0.5) rectangle (1,0.5);
            \draw ( 0, 0.5) -- ( 0,   1);
            \draw ( 0,-0.5) -- ( 0,  -1);
            \draw (-1, 0)   -- (-1.5, 0);
            \draw ( 1, 0)   -- ( 1.5, 0);
        \end{scope}
        \end{scope}
    }
}
\end{tikzpicture}
\caption{A visualization of the tensor network representing $E(\Omega_R,t)$ for a two-site operator $\Omega_R$, i.e., $R = \{ x, y \}$ with $x, y \in \Z^2$ (black dots). The boundary conditions are periodic and the side lengths tend towards infinity. To avoid cluttering, the overall time evolution effected by ternary unitary operators is represented as cuboids. $\mathbb{U}^t$ is the red and $\mathbb{U}^{-t}$ the blue tensor.}
\label{fig:sPEPS-set-up}
\end{figure}

A consequence of this theorem is that we can restrict the following discussion to PEPS that consist of local tensors that generate an MPU. We want to study their behavior under the time evolution in Eq.~\eqref{eq:time_step}, i.e., layers of ternary unitary gates. We choose the position of our local tensors $\Lambda$ such that the first four-particle ternary unitary gates act on plaquettes where each site is described by a different tensor $\Lambda$, see Fig.~\ref{fig:speps_first_layer} for reference. We can use our assumptions to simplify the tensor network in Fig.~\ref{fig:sPEPS-set-up} that represents the dynamics of an expectation value of a local operator $\Omega_R$ for a bounded $R \subset \Z^2$ in the thermodynamic limit:
\begin{equation}\label{eq:2D-expval}
\begin{split}
E(\Omega_R, t) &= \lim_{L_2 \rightarrow \infty} \lim_{L_1 \rightarrow \infty} \bra{\Psi_{L_1 L_2} (t)} \Omega_R \ket{\Psi_{L_1 L_2} (t)} \\
 &= \lim_{L_2 \rightarrow \infty} \lim_{L_1 \rightarrow \infty} \bra{\Psi_{L_1 L_2}} \mathbb{U}^{-t} \Omega_R \mathbb{U}^t \ket{\Psi_{L_1 L_2}},
\end{split}
\end{equation}
where the explicit dependence on $\Lambda$ was left out.
If $\Omega_R$ is a single-site operator, we can use a graphical proof to see that $E(\Omega_R,t) = \Tr[\Omega_R]$ for all $t$. Therefore we choose $\Omega_R$ to be a two-site operator to investigate further. Assume $\Omega_R$ is composed of two single site operators $a_x^\alpha, a_y^\beta$ that are taken from the same Hilbert-Schmidt-orthonormal set $\{ a^\alpha\}_{\alpha =0}^{d^2-1}$ as before. An immediate consequence of the state normalization is $E(a_x^0, a_y^0, t) = 1$. Therefore and as we already know the result for single-site operators, we may assume $\alpha,\beta \neq 0$. We further define $r = x - y$ and assume $r_1,r_2 \geq 0$. Another quantity, which the final result depends on, is
\begin{equation}\label{eq:Delta}
\Delta = \Delta(x,y) = \min_{\substack{\tilde{x} \in p_x\\ \tilde{y} \in p_y}} \abs{\tilde{x}_2 - \tilde{y}_2} ,
\end{equation}
where $p_z := p(2 \lceil \frac{z_1}{2} \rceil-1, 2\lceil \frac{z_2}{2} \rceil -1)$ is a plaquette of four sites. For convenience of representation, we assume further that our local tensors are actually \emph{simple} tensors. This means that there exist additional tensors $\lambda, \gamma$ such that \cite{Cirac2017}
\begin{align}
\begin{tikzpicture}[baseline=(current bounding box.center), scale=0.5]
\foreach \i in {0}{
    \begin{scope}[shift={(2*\i,0,0)}]
    \filldraw[fill=lightgray] (-0.5,-1.5) rectangle (0.5,-0.5);
    \filldraw[fill=lightgray] (-0.5, 1.5) rectangle (0.5, 0.5);
    \draw (0,-0.5) -- (0,0.5);
    \draw (0,-1.5) -- (0,-2);
    \draw (0,1.5) -- (0,2);
    \draw (-0.5,-1) -- (-1,-1);
    \draw (0.5,-1) -- (1,-1);
    \draw (-0.5,1) -- (-1,1);
    \draw (0.5,1) -- (1,1);
    \node[anchor=south] at (0,-1.5){$\Lambda$};
    \node[anchor=south] at (0,0.5){$\Lambda^*$};
    \end{scope}
}
\draw (1,1) -- (1,-1);
\draw (-1,1) -- (-1,-1);
\filldraw (1,0) circle (0.15);
\node[anchor=west] at (1.2,0){$\gamma$};
\filldraw (-1,0) circle (0.15);
\node[anchor=east] at (-1.2,0){$\lambda$};
\end{tikzpicture}
& = \, \,
\begin{tikzpicture}[baseline=(current bounding box.center), scale=0.5]
\draw (0,-2) -- (0,2);
\end{tikzpicture} \qquad \text{and} \label{eq:simple_prop_1} \\
\begin{tikzpicture}[baseline=(current bounding box.center), scale=0.5]
\foreach \i in {0,1}{
    \begin{scope}[shift={(2*\i,0,0)}]
    \filldraw[fill=lightgray] (-0.5,-1.5) rectangle (0.5,-0.5);
    \filldraw[fill=lightgray] (-0.5, 1.5) rectangle (0.5, 0.5);
    \draw (0,-0.5) -- (0,0.5);
    \draw (0,-1.5) -- (0,-2);
    \draw (0,1.5) -- (0,2);
    \draw (-0.5,-1) -- (-1,-1);
    \draw (0.5,-1) -- (1,-1);
    \draw (-0.5,1) -- (-1,1);
    \draw (0.5,1) -- (1,1);
    \node[anchor=south] at (0,-1.5){$\Lambda$};
    \node[anchor=south] at (0,0.5){$\Lambda^*$};
    \end{scope}
}
\end{tikzpicture}
& =
\begin{tikzpicture}[baseline=(current bounding box.center), scale=0.5]
\foreach \i in {0,1.5}{
    \begin{scope}[shift={(2*\i,0,0)}]
    \filldraw[fill=lightgray] (-0.5,-1.5) rectangle (0.5,-0.5);
    \filldraw[fill=lightgray] (-0.5,1.5) rectangle (0.5,0.5);
    \draw (0,-0.5) -- (0,0.5);
    \draw (0,-1.5) -- (0,-2);
    \draw (0,1.5) -- (0,2);
    \draw (-0.5,-1) -- (-1,-1);
    \draw (0.5,-1) -- (1,-1);
    \draw (-0.5,1) -- (-1,1);
    \draw (0.5,1) -- (1,1);
    \node[anchor=south] at (0,-1.5){$\Lambda$};
    \node[anchor=south] at (0, 0.5){$\Lambda^*$};
    \end{scope}
}
\draw (1,1) -- (1,-1);
\draw (2,1) -- (2,-1);
\filldraw (1,0) circle (0.15);
\node[anchor=west] at (0.1,0){$\gamma$};
\filldraw (2,0) circle (0.15);
\node[anchor=east] at (2.9,0){$\lambda$};
\end{tikzpicture} \label{eq:simple_prop_2}
\end{align}
We will now jump to the final result, for details on the derivation refer to the supplemental material. Note that the proof can easily be extended to general MPUs, since after blocking a finite number of them, the blocked tensor will be simple \cite{Cirac2017}. According to our proof find that
\begin{multline}
E\!\left(a_x^\alpha, a_y^\beta,t \right) = \delta_{x_2 \text{mod}2,0}\delta_{r_1\text{mod}2,1} \Theta \left(r_1 -4t-1\right) \cdot \\
\Theta \left(4t + \delta L_2 -\Delta \right) \tilde{E}\!\left(a_x^\alpha, a_y^\beta,t \right),
\end{multline}
where $\Theta$ denotes the Heaviside-function. To write a comprehensible expression for non-zero $\tilde{E}$ we need to define a couple of new operators. The first two are defined on a $2\times 2F$ lattice for $F\in \mathbb{N}_1$. To shorten the expressions, we will use the notation
\begin{multline}
[(a,b),(c,d)] := \\ \{a,a+1, \dots, b-1, b \} \times \{c, c+1, \dots, d-1, d\}
\end{multline}
and shorten this to $[a,(c,d)]$ for $a=b$, analogously for $c=d$. As an initial condition define $\mathfrak{M}_{F,\nu}$ for $F=1$ as
\begin{multline}
\mathfrak{M}_{1,\nu}(a) = \frac{1}{d^2} \Tr_{[\bar{\nu},(0,1)]} \bigg[U^\dagger_{p(0,0)} a_{[\nu,(0,1)]} U_{p(0,0)} \bigg] ,
\end{multline}
where $\nu \in \{0,1\}$ and $\bar{\nu} = (\nu +1)\text{ mod }2$. As a tensor network diagram this becomes
\begin{equation}
\begin{tikzpicture}[baseline=(current bounding box.center), scale=0.375]
\pgfsetxvec{\pgfpoint{1cm}{0cm}}
\pgfsetyvec{\pgfpoint{0.4cm}{0.3cm}}
\pgfsetzvec{\pgfpoint{0cm}{1cm}}
\node[anchor=east] at (-2,0,0){$\mathfrak{M}_{1,1} (a) = \frac{1}{d^2}$};
\foreach \i in {0}{
    \begin{scope}[shift={(0,4*\i,-2)}]
        \draw (-1,-1,-1) -- (-1,-1,-2);
        \draw (-1,1,-1) -- (-1,1,-2);
        \draw (1,-1,-1) -- (1,-1,-2) -- (1.5,-1,-1.5);
        \draw (1,1,-1) -- (1,1,-2) -- (1.5,1,-1.5);
        \drawcube{2}{lightred};
        \draw (1,-1,1) -- (1,-1,2);
        \draw (1,1,1) -- (1,1,2);
        \draw (-1,-1,1) -- (-1,-1,2);
        \draw (-1,1,1) -- (-1,1,2);
    \end{scope}
}
\foreach \i in {0}{
    \begin{scope}[shift={(0,4*\i,2)}]
        \draw (1,-1,-1) -- (1,-1,-2);
        \draw (1,1,-1) -- (1,1,-2);
        \draw (-1,-1,-1) -- (-1,-1,-2);
        \draw (-1,1,-1) -- (-1,1,-2);
        \drawcube{2}{lightblue};
        \draw (-1,-1,1) -- (-1,-1,2);
        \draw (-1,1,1) -- (-1,1,2);
        \draw (1,-1,1) -- (1,-1,2) -- (1.5,-1,1.5);
        \draw (1,1,1) -- (1,1,2) -- (1.5,1,1.5);
    \end{scope}
}
\begin{scope}[canvas is yz plane at x=1]
    \filldraw (-1,-0.25) rectangle (1,0.25);
    \node[anchor=west] at (1,0){$a$};
\end{scope}
\end{tikzpicture} .
\end{equation}
For $F>1$ define
\begin{equation}
\mathfrak{M}_{F,\nu} (a) = \bigotimes_{i=0}^{F-1} \mathfrak{M}_{1,\nu} \left( a \big|_{[\nu,(2i,2i+1)]} \right) .
\end{equation}
As a tensor network diagram this looks like
\begin{equation}
\begin{tikzpicture}[baseline=(current bounding box.center), scale=0.375]
\pgfsetxvec{\pgfpoint{1cm}{0cm}}
\pgfsetyvec{\pgfpoint{0.4cm}{0.3cm}}
\pgfsetzvec{\pgfpoint{0cm}{1cm}}
\node[anchor=east] at (-2,0,0){$\mathfrak{M}_{3,1} (a) = \frac{1}{d^6}$};
\foreach \i in {2,1,0}{
    \begin{scope}[shift={(0,4*\i,-2)}]
        \draw (-1,-1,-1) -- (-1,-1,-2);
        \draw (-1,1,-1) -- (-1,1,-2);
        \draw (1,-1,-1) -- (1,-1,-2) -- (1.5,-1,-1.5);
        \draw (1,1,-1) -- (1,1,-2) -- (1.5,1,-1.5);
        \drawcube{2}{lightred};
        \draw (1,-1,1) -- (1,-1,2);
        \draw (1,1,1) -- (1,1,2);
        \draw (-1,-1,1) -- (-1,-1,2);
        \draw (-1,1,1) -- (-1,1,2);
    \end{scope}
}
\foreach \i in {2,1,0}{
    \begin{scope}[shift={(0,4*\i,2)}]
        \draw (1,-1,-1) -- (1,-1,-2);
        \draw (1,1,-1) -- (1,1,-2);
        \draw (-1,-1,-1) -- (-1,-1,-2);
        \draw (-1,1,-1) -- (-1,1,-2);
        \drawcube{2}{lightblue};
        \draw (-1,-1,1) -- (-1,-1,2);
        \draw (-1,1,1) -- (-1,1,2);
        \draw (1,-1,1) -- (1,-1,2) -- (1.5,-1,1.5);
        \draw (1,1,1) -- (1,1,2) -- (1.5,1,1.5);
    \end{scope}
}
\begin{scope}[canvas is yz plane at x=1]
    \filldraw (-1,-0.4) rectangle (9,0.4);
    \node[anchor=west] at (9,0){$a$};
\end{scope}
\end{tikzpicture} .
\end{equation}
Using these two definitions we write the ``tilted triangles'' of operators $U$ as
\begin{equation}
\mathcal{D}(a,t,\nu) = \comp_{F=1}^{2t} \mathfrak{M}_{F,\nu} \left( a \right),
\end{equation}
where $\circ$ denotes the composition of maps. Finally we define on an $2\times F$ grid for $F\geq 1$ the operator
\begin{multline}
P_F (a) = \Tr_\eta \Bigg[ \lambda \cdot \left( \prod_{i=0}^{F-1} \Lambda^*_{[(0,1),i]} \right) \cdot \gamma \, \maltese\\
 a_{[(0,1),(0,F-1)]} \, \maltese \prod_{i=0}^{F-1} \Lambda_{[(0,1),i]} \Bigg],
\end{multline}
where the conventional product and $\maltese$ denote the contraction of the $\eta$-legs and physical legs, respectively, and $\Tr_\eta$ symbolizes the contraction of the open legs of $\lambda$ and $\gamma$ with the remaining open $\eta$-legs of the outermost $\Lambda$. We can once more look at the corresponding tensor network diagram:
\begin{equation}
\begin{tikzpicture}[baseline=(current bounding box.center), scale=0.375]
\pgfsetxvec{\pgfpoint{1cm}{0cm}}
\pgfsetyvec{\pgfpoint{0.4cm}{0.3cm}}
\pgfsetzvec{\pgfpoint{0cm}{1cm}}
\node[anchor=east] at (-2,0,0){$P_F(a)=$};
\begin{scope}[canvas is yz plane at x=0]
    \draw (7,-4) to[out=30 ,  in=-30] (7,4);
\end{scope}
\begin{scope}[canvas is xz plane at y=9]
 \filldraw (0,0) circle (0.2);
\end{scope}
\node[anchor=west] at (0,9,0){$\lambda$};
\foreach \i in {3,2,1,0}{
    \begin{scope}[shift={(0,2*\i,-4)}]
        \begin{scope}[canvas is xy plane at z=0]
            \filldraw[fill=lightgray, fill opacity=0.5] (-1,-0.5) rectangle (1,0.5);
            \draw (-1,0) -- (-1.5,0);
            \draw (1,0) -- (1.5,0);
            \draw (0,-0.5) -- (0,-1);
            \draw (0,0.5) -- (0,1);
        \end{scope}
        \draw (-0.5,0,0) -- (-0.5,0,4);
        \draw (0.5,0,0) -- (0.5,0,4);
    \end{scope}
    \begin{scope}[shift={(0,2*\i,4)}]
        \begin{scope}[canvas is xy plane at z=0]
            \filldraw[fill=gray] (-1,-0.5) rectangle (1,0.5);
            \draw (-1,0) -- (-1.5,0);
            \draw (1,0) -- (1.5,0);
            \draw (0,-0.5) -- (0,-1);
            \draw (0,0.5) -- (0,1);
        \end{scope}
        \draw (-0.5,0,0) -- (-0.5,0,-4);
        \draw (0.5,0,0) -- (0.5,0,-4);
    \end{scope}
}
\begin{scope}[canvas is yz plane at x=0.5]
    \filldraw[fill opacity = 0.25] (0,-0.5) rectangle (6,0.5);
\end{scope}
\begin{scope}[canvas is xy plane at z=0.5]
    \filldraw[fill opacity = 0.25] (-0.5,0) rectangle (0.5,6);
\end{scope}
\begin{scope}[canvas is xz plane at y=0]
    \filldraw[fill opacity = 0.25] (-0.5,-0.5) rectangle (0.5,0.5);
\end{scope}
\draw[dashed] (-0.5,0,-0.5) -- (-0.5,6,-0.5) -- (0.5,6,-0.5);
\draw[dashed] (-0.5,6,-0.5) -- (-0.5,6,0.5);
\node[anchor=south] at (0.12,6,0.5){$a$};
\begin{scope}[canvas is yz plane at x=0]
    \draw (-1,-4) to[out=120 ,  in=-120] (-1,4);
\end{scope}
\begin{scope}[canvas is xz plane at y=-2.2]
    \filldraw (0,0) circle (0.2);
\end{scope}
\node[anchor=east] at (0,-2.2,0){$\gamma$};
\end{tikzpicture} .
\end{equation}
Further we denote $P_F = P_F (\mathbbm{1})$ and define $\ell_1 = \frac{r_1-4t+1}{2}$, $\delta \ell_2 = \left\lceil \frac{\ell_1}{2} \right\rceil$, and $\nu \in \{1,2\}$ as the parity of $\ell_1$. Using these definitions we can express the remaining local tensors to which no unitary operator is applied to as
\begin{multline}
\mathcal{S}(\ell_1) = \left[ \prod_{F=\delta \ell_2}^1 P_{2t +F}\right]
\left[ P_{2t+\delta \ell_2} \right]^\nu \left[ \prod_1^{F=\delta \ell_2 -1} P_{2t +F} \right]
\end{multline}
and finally write for $\ell_1 > 2$
\begin{multline}\label{eq:res_zero}
\tilde{E}\!\left( a_x^\alpha, a_y^\beta, t \right) =\\
\Tr_{\mu} \Bigg[ P_{2t+2} \left( \mathcal{D}\left(a^\beta_y , t, 0 \right) \right) \mathcal{S}(\ell_1) \,  P_{2t+2} \left( \mathcal{D}\left(a^\alpha_x , t, 1 \right) \right)\Bigg],
\end{multline}
where $\Tr_\mu$ connects the open $\mu$-legs of a local tensor $\Lambda$ to the corresponding leg of the adjoint. Additionally we get 
\begin{multline}\label{eq:res_zero}
\tilde{E}\!\left( a_x^\alpha, a_y^\beta, t \right) =\\
\Tr_{\mu} \Bigg[ P_{2t+2} \left( \mathcal{D}\left(a^\beta_y , t, 0 \right) \right) \cdot  P_{2t+2} \left( \mathcal{D}\left(a^\alpha_x , t, 1 \right) \right)\Bigg]
\end{multline}
for $\ell_1 = 2$ and
\begin{multline}\label{eq:res_zero}
\tilde{E}\!\left( a_x^\alpha, a_y^\beta, t \right) =\\
\Tr_{\mu} \Bigg[ P_{2t+2} \left( \mathcal{D}\left(a^\beta_y , t, 0 \right) \otimes \mathcal{D}\left(a^\alpha_x , t, 1 \right) \right)\Bigg]
\end{multline}
for $\ell_1=1$. The final result for $\Delta=0$, $r_1=7$ and $t=1$ in the form of a tensor network diagram is given in Fig.~\ref{fig:tilted_final}.

\begin{figure}
\begin{tikzpicture}[baseline=(current bounding box.center), scale=0.375]
\pgfsetxvec{\pgfpoint{1cm}{0cm}}
\pgfsetyvec{\pgfpoint{0.4cm}{0.3cm}}
\pgfsetzvec{\pgfpoint{0cm}{1cm}}

\begin{scope}[canvas is yz plane at x=-1.5]
    \draw (11,-7) to[out=30 ,  in=-30] (11,7);
\end{scope}
\begin{scope}[canvas is xz plane at y=14.5]
 \filldraw (-1.5,0) circle (0.15);
\end{scope}
\node[anchor=east] at (-1.4,14.5,0){$\lambda$};

\begin{scope}[canvas is yz plane at x=1.5]
    \draw (11,-7) to[out=30 ,  in=-30] (11,7);
\end{scope}
\begin{scope}[canvas is xz plane at y=14.5]
 \filldraw (1.5,0) circle (0.15);
\end{scope}
\node[anchor=east] at (1.6,14.5,0){$\lambda$};

\foreach \i in {0,...,5}{
    \begin{scope}[shift={(0,2*\i,-7)}]
        \begin{scope}[canvas is xy plane at z=0]
            \filldraw[fill=lightgray, fill opacity=0.5] (-2.5,-0.5) rectangle (-0.5,0.5);
            \filldraw[fill=lightgray, fill opacity=0.5] (2.5,-0.5) rectangle (0.5,0.5);
            \draw (-2.5,0) -- (-3,0);
            \draw (-0.5,0) -- (0.5,0);
            \draw (2.5,0) -- (3,0);
            \draw (-1.5,-0.5) -- (-1.5,-1);
            \draw (-1.5,0.5) -- (-1.5,1);
            \draw (1.5,-0.5) -- (1.5,-1);
            \draw (1.5,0.5) -- (1.5,1);
        \end{scope}
    \end{scope}
}
\foreach \i in {7,3}{
    \begin{scope}[shift={(0,\i,-5)}]
        \begin{scope}[shift={(-3,0,0)}]
            \draw (1,-1,-1) -- (1,-1,-2);
            \draw (1,1,-1) -- (1,1,-2);
            \drawcube{2}{lightred};
        \end{scope}
        \begin{scope}[shift={(3,0,0)}]
            \draw (-1,-1,-1) -- (-1,-1,-2);
            \draw (-1,1,-1) -- (-1,1,-2);
            \drawcube{2}{lightred};
        \end{scope}
    \end{scope}
}
\begin{scope}[shift={(0,5,-2)}]
    \begin{scope}[shift={(-5,0,0)}]
        \draw (1,-1,-1) -- (1,-1,-2);
        \draw (1,1,-1) -- (1,1,-2);
        \drawcube{2}{lightred};
        \draw (-1,-1,1) -- (-1,-1,2);
        \draw (-1,1,1) -- (-1,1,2);
    \end{scope}
    \begin{scope}[shift={(5,0,0)}]
        \draw (-1,-1,-1) -- (-1,-1,-2);
        \draw (-1,1,-1) -- (-1,1,-2);
        \drawcube{2}{lightred};
        \draw (1,-1,1) -- (1,-1,2);
        \draw (1,1,1) -- (1,1,2);
    \end{scope}
\end{scope}

\begin{scope}[shift={(0,5,2)}]
    \begin{scope}[shift={(-5,0,0)}]
        \draw (-1,-1,-1) -- (-1,-1,-2);
        \draw (-1,1,-1) -- (-1,1,-2);
        \drawcube{2}{lightblue};
        \draw (1,-1,1) -- (1,-1,2);
        \draw (1,1,1) -- (1,1,2);
    \end{scope}
    \begin{scope}[shift={(5,0,0)}]
        \draw (1,-1,-1) -- (1,-1,-2);
        \draw (1,1,-1) -- (1,1,-2);
        \drawcube{2}{lightblue};
        \draw (-1,-1,1) -- (-1,-1,2);
        \draw (-1,1,1) -- (-1,1,2);
    \end{scope}
\end{scope}
\foreach \i in {7,3}{
    \begin{scope}[shift={(0,\i,5)}]
        \begin{scope}[shift={(-3,0,0)}]
            \drawcube{2}{lightblue};
            \draw (1,-1,1) -- (1,-1,2);
            \draw (1,1,1) -- (1,1,2);
        \end{scope}
        \begin{scope}[shift={(3,0,0)}]
            \drawcube{2}{lightblue};
            \draw (-1,-1,1) -- (-1,-1,2);
            \draw (-1,1,1) -- (-1,1,2);
        \end{scope}
    \end{scope}
}
\foreach \i in {0,...,5}{
    \begin{scope}[shift={(0,2*\i,7)}]
        \begin{scope}[canvas is xy plane at z=0]
            \filldraw[fill=gray, fill opacity=1] (-2.5,-0.5) rectangle (-0.5,0.5);
            \filldraw[fill=gray, fill opacity=1] ( 2.5,-0.5) rectangle ( 0.5,0.5);
            \draw (-2.5,0) -- (-3,0);
            \draw (-0.5,0) -- (0.5,0);
            \draw (2.5,0) -- (3,0);
            \draw (-1.5,-0.5) -- (-1.5,-1);
            \draw (-1.5,0.5) -- (-1.5,1);
            \draw (1.5,-0.5) -- (1.5,-1);
            \draw (1.5,0.5) -- (1.5,1);
        \end{scope}
    \end{scope}
}
\begin{scope}[canvas is xz plane at y=6]
    \filldraw (6,0) circle (0.15);
    \node[anchor=west] at (6,0){$a_x^\alpha$};
\end{scope}
\begin{scope}[canvas is xz plane at y=4]
    \filldraw (-6,0) circle (0.15);
    \node[anchor=east] at (-6,0){$a_y^\beta$};
\end{scope}

\begin{scope}[canvas is yz plane at x=-1.5]
    \draw (-1,-7) to[out=120 ,  in=-120] (-1,7);
\end{scope}
\begin{scope}[canvas is xz plane at y=-2.8]
    \filldraw (-1.6,0) circle (0.15);
\end{scope}
\node[anchor=west] at (-1.5,-2.8,0){$\gamma$};

\begin{scope}[canvas is yz plane at x=1.5]
    \draw (-1,-7) to[out=120 ,  in=-120] (-1,7);
\end{scope}
\begin{scope}[canvas is xz plane at y=-2.8]
    \filldraw (1.4,0) circle (0.15);
\end{scope}
\node[anchor=west] at (1.5,-2.8,0){$\gamma$};
\end{tikzpicture}
\caption{The tensor network that remains after simplification for the example $\Delta=0$, $r_1=7$ and $t=1$. The legs of the unitaries as well as the physical and $\alpha$-legs of the $\Lambda$-tensors are contracted with the corresponding leg of their adjoint. The tilted triangles in this case each consist of three ternary unitary operators.}
\label{fig:tilted_final}
\end{figure}
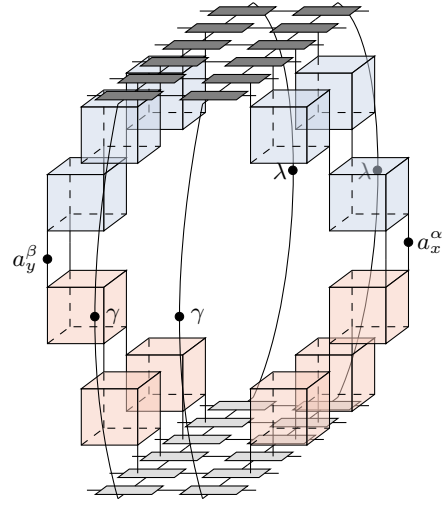

\paragraph{Numerical algorithm and simulations.}

The evaluation of equal-time correlation requires the contraction of tensor networks as visualized in Fig.~\ref{fig:tilted_final}. We split this task into two steps:
\begin{enumerate}
\item The conjugations by ternary unitary gates of the local operators $a_x^{\alpha}$ and $a_y^{\beta}$ are represented as two MPOs, respectively. Specifically, we start from a MPO representation of the identity with length equal to the extent in $x_2$-direction, substitute the local operators $a_x^{\alpha}$ and $a_y^{\beta}$, and then apply the maps $\mathfrak{M}_{1,\nu}$ in a TEBD-type brickwall pattern. Since the maps $\mathfrak{M}_{1,\nu}$ are contractive, one expects only a modest increase of entanglement. The two final MPOs are sandwiched between the PEPS tensors on the left and right boundary.
\item The PEPS tensors are first contracted along the physical legs with their conjugated copy, while inserting the MPOs from step 1 on the left and right boundary, and MPU ring transfer states on the top and bottom boundary. This leads to a two-dimensional grid of new tensors connected by virtual bonds in each of the four spatial directions. Finally, this network is then contracted row by row or column by column.
\end{enumerate}
We remark that the last substep is affected by the ``curse of dimensionality'' for increasing distance $\abs{x_1 - y_1}$.

\begin{figure}[!ht]
\centering
\includegraphics[width=0.65\columnwidth]{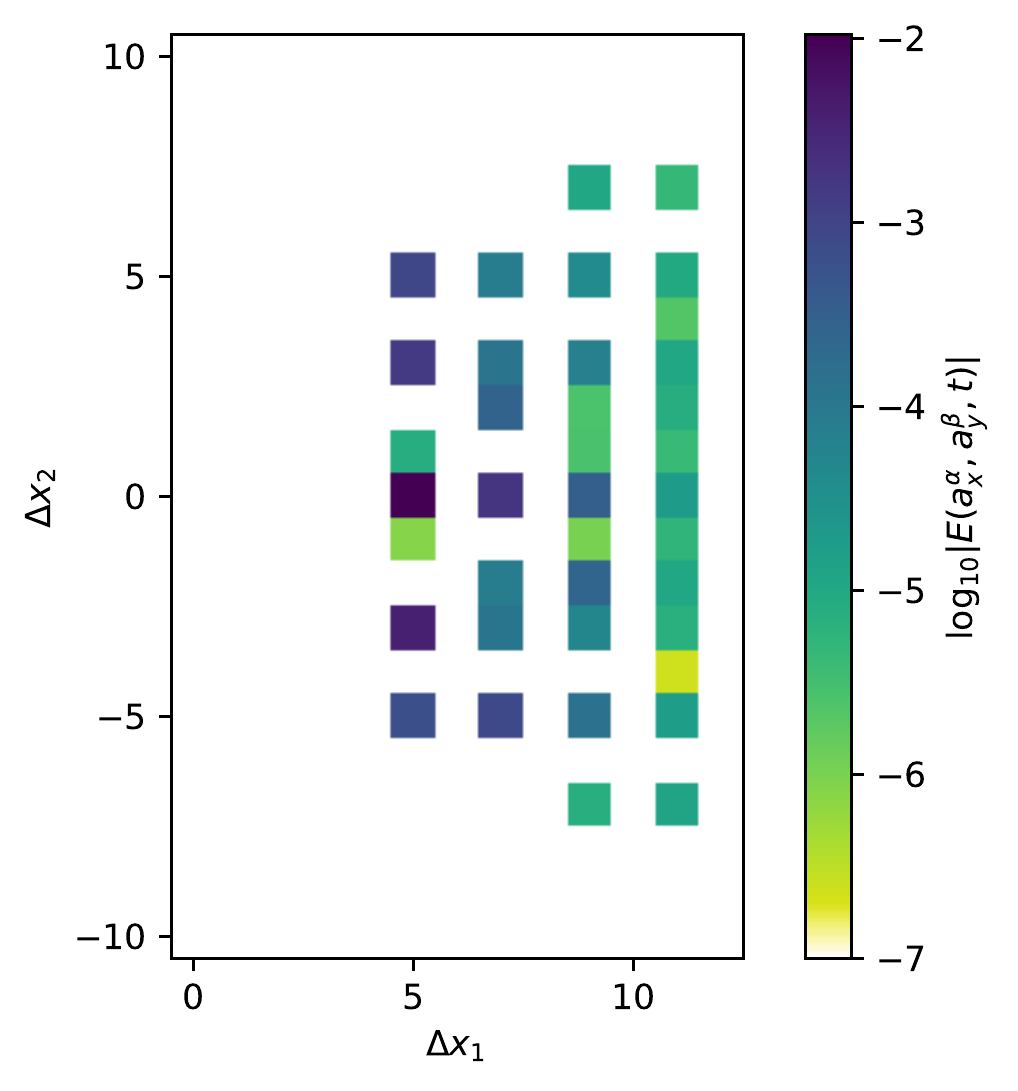}
\caption{Equal-time correlation functions for $t = 1$ and a random simple MPU tensor $\Lambda$ defining the sPEPS, a uniform ternary unitary gate of the form \eqref{eq:four_dual_construction}, and random traceless local operators $a_x^{\alpha}$ and $a_y^{\beta}$.}
\label{fig:equal_time_corr_t1}
\end{figure}

Fig.~\ref{fig:equal_time_corr_t1} visualizes equal-time correlation functions for $t = 1$, obtained by contracting the tensor network architecture in Fig.~\ref{fig:tilted_final}. The correlation was taken as the average over two neighboring $x_2$-positions to account for even-odd-effects. The distance $x_1 - y_1$ has to be odd for the correlation function to be non-zero. Note that we used ternary unitaries in our numerical computations, even though the derivation to find a finite tensor network diagram of \eqref{eq:2D-expval} only requires the four-particle gates to be dual  unitary with respect to the $x_1$-direction. However, the use of ternary unitaries causes a further pattern for which $E( a^\alpha_x, a^\beta_y, t) =0$. A more detailed explanation of this can be found in the supplemental material. In general the computed values fit the theoretical considerations, where the correlation reduces asymptotically for larger distances and is always zero if $\Delta x_2$ is above a certain threshold with respect to $\Delta x_1$.

\begin{figure}[!ht]
\centering
\includegraphics[width=0.8\columnwidth]{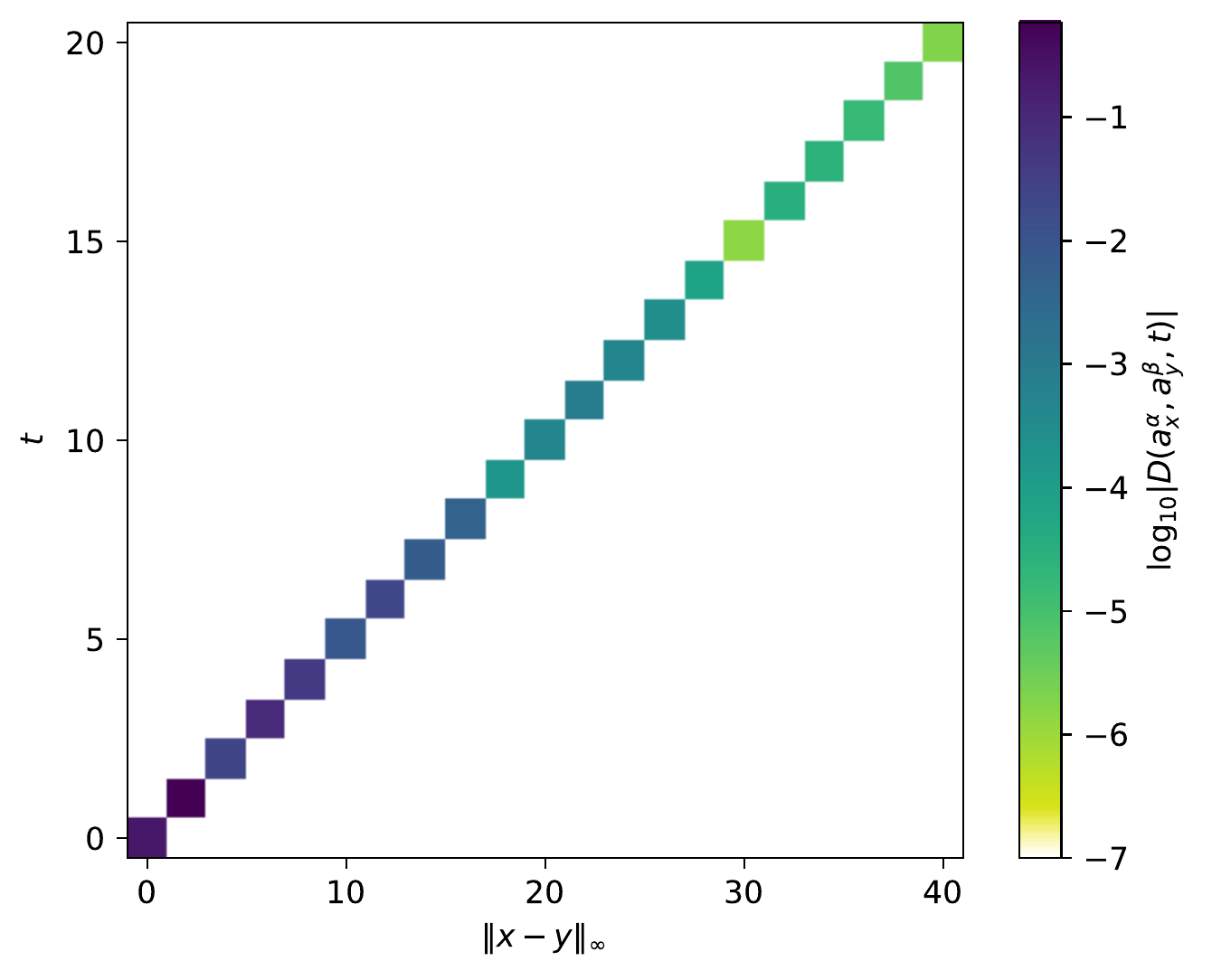}
\caption{The dynamical correlation function for a uniform ternary unitary gate of the form \eqref{eq:four_dual_construction}, and random traceless local operators $a_x^{\alpha}$ and $a_y^{\beta}$.}
\label{fig:dynamical_corr_random}
\end{figure}

We also computed the dynamical correlation function \eqref{eq:corr_function} by interpreting the network \eqref{eq:corr_function_ray_diagram} as the expectation value of a $M^n$ with respect to the vectors $a_x^\alpha$ and $a_y^\beta$ for the non-zero values. The results can be seen in Fig.~\ref{fig:dynamical_corr_random}. While the correlation does not fall monotonously with time (or distance), it does so asymptotically. This fits with the results one would expect according to \eqref{eq:corr_function_eigenvalues} and the results obtained in \cite{Claeys2021} for dual unitary operators.

\paragraph{Discussion and outlook.}
First note that the results obtained for the expectation value dynamics is a generalization of the results found in \cite{Suzuki2022}. They confined themselves to solvable states with $\chi_2 = 1$, which admits non-zero $E$ only in the case of $r_2=0$. Since the MPU are a well known object, there are many properties that might be used to analyse the solvable PEPS further. However, one might not consider the MPU the best Ansatz for a solvable state, since the treatment of the two spatial dimensions is highly asymmetric. Notably the limits in \eqref{eq:2D-expval} do not commute and in our derivation it is sufficient to assume the operators $U$ to be dual unitary in $x_1$-direction. Apart from the solvable states, a new class of four-particle gates called ternary unitary gates was introduced in this work. While we gave two possible ways to construct them from dual unitaries, it would be interesting to find a more general parametrization of them. Notably not only the physically relevant dual unitaries are significant, but the methods used to find them yielded the solution to a famous open problem of quantum information \cite{Rather2022, Zyczkowski2022}. Therefore one could hope to find interesting results while further analyzing the class of $(2+1)$-dimensional ternary unitaries. Other aspects to be explored could take inspiration from the works of recent years considering dual unitaries \cite{Zhou2022, Rather2020, Kos2021, Claeys2021}. We expect that two-spatial dimensions admit new and interesting phenomena but also give rise to different challenges.
\newline
\acknowledgments
We thank Frank Pollmann and Toby Cubitt for inspiring discussions. The research is part of the Munich Quantum Valley, which is supported by the Bavarian state government with funds from the Hightech Agenda Bayern Plus.

\bibliography{references}

\appendix

\newpage

{\Large Supplementary Information}

\section{The spatial matrix product}
In the main part we defined two new matrix products $\circ_1$ and $\circ_2$ to be matrix multiplication along the $x_1$- and $x_2$-direction, respectively. In this part, we want to give a proper definition and show how this definition relates to the unitary condition \eqref{eq:unitary_x1}. As a reference, note that the usual matrix product along the time-direction is graphically represented as
\begin{equation}
\begin{tikzpicture}[>=stealth, scale=0.75]
\pgfsetxvec{\pgfpoint{1cm}{0cm}}
\pgfsetyvec{\pgfpoint{0.4cm}{0.3cm}}
\pgfsetzvec{\pgfpoint{0cm}{1cm}}
\node[anchor=east] at ( -1, 0, 0) {$U^\dagger U \, =$};
\begin{scope}[shift={(0, 0, 0)}]
\begin{scope}[shift={( 0, 0,-0.85)}]
\drawcube{1}{lightred};
\draw (-0.5,-0.5,-0.5) -- (-0.7,-0.7,-0.7);
\draw ( 0.5,-0.5,-0.5) -- ( 0.7,-0.7,-0.7);
\draw (-0.5, 0.5,-0.5) -- (-0.7, 0.7,-0.7);
\draw ( 0.5, 0.5,-0.5) -- ( 0.7, 0.7,-0.7);
\end{scope}
\begin{scope}[shift={( 0, 0, 0.85)}]
\drawcube{1}{lightblue};
\draw (-0.5,-0.5, 0.5) -- (-0.7,-0.7, 0.7);
\draw ( 0.5,-0.5, 0.5) -- ( 0.7,-0.7, 0.7);
\draw (-0.5, 0.5, 0.5) -- (-0.7, 0.7, 0.7);
\draw ( 0.5, 0.5, 0.5) -- ( 0.7, 0.7, 0.7);
\end{scope}
% connecting legs
\begin{scope}[canvas is plane={O(-0.5,-0.5,-0.35)x(-0.7071,-0.7071,0)y( 0, 0, 1)}]
\draw ( 0, 0) to [out=45, in=-45] ( 0, 0.7);
\end{scope}
\begin{scope}[canvas is plane={O( 0.5,-0.5,-0.35)x( 0.7071,-0.7071,0)y( 0, 0, 1)}]
\draw ( 0, 0) to [out=45, in=-45] ( 0, 0.7);
\end{scope}
\begin{scope}[canvas is plane={O(-0.5, 0.5,-0.35)x(-0.7071, 0.7071,0)y( 0, 0, 1)}]
\draw ( 0, 0) to [out=45, in=-45] ( 0, 0.7);
\end{scope}
\begin{scope}[canvas is plane={O( 0.5, 0.5,-0.35)x( 0.7071, 0.7071,0)y( 0, 0, 1)}]
\draw ( 0, 0) to [out=45, in=-45] ( 0, 0.7);
\end{scope}
\end{scope}
\end{tikzpicture} .
\end{equation}
This is equivalent to viewing $U$ as a matrix with elements
\begin{equation}
U^{a_1 a_2 a_3 a_4}_{a_5 a_6 a_7 a_8},
\end{equation}
where the number on the index corresponds to the enumeration of the legs in Eq.~\eqref{eq:IndexConvention}. The upper indices form the combined row index of the matrix and the lower ones the column index. The multiplication along the $x_1$-direction is represented as
\begin{equation}\label{eq:SpatialMatrixProduct}
\begin{tikzpicture}[>=stealth, scale=0.75, baseline=(current bounding box.center)]
\pgfsetxvec{\pgfpoint{1cm}{0cm}}
\pgfsetyvec{\pgfpoint{0.4cm}{0.3cm}}
\pgfsetzvec{\pgfpoint{0cm}{1cm}}
\node[anchor=east] at ( -5.5, 0, 0) {$U^{\dagger_1} \circ_1 U \, = $};
\begin{scope}[shift={(-3.5,0,0)}]
\begin{scope}[shift={(-0.85,0,0)}]
\drawcube{1}{lightred}
\draw (-0.5,-0.5,-0.5) -- (-0.7,-0.7,-0.7);
\draw (-0.5, 0.5,-0.5) -- (-0.7, 0.7,-0.7);
\draw (-0.5,-0.5, 0.5) -- (-0.7,-0.7, 0.7);
\draw (-0.5, 0.5, 0.5) -- (-0.7, 0.7, 0.7);
\end{scope}
\begin{scope}[shift={(0.85,0,0)}]
\drawcube{1}{lightgreen}
\draw ( 0.5,-0.5,-0.5) -- ( 0.7,-0.7,-0.7);
\draw ( 0.5, 0.5,-0.5) -- ( 0.7, 0.7,-0.7);
\draw ( 0.5,-0.5, 0.5) -- ( 0.7,-0.7, 0.7);
\draw ( 0.5, 0.5, 0.5) -- ( 0.7, 0.7, 0.7);
\end{scope}
% connecting legs
\begin{scope}[canvas is plane={O(-0.35,-0.5,-0.5)x( 0,-0.7071,-0.7071)y( 1, 0, 0)}]
\draw ( 0, 0) to [out=45, in=-45] ( 0, 0.7);
\end{scope}
\begin{scope}[canvas is plane={O(-0.35,0.5,-0.5)x( 0,0.7071,-0.7071)y( 1, 0, 0)}]
\draw ( 0, 0) to [out=45, in=-45] ( 0, 0.7);
\end{scope}
\begin{scope}[canvas is plane={O(-0.35,0.5,0.5)x( 0,0.7071,0.7071)y( 1, 0, 0)}]
\draw ( 0, 0) to [out=45, in=-45] ( 0, 0.7);
\end{scope}
\begin{scope}[canvas is plane={O(-0.35,-0.5,0.5)x( 0,-0.7071,0.7071)y( 1, 0, 0)}]
\draw ( 0, 0) to [out=-45, in=-45] ( 0, 0.7);
\end{scope}
\end{scope}
\end{tikzpicture} ,
\end{equation}
where we view $U$ as a matrix with elements
\begin{equation}
\tilde{U}^{a_2 a_4 a_6 a_8}_{a_1 a_3 a_5 a_7}  = U^{a_1 a_2 a_3 a_4}_{a_5 a_6 a_7 a_8}
\end{equation}
and $U^{\dagger_1}$ (green cube) is the complex conjugate of $U$ with respect to the new matrix product, i.e., the left indices $1, 3, 5, 7$ are exchanged with the right indices $2, 4, 6, 8$. Thus
\begin{equation}
\begin{split}
\left( U^{\dagger_1} \circ_1 U \right)_{a_1 a_3 a_5 a_7}^{b_1 b_3 b_5 b_7} &= \sum_{p q \ell k} \left( U ^{\dagger_1} \right)_{p q k \ell}^{b_1 b_3 b_5 b_7} \tilde{U}_{a_1 a_3 a_5 a_7}^{p q k \ell} \\
&= \sum_{p q k \ell} \left( \tilde{U}_{b_1 b_3 b_5 b_7}^{p q k \ell} \right)^* \tilde{U}_{a_1 a_3 a_5 a_7}^{p q k \ell} \\
&= \sum_{p q k \ell} \left( U_{b_5 k b_7 \ell}^{b_1 p b_3 q} \right)^* U_{a_5 k a_7 \ell}^{a_1 p a_3 q} \\
&= \sum_{p q k \ell} \left( U^\dagger \right)_{b_1 p b_3 q}^{b_5 k b_7 \ell} U_{a_5 k a_7 \ell}^{a_1 p a_3 q} .
\end{split}
\end{equation}
This corresponds to the graphical tensor diagram
\begin{equation}
\begin{tikzpicture}[>=stealth, scale=0.75, baseline=(current bounding box.center)]
\pgfsetxvec{\pgfpoint{1cm}{0cm}}
\pgfsetyvec{\pgfpoint{0.4cm}{0.3cm}}
\pgfsetzvec{\pgfpoint{0cm}{1cm}}
\begin{scope}[shift={(-4.5,0,0)}]
\begin{scope}[shift={(-0.85,0,0)}]
\drawcube{1}{lightred};
\draw ( -0.5,-0.5,-0.5) -- (-0.7,-0.7,-0.7);
\node[anchor=east] at (-0.7,-0.7,-0.7){\tiny $a_5$};
\draw ( -0.5,0.5,-0.5) -- ( -0.7,0.7,-0.7);
\node[anchor=east] at (-0.5,0.7,-0.6){\tiny $a_7$};
\draw ( -0.5,-0.5,0.5) -- ( -0.7,-0.7,0.7);
\node[anchor=east] at (-0.7,-0.7,0.7){\tiny $a_1$};
\draw ( -0.5,0.5,0.5) -- ( -0.7,0.7,0.7);
\node[anchor=east] at (-0.7,0.7,0.7){\tiny $a_3$};
\end{scope}
\begin{scope}[shift={(0.85,0,0)}]
\drawcube{1}{lightgreen};
\draw ( 0.5,-0.5,-0.5) -- ( 0.7,-0.7,-0.7);
\node[anchor=west] at ( 0.7,-0.7,-0.7){\tiny $b_5$};
\draw ( 0.5,0.5,-0.5) -- ( 0.7,0.7,-0.7);
\node[anchor=west] at ( 0.7,0.7,-0.7){\tiny $b_7$};
\draw ( 0.5,-0.5,0.5) -- ( 0.7,-0.7,0.7);
\node[anchor=west] at ( 0.6,-0.7,0.55){\tiny $b_1$};
\draw ( 0.5,0.5,0.5) -- ( 0.7,0.7,0.7);
\node[anchor=west] at ( 0.7,0.7,0.7){\tiny $b_3$};
\end{scope}
% connecting legs
\begin{scope}[canvas is plane={O(-0.35,-0.5,-0.5)x( 0,-0.7071,-0.7071)y( 1, 0, 0)}]
\draw ( 0, 0) to [out=45, in=-45] ( 0, 0.7);
\end{scope}
\begin{scope}[canvas is plane={O(-0.35,0.5,-0.5)x( 0,0.7071,-0.7071)y( 1, 0, 0)}]
\draw ( 0, 0) to [out=45, in=-45] ( 0, 0.7);
\end{scope}
\begin{scope}[canvas is plane={O(-0.35,0.5,0.5)x( 0,0.7071,0.7071)y( 1, 0, 0)}]
\draw ( 0, 0) to [out=45, in=-45] ( 0, 0.7);
\end{scope}
\begin{scope}[canvas is plane={O(-0.35,-0.5,0.5)x( 0,-0.7071,0.7071)y( 1, 0, 0)}]
\draw ( 0, 0) to [out=-45, in=-45] ( 0, 0.7);
\end{scope}
\node[anchor=north] at ( 0,   0.7, 1)  {\tiny $q$};
\node[anchor=north] at ( 0.3,-0.7, 1)  {\tiny $p$};
\node[anchor=north] at (-0.3, 0.7,-0.5){\tiny $\ell$};
\node[anchor=north] at ( 0,  -0.7,-0.5){\tiny $k$};
\end{scope}

\node[anchor = east] at (-1.8,0,0){$=$};

\begin{scope}[shift={(0, 0, 0)}]
\begin{scope}[shift={( 0, 0,-0.85)}]
\drawcube{1}{lightred};
\draw ( -0.5,-0.5,-0.5) -- (-0.7,-0.7,-0.7);
\node[anchor=east] at (-0.7,-0.7,-0.7){\tiny $a_5$};
\draw ( -0.5,0.5,-0.5) -- ( -0.7,0.7,-0.7);
\node[anchor=east] at (-0.5,0.7,-0.6){\tiny $a_7$};
\draw ( -0.5,-0.5,0.5) -- ( -0.7,-0.7,0.7);
\node[anchor=east] at (-0.7,-0.7,0.7){\tiny $a_1$};
\draw ( -0.5,0.5,0.5) -- ( -0.7,0.7,0.7);
\node[anchor=east] at (-0.45,0.45,0.7){\tiny $a_3$};
\end{scope}
\begin{scope}[shift={( 0, 0, 0.85)}]
\drawcube{1}{lightblue};
\draw (-0.5,-0.5,-0.5) -- (-0.7,-0.7,-0.7);
\node[anchor=east] at (-0.7,-0.7,-0.7){\tiny $b_1$};
\draw (-0.5, 0.5,-0.5) -- (-0.7, 0.7,-0.7);
\node[anchor=east] at (-0.5,0.7,-0.6){\tiny $b_3$};
\draw (-0.5,-0.5, 0.5) -- (-0.7,-0.7, 0.7);
\node[anchor=east] at (-0.7,-0.7,0.7){\tiny $b_5$};
\draw (-0.5, 0.5, 0.5) -- (-0.7, 0.7, 0.7);
\node[anchor=east] at (-0.7,0.7,0.7){\tiny $b_7$};
\end{scope}
% connecting legs
\begin{scope}[canvas is plane={O( 0.5,-0.5,-0.35)x( 0.7071,-0.7071,0)y( 0, 0, 1)}]
\draw ( 0, 0) to [out=45, in=-45] ( 0, 0.7);
\end{scope}
\begin{scope}[canvas is plane={O( 0.5,-0.5,-1.35)x( 0.7071,-0.7071,0)y( 0, 0, 1)}]
\draw ( 0, 0) to [out=-45, in=45, looseness=1.5] ( 0, 2.7);
\end{scope}
\begin{scope}[canvas is plane={O( 0.5, 0.5,-0.35)x( 0.7071, 0.7071,0)y( 0, 0, 1)}]
\draw ( 0, 0) to [out=45, in=-45] ( 0, 0.7);
\end{scope}
\begin{scope}[canvas is plane={O( 0.5, 0.5,-1.35)x( 0.7071, 0.7071,0)y( 0, 0, 1)}]
\draw ( 0, 0) to [out=-45, in=45, looseness=1.5] ( 0, 2.7);
\end{scope}
\node[anchor=west] at (0.6,0.6,0){\tiny $q$};
\node[anchor=west] at (0.2,-0.6,0.2){\tiny $p$};
\node[anchor=west] at (1.1,1,0){\tiny $k$};
\node[anchor=west] at (0.6,-0.7,0.2){\tiny $l$};
\end{scope}
\end{tikzpicture} ,
\end{equation}
An analogous derivation holds for $\circ_2$.

\section{Algebraic derivation of light ray correlations}

We will now algebraically derive the light ray form of the tensor network \eqref{eq:corr_function_ray_diagram} representing $D^{\alpha \beta}$. Following the arguments in the main text, we assume that $\alpha \neq 0$ and $\beta \neq 0$. We first determine which operators do not immediately cancel via the usual unitary condition with their adjoint. It will turn be useful to do this by finding the set of space-time points these operators are applied to. We will denote this set by $\mathcal{LC}$. Furthermore, $\mathcal{LC} (\tau)$ will be the set of sites $z$ for which $(z,\tau) \in \mathcal{LC}$.

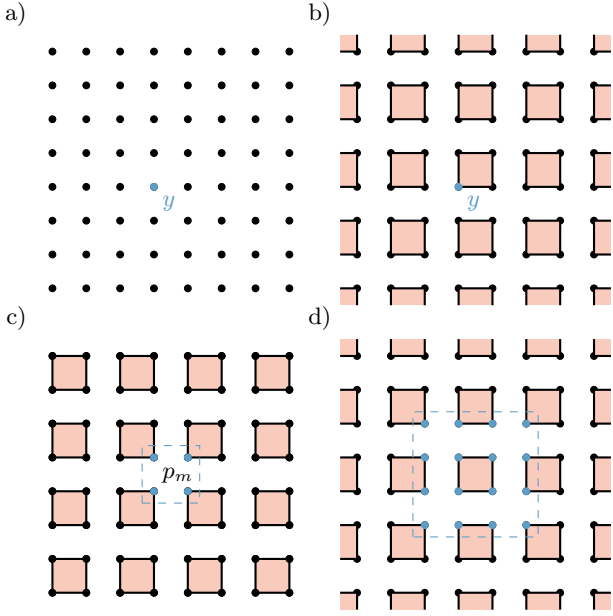
\begin{figure}[!ht]
\begin{tikzpicture}[>=stealth, scale=0.45]
%% tau =t
\node[anchor=south east] at (-0.5,7.5){a)};

\foreach \i in {0,...,7} {
    \foreach \j in {0,...,7}{
        \filldraw[black] (\i,\j) circle (0.1);
    }
}

\filldraw[mcyan] (3,3) circle (0.1)node[anchor=north west]{$y$};
%%

%% tau = t-1/2
\begin{scope}[shift={(9,0)}]
\node[anchor=south east] at (-0.5,7.5){b)};

% Lattice
\foreach \i in {0,...,7} {
    \foreach \j in {0,...,7}{
        \filldraw[black] (\i,\j) circle (0.1);
    }
}

%Operators
\foreach \i in {1,...,3} {
    \foreach \j in {1,...,3} {
        \draw[thick, fill=lightred] (2*\i-1,2*\j-1) rectangle (2*\i,2*\j);
    }
}
\foreach \i in {1,...,3} {
    \fill[lightred] (2*\i-1,0) rectangle (2*\i,-0.5);
    \fill[lightred] (2*\i-1,7) rectangle (2*\i, 7.5);
    \draw[thick]    (2*\i-1,-0.5) -- (2*\i-1,0) -- (2*\i,0) -- (2*\i,-0.5);
    \draw[thick]    (2*\i-1, 7.5) -- (2*\i-1,7) -- (2*\i,7) -- (2*\i, 7.5);
}
\foreach \j in {1,...,3} {
    \fill[lightred] (0,2*\j-1) rectangle (-0.5,2*\j);
    \fill[lightred] (7,2*\j-1) rectangle ( 7.5,2*\j);
    \draw[thick]    (-0.5,2*\j-1) -- (0,2*\j-1) -- (0,2*\j) -- (-0.5,2*\j);
    \draw[thick]    ( 7.5,2*\j-1) -- (7,2*\j-1) -- (7,2*\j) -- ( 7.5,2*\j);
}
\fill[lightred] ( 0,   0) rectangle (-0.5,-0.5);
\fill[lightred] ( 0,   7) rectangle (-0.5, 7.5);
\fill[lightred] ( 7,   0) rectangle ( 7.5,-0.5);
\fill[lightred] ( 7,   7) rectangle ( 7.5, 7.5);
\draw[thick]    (-0.5, 0  ) -- ( 0, 0) -- ( 0,  -0.5);
\draw[thick]    (-0.5, 7  ) -- ( 0, 7) -- ( 0,   7.5);
\draw[thick]    ( 7,  -0.5) -- ( 7, 0) -- ( 7.5, 0  );
\draw[thick]    ( 7,   7.5) -- ( 7, 7) -- ( 7.5, 7  );

%Blocked sites
\filldraw[mcyan] (3,3) circle (0.1)node[anchor=north west]{$y$};
\end{scope}
%%

%% tau = t-1
\begin{scope}[shift = {(0,-9)}]
\node[anchor=south east] at (-0.5,7.5){c)};

% Lattice
\foreach \i in {0,...,7} {
    \foreach \j in {0,...,7}{
        \filldraw[black] (\i,\j) circle (0.1);
    }
}

% Operators
\foreach \i in {0,...,3}{
    \foreach \j in {0,...,3}{
        \draw[thick, fill=lightred] (2*\i,2*\j) rectangle (2*\i+1,2*\j+1);
    }
}
\foreach \i in {0,...,7} {
    \foreach \j in {0,...,7}{
        \filldraw[black] (\i,\j) circle (0.1);
    }
}

% Blocked Sites
\foreach \i in {3,...,4} {
    \foreach \j in {3,...,4}{
        \filldraw[mcyan] (\i,\j) circle (0.1);
    }
}
\draw[dashed,mcyan] ( 2.65, 2.65) rectangle ( 4.35, 4.35);
\node[anchor=south west] at (3,3){$p_m$};

\end{scope}
%%

%% tau = t-3/2
\begin{scope}[shift = {(9,-9)}]
\node[anchor=south east] at (-0.5,7.5){d)};

% Lattice
\foreach \i in {0,...,7} {
    \foreach \j in {0,...,7}{
        \filldraw[black] (\i,\j) circle (0.1);
    }
}
%Operators
\foreach \i in {1,...,3} {
    \foreach \j in {1,...,3} {
        \draw[thick, fill=lightred] (2*\i-1,2*\j-1) rectangle (2*\i,2*\j);
    }
}
\foreach \i in {1,...,3} {
    \fill[lightred] (2*\i-1,0) rectangle (2*\i,-0.5);
    \fill[lightred] (2*\i-1,7) rectangle (2*\i, 7.5);
    \draw[thick]    (2*\i-1,-0.5) -- (2*\i-1,0) -- (2*\i,0) -- (2*\i,-0.5);
    \draw[thick]    (2*\i-1, 7.5) -- (2*\i-1,7) -- (2*\i,7) -- (2*\i, 7.5);
}
\foreach \j in {1,...,3} {
    \fill[lightred] (0,2*\j-1) rectangle (-0.5,2*\j);
    \fill[lightred] (7,2*\j-1) rectangle ( 7.5,2*\j);
    \draw[thick]    (-0.5,2*\j-1) -- (0,2*\j-1) -- (0,2*\j) -- (-0.5,2*\j);
    \draw[thick]    ( 7.5,2*\j-1) -- (7,2*\j-1) -- (7,2*\j) -- ( 7.5,2*\j);
}
\fill[lightred] ( 0,   0) rectangle (-0.5,-0.5);
\fill[lightred] ( 0,   7) rectangle (-0.5, 7.5);
\fill[lightred] ( 7,   0) rectangle ( 7.5,-0.5);
\fill[lightred] ( 7,   7) rectangle ( 7.5, 7.5);
\draw[thick]    (-0.5, 0  ) -- ( 0, 0) -- ( 0,  -0.5);
\draw[thick]    (-0.5, 7  ) -- ( 0, 7) -- ( 0,   7.5);
\draw[thick]    ( 7,  -0.5) -- ( 7, 0) -- ( 7.5, 0  );
\draw[thick]    ( 7,   7.5) -- ( 7, 7) -- ( 7.5, 7  );

% Blocked sites
\foreach \i in {2,...,5} {
    \foreach \j in {2,...,5}{
        \filldraw[mcyan] (\i,\j) circle (0.1);
    }
}
\draw[dashed,mcyan] ( 1.65, 1.65) rectangle ( 5.35, 5.35);
\end{scope}
\end{tikzpicture}
\caption{A visualization of the step by step method in a top-down view. a) At time $t$ the operator $a_y^\beta$ is applied and blocks the site $y$ marked in cyan. b) The layer $\mathbb{U}_{oo}$, marked in red, is applied at time $t-\frac{1}{2}$ and connected to a layer $\mathbb{U}_{oo}^{-1}$ above the $t$-layer. Only on the blocked site $y$ they are not trivially connected. c) After cancellations due to the conventional unitary property, the operator applied to $p_m$ marked in cyan at time $t-\frac{1}{2}$ remains, blocking all four sites for the layer $\mathbb{U}_{ee}$ applied at the earlier time $t-1$. This results in more operators not trivially connected to the corresponding layer $\mathbb{U}_{ee}^{-1}$. d) The set of blocked sites for the next earlier time $t-\frac{3}{2}$ increases even further. Here the set $\mathcal{LC}(t-\frac{3}{2})$ is marked in cyan.}
\label{fig:StepbyStepVis}
\end{figure}

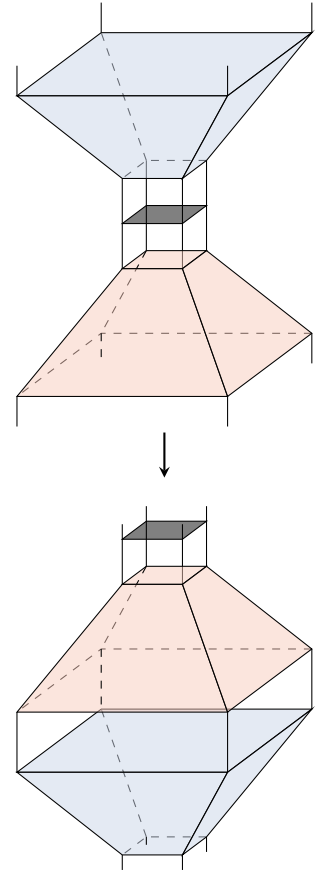
\begin{figure}
\begin{tikzpicture}[>=stealth, scale=0.4]
\pgfsetxvec{\pgfpoint{1cm}{0cm}}
\pgfsetyvec{\pgfpoint{0.4cm}{0.3cm}}
\pgfsetzvec{\pgfpoint{0cm}{1cm}}

\begin{scope}[shift={( 0, 0, 8)}]
% operators
\begin{scope}[canvas is xy plane at z=0]
\draw[fill=gray] (-1,-1) rectangle ( 1, 1);
\end{scope}
\draw (-1,-1,-1.5) -- (-1,-1, 1.5);
\draw (-1, 1,-1.5) -- (-1, 1, 1.5);
\draw ( 1,-1,-1.5) -- ( 1,-1, 1.5);
\draw ( 1, 1,-1.5) -- ( 1, 1, 1.5);
% upper blue pyramid
\draw[dashed] ( 1, 1, 1.5) -- (-1, 1, 1.5) -- (-1,-1, 1.5);
\draw[dashed] (-1, 1, 1.5) -- (-3.5, 3.5, 5);
\begin{scope}[canvas is xy plane at z=5]
\draw[fill=lightblue, fill opacity=0.5] (-3.5,-3.5) rectangle ( 3.5, 3.5);
\end{scope}
\draw[fill=lightblue, fill opacity=0.5] ( 1,-1, 1.5) -- ( 3.5,-3.5, 5) -- (-3.5,-3.5, 5) -- (-1,-1, 1.5) -- cycle;
\draw[fill=lightblue, fill opacity=0.5] ( 1, 1, 1.5) -- ( 3.5, 3.5, 5) -- ( 3.5,-3.5, 5) -- ( 1,-1, 1.5) -- cycle;
\draw (-3.5,-3.5, 5) -- (-3.5,-3.5, 6);
\draw (-3.5, 3.5, 5) -- (-3.5, 3.5, 6);
\draw ( 3.5,-3.5, 5) -- ( 3.5,-3.5, 6);
\draw ( 3.5, 3.5, 5) -- ( 3.5, 3.5, 6);
% lower red pyramid
\begin{scope}[canvas is xy plane at z=-1.5]
\draw[fill=lightred, fill opacity=0.5] (-1,-1) rectangle ( 1, 1);
\end{scope}
\draw[dashed] ( 3.5, 3.5,-5) -- (-3.5, 3.5,-5) -- (-3.5,-3.5,-5);
\draw[dashed] (-1, 1,-1.5) -- (-3.5, 3.5,-5);
\draw[fill=lightred, fill opacity=0.5] ( 1,-1,-1.5) -- ( 3.5,-3.5,-5) -- (-3.5,-3.5,-5) -- (-1,-1,-1.5) -- cycle;
\draw[fill=lightred, fill opacity=0.5] ( 1, 1,-1.5) -- ( 3.5, 3.5,-5) -- ( 3.5,-3.5,-5) -- ( 1,-1,-1.5) -- cycle;
\draw         (-3.5,-3.5,-5) -- (-3.5,-3.5,-6);
\draw[dashed] (-3.5, 3.5,-5) -- (-3.5, 3.5,-6);
\draw         ( 3.5,-3.5,-5) -- ( 3.5,-3.5,-6);
\draw         ( 3.5, 3.5,-5) -- ( 3.5, 3.5,-6);
\end{scope}

% arrow
\draw[->, thick] ( 0, 0, 0.75) -- ( 0, 0,-0.75);

\begin{scope}[shift={( 0, 0,-8.5)}]
% operators
\begin{scope}[canvas is xy plane at z=6]
\draw[fill=gray] (-1,-1) rectangle ( 1, 1);
\end{scope}
\draw (-1,-1, 4.5) -- (-1,-1, 6.5);
\draw (-1, 1, 4.5) -- (-1, 1, 6.5);
\draw ( 1,-1, 4.5) -- ( 1,-1, 6.5);
\draw ( 1, 1, 4.5) -- ( 1, 1, 6.5);
% lower blue pyramid
\draw[dashed] ( 1, 1,-4.5) -- (-1, 1,-4.5) -- (-1,-1,-4.5);
\draw[dashed] (-1, 1,-4.5) -- (-3.5, 3.5,-1);
\begin{scope}[canvas is xy plane at z=-1]
\draw[fill=lightblue, fill opacity=0.5] (-3.5,-3.5) rectangle ( 3.5, 3.5);
\end{scope}
\draw[fill=lightblue, fill opacity=0.5] ( 1,-1,-4.5) -- ( 3.5,-3.5,-1) -- (-3.5,-3.5,-1) -- (-1,-1,-4.5) -- cycle;
\draw[fill=lightblue, fill opacity=0.5] ( 1, 1,-4.5) -- ( 3.5, 3.5,-1) -- ( 3.5,-3.5,-1) -- ( 1,-1,-4.5) -- cycle;
\draw         (-1,-1,-4.5) -- (-1,-1,-5);
\draw[dashed] (-1, 1,-4.5) -- (-1, 1,-5);
\draw         ( 1,-1,-4.5) -- ( 1,-1,-5);
\draw         ( 1, 1,-4.5) -- ( 1, 1,-5);
% upper red pyramid
\begin{scope}[canvas is xy plane at z=4.5]
\draw[fill=lightred, fill opacity=0.5] (-1,-1) rectangle ( 1, 1);
\end{scope}
\draw[dashed] ( 3.5, 3.5, 1) -- (-3.5, 3.5, 1) -- (-3.5,-3.5, 1);
\draw[dashed] (-1, 1, 4.5) -- (-3.5, 3.5, 1);
\draw[fill=lightred, fill opacity=0.5] ( 1,-1, 4.5) -- ( 3.5,-3.5, 1) -- (-3.5,-3.5, 1) -- (-1,-1, 4.5) -- cycle;
\draw[fill=lightred, fill opacity=0.5] ( 1, 1, 4.5) -- ( 3.5, 3.5, 1) -- ( 3.5,-3.5, 1) -- ( 1,-1, 4.5) -- cycle;
% legs between pyramids
\draw         (-3.5,-3.5, 1) -- (-3.5,-3.5,-1);
\draw[dashed] (-3.5, 3.5, 1) -- (-3.5, 3.5,-1);
\draw         ( 3.5,-3.5, 1) -- ( 3.5,-3.5,-1);
\draw         ( 3.5, 3.5, 1) -- ( 3.5, 3.5,-1);
\end{scope}
\end{tikzpicture}
\caption{Visualization of the pyramid-shaped light-cones. As we look at a trace, we can utilize its cyclic property and shift along the time direction. $a_y^\beta$ is applied to any of the four sites in the gray square, corresponding to time $\tau = t$. $a_x^\alpha$ is applied somewhere in between the bases of the two pyramids.}
\label{fig:pyramidshift}
\end{figure}

To determine $\mathcal{LC}$ we will go layer by layer. The following process for the first two steps is visualized in Fig.~\ref{fig:StepbyStepVis}. First look a the layer at time $\tau = t$. Here $a_y^\beta$ is applied as well as $\mathbb{U}_{\text{oo}}^\dagger$. At time $\tau = t - \frac{1}{2}$, a layer $\mathbb{U}_{\text{oo}}$ follows. Now using the unitary-condition in time direction, most of the unitaries $U$ will cancel with their adjoint $U^\dagger$, except for the one applied on the plaquette $p_m := p(2 \lceil \frac{y_1}{2} \rceil-1, 2\lceil \frac{y_2}{2} \rceil -1)$. In the next earlier time step $\tau = t - 1$, all unitaries that are applied to at least one site in $p_m$ will be blocked to cancel with their adjoint and will in turn block even more sites. Due to the odd-even application of the unitaries, the set blocked sites will increase in each direction by one row with each time-step. $\mathcal{LC}(\tau)$ is equal to the set of what we called blocked sites at time $\tau$. We then get the recursion relation
\begin{equation}
\mathcal{LC}(\tau ) = \mathcal{LC}\left(\tau+\frac{1}{2}\right) \cup \left\{ z \, \Big| \min_{\tilde{y}\in \mathcal{LC}(\tau +\frac{1}{2})} \norm{ z - \tilde{y} }_\infty = 1 \right\},
\end{equation}
with the initial condition $\mathcal{LC}\left( t-\frac{1}{2}\right) = p_m$. So any operator that at time $\tau$ is applied to sites in $\mathcal{LC}(\tau)$ will not cancel with its adjoint. We can find a solution to the above recursion relation as
\begin{equation}
\label{eq:LCy}
\mathcal{LC}(\tau ) = \left\{ z \, \Big| \min_{\tilde{y}\in p_m} \norm{z -\tilde{y}}_\infty \leq 2t - 1 -2\tau \right\}.
\end{equation}
Thus one arrives at a light-cone given by
\begin{multline}
\mathcal{LC} = \bigg\{ (z,\tau) \in \{ 0, \cdots , L-1 \}^2 \times \{ 0, \tfrac{1}{2}, \cdots , t-\tfrac{1}{2} \}) | \\
\min_{\tilde{y}\in p_m} \norm{z -\tilde{y}}_\infty \leq 2t-1-2\tau \bigg\}.
\end{multline}
This set has a square pyramid shape with a base length of $4t$ sites. Every unitary $U_{(p,\tau)}$ that is applied to a plaquette $p \subset \mathcal{LC}(\tau)$ of sites at time $\tau$ will not cancel with $U^\dagger_{(p,2t-\tau)}$. From this we immediately get the result
\begin{equation}
D^{\alpha \beta}(x,y,t) = 0 \quad \text{if} \quad x \notin \mathcal{LC}(0).
\end{equation}
So far we have solely used the conventional unitary property of the gates and ended up with a light cone structure resembling two pyramids with flat tops. This is visualized in Fig.~\ref{fig:pyramidshift}. Now assume that $L \geq 4t$ and that the operators are ternary unitaries. We may characterize the sides of our light-cone by their distance to two sites in $p_m = \{y^1,y^2,y^3,y^4 \}$ as
\begin{multline}
S_{ij} (\tau) = S_{ji} (\tau) = \bigg\{ z \in \mathcal{LC}(\tau) \, \Big|\\ \, \norm{z - y^i}_\infty = \norm{z - y^j}_\infty = 2t-2\tau \bigg\}.
\end{multline}
In a similar way, we can also characterize the corners of the base $z^i$ via
\begin{equation}
\abs{z^i_j - y^i_j} = 2t \quad \forall j \in \{ 1,2 \}.
\end{equation}
Intuitively they are the corner and side furthest away from the given sites in $p_m$. Now assume that $x \neq z^i$ with $y = y^i$. First we make use of the cyclicity of the trace
\begin{equation}
\Tr [ a_x^\alpha \mathbb{U}^{-t} a_y^\beta \mathbb{U}^t] = \Tr [a_y^\beta \mathbb{U}^t a_x^\alpha \mathbb{U}^{-t} ]
\end{equation}
and assign the space-time points which the adjoints are applied to negative times $\tau \in \{ -\frac{1}{2} , \cdots , -t\}$. By assumption, we can always find a side $S_{ij}(0)$, such that $x \notin S_{ij}(0)$. Due to the trace the ternary unitaries and their adjoints are connected via the identity through the sites in $S_{ij}(0)$. Thus we can apply the appropriate unitary condition in space, i.e.,
\begin{equation}
\label{eq:SideVis}
\begin{tikzpicture}[baseline=(current bounding box.center), scale=0.2]
\pgfsetxvec{\pgfpoint{1cm}{0cm}}
\pgfsetyvec{\pgfpoint{0.4cm}{0.3cm}}
\pgfsetzvec{\pgfpoint{0cm}{1cm}}
\foreach \i in {0, 1}
{
    \begin{scope}[shift={(6*\i,0,0)}]
    \drawcube{2}{lightblue};
    \draw (-1, 1,1) -- (-1,1,3);
    \draw (-1, -1, 1) -- (-1,-1, 3);
    \draw ( 1, -1, 1) -- ( 1, -1, 3);
    \draw ( 1, 1,1) -- ( 1,1,3);
    \pgfsetxvec{\pgfpoint{1cm}{0cm}}
    \pgfsetyvec{\pgfpoint{0cm}{1cm}}
    \pgfsetzvec{\pgfpoint{0.4cm}{0.3cm}}
    \filldraw[fill=lime] (-1, 2, -1) circle (0.15);
    \filldraw[fill=lime] ( 1, 2, -1) circle (0.15);
    \pgfsetxvec{\pgfpoint{1cm}{0cm}}
    \pgfsetyvec{\pgfpoint{0.4cm}{0.3cm}}
    \pgfsetzvec{\pgfpoint{0cm}{1cm}}
    \draw (1,1,-1) -- (2,2,-5);
    \draw (-1,-1, -1) -- (-1,-2, -2) -- (-1, -2, 6) -- (-1, -1, 5);
    \draw ( 1,-1, -1) -- ( 1,-2, -2) -- ( 1, -2, 6) -- ( 1, -1, 5);
    \end{scope}
    \begin{scope}[shift={(3+6*\i,3,-6)}]
    \drawcube{2}{lightblue};
    \draw (1,-1,1) -- (2,-2,5);
    \draw (-1,-1,-1) -- (-1,-1,-2) -- (-1,-1.5,-1.5);
    \draw (1,-1,-1) -- (1,-1,-2) -- (1,-1.5,-1.5);
    \end{scope}
    \begin{scope}[shift={(6*\i,0,4)}]
    \drawcube{2}{lightred};
    \draw (1,1,1) -- (2,2,5);
    \end{scope}
    \begin{scope}[shift={(3+6*\i,3,10)}]
    \drawcube{2}{lightred};
    \draw (1,-1,-1) -- (2,-2,-5);
    \draw (-1,-1,1) -- (-1,-1,2) -- (-1,-1.5,1.75);
    \draw (1,-1,1) -- (1,-1,2) -- (1,-1.5,1.75);
    \end{scope}
}
\begin{scope}
\draw (-1,1,-1) -- (-1,1,-2.5) -- (-1.25,1,-2);
\end{scope}
\begin{scope}[shift={(0,0,4)}]
\draw (-1,1,1) -- (-1,1,2) -- (-1.25,1,1.5);
\end{scope}
\node at (14,0,3) {$\cdots$};
\begin{scope}[shift={(28,0,0)}]
\foreach \i in {0,1}
{
    \begin{scope}[shift={(-6*\i,0,0)}]
    \drawcube{2}{lightblue};
    \draw (1,1,1) -- (1,1,3);
    \draw (1,-1,1) -- (1,-1,3);
    \draw (-1,-1,1) -- (-1,-1,3);
    \draw (-1,1,1) -- (-1,1,3);
    \pgfsetxvec{\pgfpoint{1cm}{0cm}}
    \pgfsetyvec{\pgfpoint{0cm}{1cm}}
    \pgfsetzvec{\pgfpoint{0.4cm}{0.3cm}}
    \filldraw[fill=lime] (-1, 2, -1) circle (0.15);
    \filldraw[fill=lime] ( 1, 2, -1) circle (0.15);
    \pgfsetxvec{\pgfpoint{1cm}{0cm}}
    \pgfsetyvec{\pgfpoint{0.4cm}{0.3cm}}
    \pgfsetzvec{\pgfpoint{0cm}{1cm}}
    \draw (-1,1,-1) -- (-2,2,-5);
    \draw (1,-1,-1) -- (1,-2,-2) -- (1,-2,6) -- (1,-1,5);
    \draw (-1,-1,-1) -- (-1,-2,-2) -- (-1,-2,6) -- (-1,-1,5);
    \end{scope}
    \begin{scope}[shift={(-3-6*\i,3,-6)}]
    \drawcube{2}{lightblue};
    \draw (-1,-1,1) -- (-2,-2,5);
    \draw (1,-1,-1) -- (1,-1,-2) -- (1,-1.5,-1.5);
    \draw (-1,-1,-1) -- (-1,-1,-2) -- (-1,-1.5,-1.5);
    \end{scope}
    \begin{scope}[shift={(-6*\i,0,4)}]
    \drawcube{2}{lightred};
    \draw (-1,1,1) -- (-2,2,5);
    \end{scope}
    \begin{scope}[shift={(-3-6*\i,3,10)}]
    \drawcube{2}{lightred};
    \draw (-1,-1,-1) -- (-2,-2,-5);
    \draw (1,-1,1) -- (1,-1,2) -- (1,-1.5,1.75);
    \draw (-1,-1,1) -- (-1,-1,2) -- (-1,-1.5,1.75);
    \end{scope}
}
\begin{scope}
\draw (1,1,-1) -- (1,1,-2.5) -- (1.25,1,-2);
\end{scope}
\begin{scope}[shift={(0,0,4)}]
\draw (1,1,1) -- (1,1,2) -- (1.25,1,1.5);
\end{scope}
\end{scope}
\draw (-1.5,0,13.5) -- (-1.5,0,14.5) -- (-1.5,0,14) -- (29,0,14) -- (29,0,13.5) -- (29,0,14.5);
\node at (14.5,0,15){$4t$};
\begin{scope}[shift={(4,0,-20)}]
\foreach \i in {0,1}{
\begin{scope}[shift={(\i*6,0,0)}]
\drawcube{2}{lightblue};
\draw (-1,-1,1) -- (-1,-1,7);
\draw (1,-1,1) -- (1,-1,7);
\draw (-1,-1,-1) -- (-1,-2,-2) -- (-1,-2,10) -- (-1,-1,9);
\draw (1,-1,-1) -- (1,-2,-2) -- (1,-2,10) -- (1,-1,9);
\draw (1,1,-1) -- (1.5,2,-2);
\end{scope}
\begin{scope}[shift={(\i*6,0,8)}]
\drawcube{2}{lightred};
\draw (1,1,1) -- (1.5,2,2);
\end{scope}
}
\begin{scope}
\draw (-1,1,-1) -- (-1,1,-2.5) -- (-1.25,1,-2);
\end{scope}
\begin{scope}[shift={(6,0,0)}]
\draw (-1,1,-1) -- (-1.5,2,-2.5,2);
\end{scope}
\begin{scope}[shift={(0,0,8)}]
\draw (-1,1,1) -- (-1,1,2) -- (-1.25,1,1.5);
\end{scope}
\begin{scope}[shift={(6,0,8)}]
\draw (-1,1,1) -- (-2,2,2);
\end{scope}
\node at (10,0,5) {$\cdots$};
\begin{scope}[shift={(21,0,0)}]
\foreach \i in {0,1}{
\begin{scope}[shift={(-\i*6,0,0)}]
\drawcube{2}{lightblue};
\draw (1,-1,1) -- (1,-1,7);
\draw (-1,-1,1) -- (-1,-1,7);
\draw (1,-1,-1) -- (1,-2,-2) -- (1,-2,10) -- (1,-1,9);
\draw (-1,-1,-1) -- (-1,-2,-2) -- (-1,-2,10) -- (-1,-1,9);
\draw (-1,1,-1) -- (-1.5,2,-2.5);
\end{scope}
\begin{scope}[shift={(-\i*6,0,8)}]
\drawcube{2}{lightred};
\draw (-1,1,1) -- (-1.5,2,2);
\end{scope}
}
\begin{scope}
\draw (1,1,-1) -- (1,1,-2.5) -- (1.25,1,-2);
\end{scope}
\begin{scope}[shift={(0,0,8)}]
\draw (1,1,1) -- (1,1,2) -- (1.25,1,1.5);
\end{scope}
\begin{scope}[shift={(-6,0,8)}]
\draw (1,1,1) -- (2,2,2);
\end{scope}
\begin{scope}[shift={(-6,0,0)}]
\draw (1,1,-1) -- (2,2,-2);
\end{scope}
\end{scope}
\begin{scope}[shift={(0,0,-17)}]
\draw (-1.5,0,12.5) -- (-1.5,0,11.5) -- (-1.5,0,12) -- (24,0,12) -- (24,0,12.5) -- (24,0,11.5);
\node at (12,0,13) {$4t-2$};
\end{scope}
\node[anchor=east] at (-2,0,4){$=d^2$};
\end{scope}
\end{tikzpicture}
\end{equation}
The sites in $S_{ij}(0)$ are marked by green dots. This will serve as our base case for the following recursion. For the operators applied at time $\tau$ at least partially on sites in $S_{ij} (0)$ the adjoints will be applied at time $-\tau - \frac{1}{2}$. Due to the trace, any point of the form $(z,\tau)$ with $z\in S_{ij} (\tau)$ is connected to the event $\left( z,-\tau -\frac{1}{2}\right)$ via the identity. If the events $(z,\tau)$ and $(z,-\tau)$ are also connected via the identity, we can use one of the spatial unitary conditions to cancel the aforementioned operators with their adjoint. This will result in any point of the form $\left( \tilde{z},\tau + \frac{1}{2}\right)$ for $\tilde{z} \in S_{ij}\left( \tau +\frac{1}{2}\right)$ being connected to the point $\left( \tilde{z}, -\tau - \frac{1}{2}\right)$ via the identity, allowing for the same argument again. This can in short be visualized by
\begin{equation}
\begin{tikzpicture}[baseline=(current  bounding  box.center), scale=0.2]
\pgfsetxvec{\pgfpoint{1cm}{0cm}}
\pgfsetyvec{\pgfpoint{0.4cm}{0.3cm}}
\pgfsetzvec{\pgfpoint{0cm}{1cm}}
\foreach \i in {0,1}{
\begin{scope}[shift={(\i*6,0,0)}]
\drawcube{2}{lightblue};
\draw (-1,-1,1) -- (-1,-1,7);
\draw (1,-1,1) -- (1,-1,7);
\draw (-1,-1,-1) -- (-1,-2,-2) -- (-1,-2,10) -- (-1,-1,9);
\draw (1,-1,-1) -- (1,-2,-2) -- (1,-2,10) -- (1,-1,9);
\draw (1,1,-1) -- (1.5,2,-2);
\pgfsetxvec{\pgfpoint{1cm}{0cm}}
\pgfsetyvec{\pgfpoint{0cm}{1cm}}
\pgfsetzvec{\pgfpoint{0.4cm}{0.3cm}}
\filldraw[fill=lightgreen] (1.5,-2,2) circle (0.2);
\pgfsetxvec{\pgfpoint{1cm}{0cm}}
\pgfsetyvec{\pgfpoint{0.4cm}{0.3cm}}
\pgfsetzvec{\pgfpoint{0cm}{1cm}}
\end{scope}
\begin{scope}[shift={(\i*6,0,8)}]
\drawcube{2}{lightred};
\draw (1,1,1) -- (1.5,2,2);
\pgfsetxvec{\pgfpoint{1cm}{0cm}}
\pgfsetyvec{\pgfpoint{0cm}{1cm}}
\pgfsetzvec{\pgfpoint{0.4cm}{0.3cm}}
\filldraw[fill=lightgreen] (1.5,2,2) circle (0.2);
\pgfsetxvec{\pgfpoint{1cm}{0cm}}
\pgfsetyvec{\pgfpoint{0.4cm}{0.3cm}}
\pgfsetzvec{\pgfpoint{0cm}{1cm}}
\end{scope}
}
\begin{scope}
\draw (-1,1,-1) -- (-1,1,-2.5) -- (-1.25,1,-2);
\end{scope}
\begin{scope}[shift={(6,0,0)}]
\draw (-1,1,-1) -- (-1.5,2,-2.5);
\pgfsetxvec{\pgfpoint{1cm}{0cm}}
\pgfsetyvec{\pgfpoint{0cm}{1cm}}
\pgfsetzvec{\pgfpoint{0.4cm}{0.3cm}}
\filldraw[fill=lightgreen] (-1.5,-2.5,2) circle (0.2);
\pgfsetxvec{\pgfpoint{1cm}{0cm}}
\pgfsetyvec{\pgfpoint{0.4cm}{0.3cm}}
\pgfsetzvec{\pgfpoint{0cm}{1cm}}
\end{scope}
\begin{scope}[shift={(0,0,8)}]
\draw (-1,1,1) -- (-1,1,2) -- (-1.25,1,1.5);
\end{scope}
\begin{scope}[shift={(6,0,8)}]
\draw (-1,1,1) -- (-2,2,2);
\pgfsetxvec{\pgfpoint{1cm}{0cm}}
\pgfsetyvec{\pgfpoint{0cm}{1cm}}
\pgfsetzvec{\pgfpoint{0.4cm}{0.3cm}}
\filldraw[fill=lightgreen] (-2,2,2) circle (0.2);
\pgfsetxvec{\pgfpoint{1cm}{0cm}}
\pgfsetyvec{\pgfpoint{0.4cm}{0.3cm}}
\pgfsetzvec{\pgfpoint{0cm}{1cm}}
\end{scope}
\node[anchor=west] at (8,0,5) {$\cdots$};
\begin{scope}[shift={(20,0,0)}]
\foreach \i in {0,1}{
\begin{scope}[shift={(-\i*6,0,0)}]
\drawcube{2}{lightblue};
\draw (1,1,1) -- (1,1,7);
\draw (-1,1,1) -- (-1,1,7);
\draw (1,-1,-1) -- (1,-2,-2) -- (1,-2,10) -- (1,-1,9);
\draw (-1,-1,-1) -- (-1,-2,-2) -- (-1,-2,10) -- (-1,-1,9);
\draw (-1,1,-1) -- (-1.5,2,-2.5);
\pgfsetxvec{\pgfpoint{1cm}{0cm}}
\pgfsetyvec{\pgfpoint{0cm}{1cm}}
\pgfsetzvec{\pgfpoint{0.4cm}{0.3cm}}
\filldraw[fill=lightgreen] (-1.5,-2.5,2) circle (0.2);
\pgfsetxvec{\pgfpoint{1cm}{0cm}}
\pgfsetyvec{\pgfpoint{0.4cm}{0.3cm}}
\pgfsetzvec{\pgfpoint{0cm}{1cm}}
\end{scope}
\begin{scope}[shift={(-\i*6,0,8)}]
\drawcube{2}{lightred};
\draw (-1,1,1) -- (-1.5,2,2);
\pgfsetxvec{\pgfpoint{1cm}{0cm}}
\pgfsetyvec{\pgfpoint{0cm}{1cm}}
\pgfsetzvec{\pgfpoint{0.4cm}{0.3cm}}
\filldraw[fill=lightgreen] (-1.5,2,2) circle (0.2);
\pgfsetxvec{\pgfpoint{1cm}{0cm}}
\pgfsetyvec{\pgfpoint{0.4cm}{0.3cm}}
\pgfsetzvec{\pgfpoint{0cm}{1cm}}
\end{scope}
}
\begin{scope}
\draw (1,1,-1) -- (1,1,-2.5) -- (1.25,1,-2);
\end{scope}
\begin{scope}[shift={(0,0,8)}]
\draw (1,1,1) -- (1,1,2) -- (1.25,1,1.5);
\end{scope}
\begin{scope}[shift={(-6,0,8)}]
\draw (1,1,1) -- (2,2,2);
\pgfsetxvec{\pgfpoint{1cm}{0cm}}
\pgfsetyvec{\pgfpoint{0cm}{1cm}}
\pgfsetzvec{\pgfpoint{0.4cm}{0.3cm}}
\filldraw[fill=lightgreen] (2,2,2) circle (0.2);
\pgfsetxvec{\pgfpoint{1cm}{0cm}}
\pgfsetyvec{\pgfpoint{0.4cm}{0.3cm}}
\pgfsetzvec{\pgfpoint{0cm}{1cm}}
\end{scope}
\begin{scope}[shift={(-6,0,0)}]
\draw (1,1,-1) -- (2,2,-2);
\pgfsetxvec{\pgfpoint{1cm}{0cm}}
\pgfsetyvec{\pgfpoint{0cm}{1cm}}
\pgfsetzvec{\pgfpoint{0.4cm}{0.3cm}}
\filldraw[fill=lightgreen] (2,-2,2) circle (0.2);
\pgfsetxvec{\pgfpoint{1cm}{0cm}}
\pgfsetyvec{\pgfpoint{0.4cm}{0.3cm}}
\pgfsetzvec{\pgfpoint{0cm}{1cm}}
\end{scope}
\end{scope}
\begin{scope}[shift={(0,0,-15)}]
\node[anchor=east] at (-0.5,0,4){$=d^2$};
\foreach \i in {0,...,2}{
    \draw (3*\i,1,-1) -- (3*\i,-1,4.5) -- (3*\i,1,9);
    \pgfsetxvec{\pgfpoint{1cm}{0cm}}
    \pgfsetyvec{\pgfpoint{0cm}{1cm}}
    \pgfsetzvec{\pgfpoint{0.4cm}{0.3cm}}
    \filldraw[fill=lightgreen] (3*\i,-1,1) circle (0.2);
    \filldraw[fill=lightgreen] (3*\i,9,1) circle (0.2);
    \pgfsetxvec{\pgfpoint{1cm}{0cm}}
    \pgfsetyvec{\pgfpoint{0.4cm}{0.3cm}}
    \pgfsetzvec{\pgfpoint{0cm}{1cm}}
}
\node[anchor=west] at (6,0,5){$\cdots$};
\foreach \i in {0,...,2}{
    \draw (3*\i+9,1,-1) -- (3*\i+9,-1,4.5) -- (3*\i+9,1,9);
    \pgfsetxvec{\pgfpoint{1cm}{0cm}}
    \pgfsetyvec{\pgfpoint{0cm}{1cm}}
    \pgfsetzvec{\pgfpoint{0.4cm}{0.3cm}}
    \filldraw[fill=lightgreen] (3*\i+9,-1,1) circle (0.2);
    \filldraw[fill=lightgreen] (3*\i+9,9,1) circle (0.2);
    \pgfsetxvec{\pgfpoint{1cm}{0cm}}
    \pgfsetyvec{\pgfpoint{0.4cm}{0.3cm}}
    \pgfsetzvec{\pgfpoint{0cm}{1cm}}
}
\end{scope}
\end{tikzpicture}
\end{equation}
where the green dots mark the points of the form $\left( \tilde{z},\tau + \frac{1}{2}\right)$ and $\left( \tilde{z}, -\tau - \frac{1}{2}\right)$ for $\tilde{z} \in S_{ij}\left( \tau +\frac{1}{2}\right)$. To reduce visual cluttering, some of the legs on the operators were not drawn in both equations. If we iterate this condition $2t-1$ times, including the base case, we arrive at
\begin{equation}
\label{eq:DUfinal}
\begin{tikzpicture}[baseline=(current bounding box.center), scale=0.375]
\pgfsetxvec{\pgfpoint{1cm}{0cm}}
\pgfsetyvec{\pgfpoint{0.4cm}{0.3cm}}
\pgfsetzvec{\pgfpoint{0cm}{1cm}}
\drawcube{2}{lightblue};
\draw (-1,-1,1) -- (-1,-1,3);
\draw (1,-1,1) -- (1,-1,3);
\draw (-1,-1,-1) -- (-1,-2,-2) -- (-1,-2,6) -- (-1,-1,5);
\draw (1,-1,-1) -- (1,-2,-2) -- (1,-2,6) -- (1,-1,5);
\draw (-1,1,-1) -- (-1,1,-3) -- (-1.5,1,-2.5);
\draw (1,1,-1) -- (1,1,-3) -- (1.5,1,-2.5);
\begin{scope}[shift={(0,0,4)}]
\drawcube{2}{lightred};
\draw (-1,1,1) -- (-1,1,3) -- (-1.5,1,2.5);
\draw (1,1,1) -- (1,1,3) -- (1.5,1,2.5);
\end{scope}
\pgfsetxvec{\pgfpoint{1cm}{0cm}}
\pgfsetyvec{\pgfpoint{0cm}{1cm}}
\pgfsetzvec{\pgfpoint{0.4cm}{0.3cm}}
\begin{scope}[shift={(0,4,0)}]
\filldraw (-1,2,1) circle (0.2);
\filldraw ( 1,2,1) circle (0.2);
\end{scope}
\pgfsetxvec{\pgfpoint{1cm}{0cm}}
\pgfsetyvec{\pgfpoint{0.4cm}{0.3cm}}
\pgfsetzvec{\pgfpoint{0cm}{1cm}}
\node at (3,0,2) {$\propto$};
\begin{scope}[shift={(6,0,0)}]
\draw (-1.5,0,-0.5) -- (-1,0,-1) -- (-1,0,5) -- (-1.5,0,4.5);
\draw (1.5,0,-0.5) -- (1,0,-1) -- (1,0,5) -- (1.5,0,4.5);
\pgfsetxvec{\pgfpoint{1cm}{0cm}}
\pgfsetyvec{\pgfpoint{0cm}{1cm}}
\pgfsetzvec{\pgfpoint{0.4cm}{0.3cm}}
\filldraw (-1,2,0) circle (0.2);
\filldraw ( 1,2,0) circle (0.2);
\pgfsetxvec{\pgfpoint{1cm}{0cm}}
\pgfsetyvec{\pgfpoint{0.4cm}{0.3cm}}
\pgfsetzvec{\pgfpoint{0cm}{1cm}}
\end{scope}
\node at (12,0,2) {$\propto \Tr\left[ a_y^\beta \right] = 0.$};
\end{tikzpicture}
\end{equation}
The black dots mark the possible positions of the operator $a_y^\beta$, i.e., the two sites in $p_m$ that are furthest away from the site we started from. Visually we moved up one step of the pyramid each time we applied the unitary condition. Eq.~\eqref{eq:DUfinal} implies that
\begin{equation}
D^{\alpha \beta}(x,y,t) = 0.
\end{equation}
Finally we look at the case $x = z^i$ for $y=y^i$. So $x$ is the point in $\mathcal{LC}(0)$ furthest away from $y$. While we may still find two sides looking like \eqref{eq:SideVis} and can use the appropriate unitary condition, the final step \eqref{eq:DUfinal} cannot be made, as $a_y^\beta$ would block the required site, i.e., it would be applied on one of the two sites not marked black. But we can actually simplify the situation further. In the shifted picture, again see Fig.~\ref{fig:pyramidshift}, it is easy to see that we can once more use the unitary condition in time. In a similar manner as before, we would find the light-cone of $a_x^\alpha$ to be
\begin{multline}
\label{eq:LCx}
\mathcal{LC}\left( a_x^\alpha\right) = \bigg\{ (z,\tau) \in \{ 0, \cdots , L-1\}^2 \times \{1,\tfrac{3}{2}, \cdots , t\} \, \Big| \\ \min_{\tilde{x} \in p_n} \| z - \tilde{x} \|_\infty \leq 2\tau -1 \bigg\},
\end{multline}
with the plaquette $p_n := p( 2 \left\lfloor \frac{x_1}{2} \right\rfloor, 2 \left\lfloor \frac{x_2}{2} \right\rfloor)$. As both unitary conditions in time can be applied without one resulting in a blocking of the other, what remains are operators applied to points in
\begin{equation}
\mathcal{C} := ( \mathcal{LC}(a_y^\beta) \cap \mathcal{LC} (a_x^\alpha)) \cup \{(x,0) \} .
\end{equation}
Now assume that $x_i < y_i$ for all $i\in \{ 1,2 \}$. If that is not true for some $i$ we may relabel the sites. Since this will be the case if and only if $y_i$ is odd, we can generalize the reassignment as
\begin{equation}
R_i: z_i \mapsto (-1)^{y_i}z_i.
\end{equation}
Now let $x_n$ and $y_m$ be such that they are the closest two sites to the other plaquette in $p_n$ and $p_m$ respectively, i.e.,
\begin{equation}
\norm{x_n - y_m}_\infty = \min_{(\tilde{x}, \tilde{y}) \in p_n \times p_m} \norm{\tilde{x} - \tilde{y}} = 2t - 2 .
\end{equation}
One then finds that
\begin{align}
\label{eq:case1equations}
\abs{(x_n)_i - (y_m)_i} &= 2t-2 \\
(x_n)_i &= x_i+1 \\
(y_m)_i &= y_i-1
\end{align}
for all $i$. Thus we can treat each space direction individually. To calculate the intersection of the two light-cones, we look at two different cases. First assume $(x_n)_i \leq z_i \leq (y_m)_i$. By combining the restrictions for the sets given in \eqref{eq:LCy} and \eqref{eq:LCx} with our findings in \eqref{eq:case1equations}, we get
\begin{eqnarray}
z_i & \geq & y_i - 2t + \tau = x_i + \tau\\
z_i & \leq & x_i + \tau ,
\end{eqnarray}
which implies that
\begin{equation}
z_i = x_i + \tau .
\end{equation}
Due to periodicity in space, the second case is $y_i \leq z_i \leq x_i +L$. Using that by assumption $y_i -x_i = 2t$ and $4t \leq L$ as well as the restrictions imposed by \eqref{eq:LCy} and \eqref{eq:LCx}, we find the inequalities
\begin{eqnarray}
z_i &\leq & y_i +2t-1-\tau \leq x_i + L-1-\tau \\
z_i & \geq & x_i + L -\tau + 1 .
\end{eqnarray}
These are clearly contradictory. Since this is true for all $i\in \{ 1,2\}$ and $\tau \in \{ 1,\frac{1}{2}, \cdots , t-1 \}$, we find that
\begin{multline}
\mathcal{C} = \Bigg\{ (z,\tau ) \in \bigotimes_{i=1}^2 R_i(\{ 0, \cdots ,L-1 \}) \times \left\{0 , \cdots, t - \tfrac{1}{2} \right\} \Big| \\ z = x + \sum_{i=1}^2 2\tau e_i \Bigg\}.
\end{multline}
We also know that only operators applied to events in $\mathcal{C}$ do not reduce to the identity with their adjoint. We will now cancel unitaries with their adjoint wherever possible. The result is visualized by Eq.~\eqref{eq:corr_function_ray_diagram} in the main text.

\section{Computational cost of the M-maps}

In this section we will figure out the computational cost of contracting the tensor network in equation \eqref{eq:corr_function_ray_diagram} with respect to the time $t$ and local dimension $d$. The contraction cost to construct an $M$-map from a ternary unitary $U$ is $\mathcal{O}(d^{10})$. An $M$-map and its powers can be viewed as a 4-tensor in $\C^{d \times d \times d \times d}$. Therefore the cost of contracting powers of $M$-maps is $\mathcal{O}(d^6)$. Both contractions have to be performed $2t$-many times. Since the contraction of a power of an $M$-map with a single-site operator is $\mathcal{O}(d^4)$, the overall cost of contraction scales as $\mathcal{O}(t d^{10})$. If we assume all ternary unitaries $U$ are the same, we only need to compute the $M$-map once and thus the total computational cost scales as $\mathcal{O}(t d^6 + d^{10})$.

\section{Exploration of the M-maps}

For $d = 2$, we can further analyze and evaluate the $M$-maps defined in \eqref{eq:M_def}. Let us start with a simple example. If the ternary unitary $U$ is of the form \eqref{eq:cross_construction}, one of the dual unitaries gets canceled, and what remains is
\begin{equation}\label{1D_M_Map}
\begin{tikzpicture}[baseline=(current bounding box.center), scale=0.75]
\useasboundingbox (-5,-1.5) rectangle (1.5, 1.5);
\node at (-3.5, 0) {$M_{(1,1)}(a) = \mathcal{M}_{-}^{D}(a) = \frac{1}{d}$};
\begin{scope}[shift={(0,-0.85)}]
\filldraw[thick,fill=mred] (-0.35,-0.35) rectangle ( 0.35, 0.35);
\node at ( 0,-0.75){$D$};
% open leg
\draw (-0.35,-0.35) -- (-0.75,-0.75);
\end{scope}
\begin{scope}[shift={(0, 0.85)}]
\filldraw[thick,fill=mblue] (-0.35,-0.35) rectangle ( 0.35, 0.35);
\node at ( 0, 0.75) {$D^\dagger$};
% open leg
\draw (-0.35, 0.35) -- (-0.75, 0.75);
\end{scope}
\draw ( 0.35,-0.5) to[out= 45,in=- 45] ( 0.35, 0.5);
\draw (-0.35,-0.5) to[out=135,in=-135] (-0.35, 0.5);
\draw ( 0.35,-1.2) to[out=-45, in=45, looseness=2] ( 0.35, 1.2);
\filldraw (0.56, 0) circle (0.075) node[right] {$a$};
\end{tikzpicture} \, ,
\end{equation}
where $\mathcal{M}_{-}^{D}$ is the one-dimensional analogue defined in \cite{Bertini2019}, and $D$ the remaining dual-unitary gate from \eqref{eq:cross_construction}. In the qubit case, $D$ can be parametrized as \cite{Bertini2019}
\begin{equation}
\label{eq:dual_unitary_parametrization}
D = \left( u_+ \otimes v_+ \right) \, V(J) \, \left( u_- \otimes v_- \right),
\end{equation}
where $u_\pm$, $v_\pm \in \text{SU}(2)$, $J \in \mathbb{R}$ and
\begin{equation}
V(J) = \exp\!\left[-i\left( \frac{\pi}{4}\left( \sigma_x \otimes \sigma_x + \sigma_y \otimes \sigma_y \right) + J \sigma_z \otimes \sigma_z \right) \right],
\end{equation}
with $\sigma_k$ the Pauli matrices. It was also shown that
\begin{equation}
\mathcal{M}_{-}^{V(J)} = \diag(1, \sin(2J), \sin(2J), 1)
\end{equation}
for the choice $\left\{ a^\alpha \right\} = \{ \mathbbm{1}, \sigma_x , \sigma_y, \sigma_z \}$. We can therefore interpret $\mathcal{M}_{-}^{V(J)}$ as a contraction on the Bloch-sphere in $x$- and $y$-direction. Inserting the parametrization \eqref{eq:dual_unitary_parametrization} leads to
\begin{equation}
\begin{split}
\mathcal{M}_{-}^{D}(a) %
&= u_-^\dagger \mathcal{M}_{-}^{V(J)} \left(v_+^\dagger a v_+ \right) u_- \\
&= R_{u_-} \left( \mathcal{M}_{-}^{V(J)} \left( R_{v_+}(a) \right) \right),
\end{split}
\end{equation}
where we reinterpreted conjugations by the local unitaries as Bloch sphere rotations $R_u$.

For the construction \eqref{eq:four_dual_construction}, the map in \eqref{eq:M_example} reduces to nested applications of two dual unitary gates, denoted $D_1$ and $D_2$:
\begin{equation}
\begin{tikzpicture}[baseline=(current bounding box.center), scale=0.75]
\useasboundingbox (-7,-3.25) rectangle (2.5, 3.25);
\node at (-4, 0) {$M_{(1,1)}(a) = \frac{1}{d^2}$};
\begin{scope}[shift={(0,-0.85)}]
\filldraw[thick,fill=mred] (-0.35,-0.35) rectangle ( 0.35, 0.35);
\node at ( 0,-0.75){$D_1$};
\draw (-0.35,-0.35) -- (-0.75,-0.75);
\end{scope}
\begin{scope}[shift={(-1.1,-1.95)}]
\filldraw[thick,fill=mred] (-0.35,-0.35) rectangle ( 0.35, 0.35);
\node at ( 0,-0.75){$D_2$};
\draw (-0.35,-0.35) -- (-0.75,-0.75);
\end{scope}
\begin{scope}[shift={(0, 0.85)}]
\filldraw[thick,fill=mblue] (-0.35,-0.35) rectangle ( 0.35, 0.35);
\node at ( 0, 0.75) {$D_1^\dagger$};
\draw (-0.35, 0.35) -- (-0.75, 0.75);
\end{scope}
\begin{scope}[shift={(-1.1, 1.95)}]
\filldraw[thick,fill=mblue] (-0.35,-0.35) rectangle ( 0.35, 0.35);
\node at ( 0, 0.75) {$D_2^\dagger$};
\draw (-0.35, 0.35) -- (-0.75, 0.75);
\end{scope}
\draw ( 0.35,-0.5) to[out= 45,in=- 45] ( 0.35, 0.5);
\draw (-0.35,-0.5) to[out=135,in=-135] (-0.35, 0.5);
\draw (-1.45,-1.6) to[out=135,in=-135] (-1.45, 1.6);
\draw ( 0.35,-1.2) to[out=-45, in=45, looseness=2] ( 0.35, 1.2);
\draw (-0.75,-2.3) to[out=-45, in=45, looseness=3] (-0.75, 2.3);
\filldraw (0.56, 0) circle (0.075) node[right] {$a$};
\end{tikzpicture}.
\end{equation}
This can be rewritten as
\begin{equation}
M_{(1,1)} = R_3 \circ \mathcal{M}_{-}^{V(J_2)} \circ R_2 \circ \mathcal{M}_{-}^{V(J_1)} \circ R_1
\end{equation}
for appropriately chosen single-qubit rotations $R_j$, $j = 1, 2, 3$, and $J_1, J_2 \in \R$.

\section{Proof of the classification of solvable PEPS}
In this section we give the full proof of Theorem~\ref{thm:solvable_peps} in the main text. First let us restate a definition and result from \cite{Piroli2020}. They considered two-site shift-invariant MPS on a spin chain of even length $L$ with periodic boundary conditions
\begin{equation}
\ket{\psi_L (\tilde{N})} = \sum_{ \{ i_j \} } \Tr \left( \tilde{N}^{i_0,i_1}\dots \tilde{N}^{i_{L-2,L-1}} \right) \ket{i_0, \dots, i_{L_1}},
\end{equation}
where $\{\tilde{N}^{i,j}\}_{i,j=0}^{D-1}$ is a set of $\chi$-dimensional matrices, and $D$ the physical dimension. Such an MPS is considered solvable if
\begin{enumerate}
\item The transfer operator of $\tilde{N}$, defined as
\begin{equation}
\begin{tikzpicture}[baseline=(current  bounding  box.center), scale=0.3]
\node[anchor=east] at (1,0){$\tau ( \tilde{N} )=$};

\filldraw[fill=gray, fill opacity=0.5] (2,-3) rectangle (5,-1);
\filldraw[fill=gray, fill opacity=0.5] (2,1) rectangle (5,3);

\draw (2,-1) -- (2,1);
\draw (5,-1) -- (5,1);

\draw (2,-2) -- (1,-2);
\draw (5,-2) -- (6,-2);
\draw (2,2) -- (1,2);
\draw (5,2) -- (6,2);

\node[anchor=north] at (3.5,-1){$\tilde{N}$};
\node[anchor=north] at (3.5,3){${\tilde{N}}^*$};
\end{tikzpicture},
\end{equation}
has a unique largest eigenvalue $\lambda=1$ with algebraic multiplicity $1$.
\item There exists a non-zero $\chi$-dimensional matrix $\Sigma$ such that
\begin{equation}
\begin{tikzpicture}[baseline=(current  bounding  box.center), scale=0.3]
\useasboundingbox (0,-3) rectangle (10, 3);

\filldraw[fill=gray, fill opacity=0.5] (2,-3) rectangle (5,-1);
\filldraw[fill=gray, fill opacity=0.5] (2, 1) rectangle (5, 3);

% Physical Legs
\draw (2,-1) -- (2,-0.5);
\draw (2,1) -- (2,0.5);
\draw (5,-1) -- (5,1);

% Left virtual legs
\draw (2,-2) -- (1,-2);
\draw (2,2) -- (1,2);

% Right virtual legs
\draw (5,2) to[out=20, in=-20, looseness=2] (5,-2);

\node[anchor=north] at (3.5,-1){$\tilde{N}$};
\node[anchor=north] at (3.5,3){${\tilde{N}}^*$};

% sigma matrix
\node[diamond, draw, fill=mgreen] (d) at (7.1,0) {};
\node[anchor=west] at (7.5,0){$\Sigma$};

\node[anchor=west] at (9,0){$=$};

\begin{scope}[shift={(10,0)}]
% Physical Legs
\draw (2.5,-1) -- (2,-1) -- (2,-0.5);
\draw (2.5,-1) to[out=20, in=-20, looseness=1] (2.5,1);
\draw (2.5,1) --  (2,1) -- (2,0.5);

% Left virtual legs
\draw (2,-2) -- (1,-2);
\draw (2,-2) to[out=10, in=-10, looseness=2] (2,2);
\draw (2,2) -- (1,2);

% Sigma Matrix
\node[diamond, draw, fill=mgreen] (d) at (4.1,0) {};
\node[anchor=west] at (4.6,0){$\Sigma$};
\end{scope}
\end{tikzpicture}
\end{equation}
\end{enumerate}
A weaker version of Theorem~1 in \cite{Piroli2020} is
\begin{theorem}[Classification of solvable MPS]
\label{thm:solvable_MPS}
Any solvable MPS $\ket{\psi_L (\tilde{N}) }$ is equivalent in the thermodynamic limit to some two-site shift-invariant MPS $\ket{\psi_L (N)}$ such that
\begin{equation}\label{eq:solv_cond_MPS}
\sum_{k=1}^D N^{i,k} \left( N^{k,j} \right)^\dagger = \frac{1}{D} \delta_{i,j} \mathbbm{1}.
\end{equation}
\end{theorem}
Note that the property \eqref{eq:solv_cond_MPS} is equivalent to the matrix resulting from the combination of one physical leg with one virtual leg each being proportional to a unitary matrix, i.e.,
\begin{equation}
\begin{tikzpicture}[baseline=(current  bounding  box.center), scale=0.3]
\useasboundingbox (0,-2.5) rectangle (10, 2.5);

\filldraw[fill=gray, fill opacity=0.5] (2,-3) rectangle (5,-1);
\filldraw[fill=gray, fill opacity=0.5] (2,1) rectangle (5,3);

% Physical Legs
\draw (2,-1) -- (2,-0.5);
\draw (2,1) -- (2,0.5);
\draw (5,-1) -- (5,1);

% Left virtual legs
\draw (2,-2) -- (1,-2);
\draw (2,2) -- (1,2);

% Right virtual legs
\draw (5,2) to[out=20, in=-20, looseness=2] (5,-2);

\node[anchor=north] at (3.5,-1){$N$};
\node[anchor=north] at (3.5,3){$N^*$};

\node[anchor=west] at (7.5,0){$=\frac{1}{D}$};

\begin{scope}[shift={(9,0)}]
% Physical Legs
\draw (2.5,-1) -- (2,-1) -- (2,-0.5);
\draw (2.5,-1) to[out=20, in=-20, looseness=1] (2.5,1);
\draw (2.5,1) --  (2,1) -- (2,0.5);

% Left virtual legs
\draw (2,-2) -- (1,-2);
\draw (2,-2) to[out=10, in=-10, looseness=2] (2,2);
\draw (2,2) -- (1,2);

\end{scope}
\end{tikzpicture} .
\end{equation}
To use the above theorem in our proof, we recast our two-dimensional sPEPS $\ket{\Psi_{L_1 L_2} [\tilde{ \Lambda } ]}$ into the one-dimensional MPS $\ket{\psi_{L_1} [\tilde{\Lambda}^{L_2}]}$. Clearly the two transfer operators are the same, $T(\tilde{\Lambda}) = \tau (\tilde{\Lambda}^{L_2})$, so both have the same eigenvalues. Furthermore with $\Sigma = S$, where $S$ is to be reinterpreted as a $\chi_1^{L_2}$ dimensional matrix, the condition \eqref{eq:solvable_condition} implies condition \eqref{eq:solv_cond_MPS}. Therefore the MPS $\ket{\psi_{L_1} [\tilde{\Lambda}^{L_2}]}$ is solvable and by Theorem~\ref{thm:solvable_MPS} there exists an equivalent MPS $\ket{\psi_{L_1}[N]}$ with $N$ fulfilling \eqref{eq:solv_cond_MPS}. During the proof in \cite{Piroli2020} it was found that $\Sigma$ is strictly positive and
\begin{equation}
N^{i,j} = \Sigma^{-\frac{1}{2}} \tilde{N}^{i,j} \Sigma^{\frac{1}{2}}.
\end{equation}
We can apply this to our situation, which means
\begin{equation}
N = S^{-\frac{1}{2}} \tilde{\Lambda}^{L_2} S^{\frac{1}{2}},
\end{equation}
where we left out the physical indices for convenience. Due to the shift-invariance of our sPEPS and the periodic boundary conditions, we can choose the tensor $S$ to be a shift-invariant MPO generated by a local tensor $s$ and the same is true for $S^{\frac{1}{2}}$ and $S^{-\frac{1}{2}}$. With a slight abuse of notation, we say $S^{\frac{1}{2}}$ and $S^{-\frac{1}{2}}$ are  generated by the local tensors $s^{\frac{1}{2}}$ and $s^{-\frac{1}{2}}$ respectively. Now we can return to the two-dimensional picture and look at a two-site shift-invariant PEPS with local tensors
\begin{equation}
\Lambda = s^{-\frac{1}{2}} \tilde{\Lambda} s^{\frac{1}{2}}.
\end{equation}
Then \eqref{eq:solv_cond_MPS} with $D= d^{L_2}$ implies thats the MPO generated by $\Lambda$ is an MPU up to the scalar factor $D$, which concludes the proof.

\section{Algebraic derivation of expectation value dynamics}
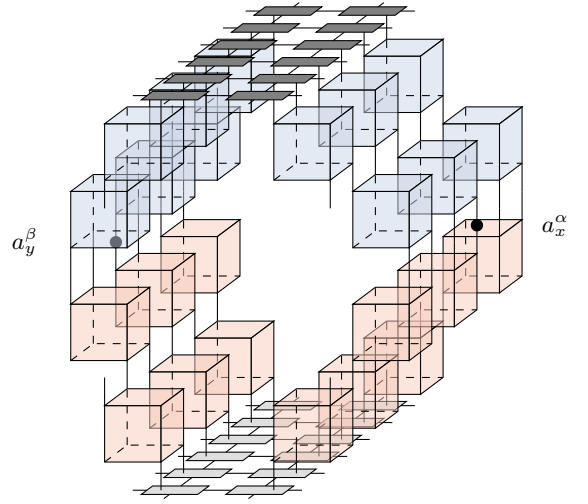
\begin{figure}
\begin{tikzpicture}[baseline=(current bounding box.center), scale=0.375]
\pgfsetxvec{\pgfpoint{1cm}{0cm}}
\pgfsetyvec{\pgfpoint{0.4cm}{0.3cm}}
\pgfsetzvec{\pgfpoint{0cm}{1cm}}

\foreach \i in {0,...,5}{
    \begin{scope}[shift={(0,2*\i,-7)}]
        \begin{scope}[canvas is xy plane at z=0]
            \filldraw[fill=lightgray, fill opacity=0.5] (-2.5,-0.5) rectangle (-0.5,0.5);
            \filldraw[fill=lightgray, fill opacity=0.5] (2.5,-0.5) rectangle (0.5,0.5);
            \draw (-2.5,0) -- (-3,0);
            \draw (-0.5,0) -- (0.5,0);
            \draw (2.5,0) -- (3,0);
            \draw (-1.5,-0.5) -- (-1.5,-1);
            \draw (-1.5,0.5) -- (-1.5,1);
            \draw (1.5,-0.5) -- (1.5,-1);
            \draw (1.5,0.5) -- (1.5,1);
        \end{scope}
    \end{scope}
}
\foreach \i in {2,1,0}{
    \begin{scope}[shift={(0,4*\i+1,-5)}]
        \begin{scope}[shift={(-3,0,0)}]
            \drawcube{2}{lightred};
            \draw (1,-1,-1) -- (1,-1,-2);
            \draw (1,1,-1) -- (1,1,-2);
            \draw (-1,-1,1) -- (-1,-1,2);
            \draw (-1,1,1) -- (-1,1,2);
        \end{scope}
        \begin{scope}[shift={(3,0,0)}]
            \drawcube{2}{lightred};
            \draw (-1,-1,-1) -- (-1,-1,-2);
            \draw (-1,1,-1) -- (-1,1,-2);
            \draw (1,-1,1) -- (1,-1,2);
            \draw (1,1,1) -- (1,1,2);
        \end{scope}
    \end{scope}
    \begin{scope}[shift={(0,4*\i+3,-2)}]
        \begin{scope}[shift={(-5,0,0)}]
            \drawcube{2}{lightred};
            \draw (-1,-1,1) -- (-1,-1,2);
            \draw (-1,1,1) -- (-1,1,2);
        \end{scope}
        \begin{scope}[shift={(5,0,0)}]
            \drawcube{2}{lightred};
            \draw (1,-1,1) -- (1,-1,2);
            \draw (1,1,1) -- (1,1,2);
        \end{scope}
    \end{scope}
}
\begin{scope}[canvas is xz plane at y=6]
    \filldraw (-6,0) circle (0.2);
    \node[anchor=west] at (-10,0){$a_y^\beta$};
\end{scope}
\foreach \i in {2,1,0}{
    \begin{scope}[shift={(0,4*\i+3,2)}]
        \begin{scope}[shift={(-5,0,0)}]
            \draw (-1,-1,-1) -- (-1,-1,-2);
            \draw (-1,1,-1) -- (-1,1,-2);
            \drawcube{2}{lightblue};
        \end{scope}
        \begin{scope}[shift={(5,0,0)}]
            \draw (1,-1,-1) -- (1,-1,-2);
            \draw (1,1,-1) -- (1,1,-2);
            \drawcube{2}{lightblue};
        \end{scope}
    \end{scope}
    \begin{scope}[shift={(0,4*\i+1,5)}]
        \begin{scope}[shift={(-3,0,0)}]
            \draw (-1,-1,-1) -- (-1,-1,-2);
            \draw (-1,1,-1) -- (-1,1,-2);
            \drawcube{2}{lightblue};
            \draw (1,-1,1) -- (1,-1,2);
            \draw (1,1,1) -- (1,1,2);
        \end{scope}
        \begin{scope}[shift={(3,0,0)}]
            \draw (1,-1,-1) -- (1,-1,-2);
            \draw (1,1,-1) -- (1,1,-2);
            \drawcube{2}{lightblue};
            \draw (-1,-1,1) -- (-1,-1,2);
            \draw (-1,1,1) -- (-1,1,2);
        \end{scope}
    \end{scope}
}
\foreach \i in {0,...,5}{
    \begin{scope}[shift={(0,2*\i,7)}]
        \begin{scope}[canvas is xy plane at z=0]
            \filldraw[fill=gray, fill opacity=1] (-2.5,-0.5) rectangle (-0.5,0.5);
            \filldraw[fill=gray, fill opacity=1] (2.5,-0.5) rectangle (0.5,0.5);
            \draw (-2.5,0) -- (-3,0);
            \draw (-0.5,0) -- (0.5,0);
            \draw (2.5,0) -- (3,0);
            \draw (-1.5,-0.5) -- (-1.5,-1);
            \draw (-1.5,0.5) -- (-1.5,1);
            \draw (1.5,-0.5) -- (1.5,-1);
            \draw (1.5,0.5) -- (1.5,1);
        \end{scope}
    \end{scope}
}
\begin{scope}[canvas is xz plane at y=8]
    \filldraw (6,0) circle (0.2);
    \node[anchor=west] at (8,0){$a_x^\alpha$};
\end{scope}
\end{tikzpicture}
\caption{The remaining tensors network after the applying the results from \cite{Piroli2020} for $r_1=7$ and $t=1$. For visual clarity, not all legs are shown. The legs that are drawn on the unitary operators as well as the $\beta$-legs of the $\Lambda$-tensors are connected periodically. All other open legs of the unitary operators are contracted with the corresponding legs of their adjoint and the same is true for the open physical and $\alpha$-legs of the $\Lambda$-tensors. We can clearly see the two tilted walls of operators mentioned in the text.}
\label{fig:tilted_opwall}
\end{figure}

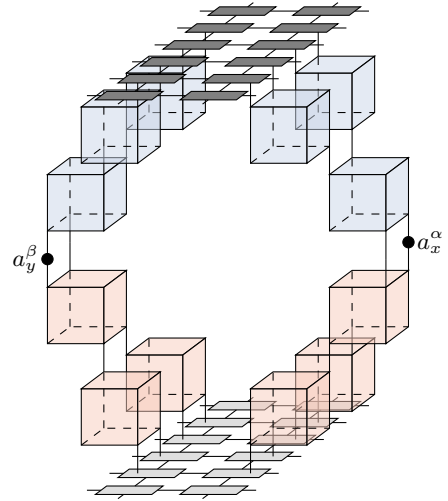
\begin{figure}
\begin{tikzpicture}[baseline=(current bounding box.center), scale=0.375]
\pgfsetxvec{\pgfpoint{1cm}{0cm}}
\pgfsetyvec{\pgfpoint{0.4cm}{0.3cm}}
\pgfsetzvec{\pgfpoint{0cm}{1cm}}

\foreach \i in {0,...,5}{
    \begin{scope}[shift={(0,2*\i,-7)}]
        \begin{scope}[canvas is xy plane at z=0]
            \filldraw[fill=lightgray, fill opacity=0.5] (-2.5,-0.5) rectangle (-0.5,0.5);
            \filldraw[fill=lightgray, fill opacity=0.5] ( 2.5,-0.5) rectangle (0.5,0.5);
            \draw (-2.5,0) -- (-3,0);
            \draw (-0.5,0) -- (0.5,0);
            \draw (2.5,0) -- (3,0);
            \draw (-1.5,-0.5) -- (-1.5,-1);
            \draw (-1.5,0.5) -- (-1.5,1);
            \draw (1.5,-0.5) -- (1.5,-1);
            \draw (1.5,0.5) -- (1.5,1);
        \end{scope}
    \end{scope}
}
\foreach \i in {7,3}{
    \begin{scope}[shift={(0,\i,-5)}]
        \begin{scope}[shift={(-3,0,0)}]
            \draw (1,-1,-1) -- (1,-1,-2);
            \draw (1,1,-1) -- (1,1,-2);
            \drawcube{2}{lightred};
        \end{scope}
        \begin{scope}[shift={(3,0,0)}]
            \draw (-1,-1,-1) -- (-1,-1,-2);
            \draw (-1,1,-1) -- (-1,1,-2);
            \drawcube{2}{lightred};
        \end{scope}
    \end{scope}
}
\begin{scope}[shift={(0,5,-2)}]
    \begin{scope}[shift={(-5,0,0)}]
        \draw (1,-1,-1) -- (1,-1,-2);
        \draw (1,1,-1) -- (1,1,-2);
        \drawcube{2}{lightred};
        \draw (-1,-1,1) -- (-1,-1,2);
        \draw (-1,1,1) -- (-1,1,2);
    \end{scope}
    \begin{scope}[shift={(5,0,0)}]
        \draw (-1,-1,-1) -- (-1,-1,-2);
        \draw (-1,1,-1) -- (-1,1,-2);
        \drawcube{2}{lightred};
        \draw (1,-1,1) -- (1,-1,2);
        \draw (1,1,1) -- (1,1,2);
    \end{scope}
\end{scope}

\begin{scope}[shift={(0,5,2)}]
    \begin{scope}[shift={(-5,0,0)}]
        \draw (-1,-1,-1) -- (-1,-1,-2);
        \draw (-1,1,-1) -- (-1,1,-2);
        \drawcube{2}{lightblue};
        \draw (1,-1,1) -- (1,-1,2);
        \draw (1,1,1) -- (1,1,2);
    \end{scope}
    \begin{scope}[shift={(5,0,0)}]
        \draw (1,-1,-1) -- (1,-1,-2);
        \draw (1,1,-1) -- (1,1,-2);
        \drawcube{2}{lightblue};
        \draw (-1,-1,1) -- (-1,-1,2);
        \draw (-1,1,1) -- (-1,1,2);
    \end{scope}
\end{scope}
\foreach \i in {7,3}{
    \begin{scope}[shift={(0,\i,5)}]
        \begin{scope}[shift={(-3,0,0)}]
            \drawcube{2}{lightblue};
            \draw (1,-1,1) -- (1,-1,2);
            \draw (1,1,1) -- (1,1,2);
        \end{scope}
        \begin{scope}[shift={(3,0,0)}]
            \drawcube{2}{lightblue};
            \draw (-1,-1,1) -- (-1,-1,2);
            \draw (-1,1,1) -- (-1,1,2);
        \end{scope}
    \end{scope}
}
\foreach \i in {0,...,5}{
    \begin{scope}[shift={(0,2*\i,7)}]
        \begin{scope}[canvas is xy plane at z=0]
            \filldraw[fill=gray, fill opacity=1] (-2.5,-0.5) rectangle (-0.5,0.5);
            \filldraw[fill=gray, fill opacity=1] ( 2.5,-0.5) rectangle (0.5,0.5);
            \draw (-2.5,0) -- (-3,0);
            \draw (-0.5,0) -- (0.5,0);
            \draw (2.5,0) -- (3,0);
            \draw (-1.5,-0.5) -- (-1.5,-1);
            \draw (-1.5,0.5) -- (-1.5,1);
            \draw (1.5,-0.5) -- (1.5,-1);
            \draw (1.5,0.5) -- (1.5,1);
        \end{scope}
    \end{scope}
}
\begin{scope}[canvas is xz plane at y=6]
    \filldraw (6,0) circle (0.2);
    \node[anchor=west] at (6,0){$a_x^\alpha$};
\end{scope}
\begin{scope}[canvas is xz plane at y=4]
    \filldraw (-6,0) circle (0.2);
    \node[anchor=east] at (-6,0){$a_y^\beta$};
\end{scope}
\end{tikzpicture}
\caption{The tensor network that remains after using the conventional unitary property for the example $r_1=7$ and $t=1$. Again the legs of the unitaries as well as the physical and $\alpha$-legs of the $\Lambda$-tensors are contracted with the corresponding leg of their adjoint. The tilted triangles in this case each consist of three ternary unitary operators. Note that this picture is shifted by $-1$ in the $x_2$-direction compared to Fig. \ref{fig:tilted_opwall}}
\label{fig:tilted_triangles}
\end{figure}

In this section we will derive the specific form the dynamics of two-point expectation values \eqref{eq:2D-expval}. To achieve this, we will once more use results from \cite{Piroli2020}. They investigated the dynamics of solvable MPS or rather their thermodynamic equivalent as defined in Theorem~\ref{thm:solvable_MPS}. The time evolution considered was a chequerboard-pattern of dual-unitary operators $\mathcal{U}$. Similar to our expectation value $E ( a_x^\alpha, a_y^\beta)$ as given in \eqref{eq:2D-expval}, they computed the quantity
\begin{equation}\label{eq:1D_expval}
\mathcal{E}(\omega_m^\alpha ,\omega_n^\beta ,t) = \lim_{L \rightarrow \infty} \bra{\Psi_L (t)} \omega_m^\alpha \otimes \omega_n^\beta \ket{\Psi_L (t)} ,
\end{equation}
where $\omega_m^\alpha$ and $\omega_n^\beta$ are Hilbert-Schmidt orthonormal single-site operators and without loss of generality, we assume $n\leq m$. We can recast our tensor network in Fig.~\ref{fig:sPEPS-set-up} into the one-dimensional case. Assuming $L_2$ is sufficiently large (we will get back to how large later on) we identify
\begin{align*}
\mathcal{U}_{k,k+1} & \leftarrow \bigotimes_{l=0}^{L_2/2} U_{p(k,2l)} & \text{ for } k \text{ even} \\
\mathcal{U}_{k,k+1} & \leftarrow \bigotimes_{l=0}^{L_2/2} U_{p(k,2l+1)} & \text{ for } k \text{ odd} \\
\omega_{x_1}^\alpha & \leftarrow a_x^\alpha & \text{analogous for } \beta,y  \\
N & \leftarrow \Lambda^{L_2}. &
\end{align*}
From the previous section, we know that the state defined by local MPU tensors interpreted as $N$ is a state fulfilling \eqref{eq:solv_cond_MPS}. Furthermore, $U$ being a ternary unitary implies $\mathcal{U}$ is a dual unitary. Therefore
\begin{equation}\label{eq:exp_equal}
\mathcal{E}\!\left( \omega_{x_1}^\alpha ,\omega_{y_1}^\beta ,t \right) = E\!\left( a_x^\alpha , a_y^\beta ,t \right).
\end{equation}
We basically look at our network from the $x_1$-$t$-plane. Note that in \cite{Piroli2020} also half-time steps are allowed, while we only allow full time-steps in accordance with our physical motivation of the time-evolution. Due to $\mathcal{U}$ being unitary, all of dual unitary operators applied outside of the sites $\{y_1-2t-1,\dots,x_1+2t+1\}$ cancel. Thus when going to the limit $L_1 \rightarrow \infty$, there will be infinitely many transfer operators $\tau (N)$ applied left of the site $m-2t-1$ and right of the site $n+2t+1$. Since $\tau (N)$ has $1$ as its unique largest eigenvalue, we can replace these infinite transfer operators by the identity connecting the corresponding virtual leg of the tensors $N$ with the virtual leg of $N^*$ \cite{Piroli2020}. Now we only have to consider the remaining sites $\{m-2t-1,\dots,n+2t+1\}$. Due to even odd effects, the usual unitary condition and the solvable condition \eqref{eq:solv_cond_MPS} there are a lot of coordinates $m,n$ for which $\mathcal{E}$ and by \eqref{eq:exp_equal} also $E$ is $0$. As we assumed $r_1 > 0$, this can be expressed as \cite{Piroli2020}
\begin{multline}
E (a_x^\alpha ,a_y^\beta ,t) = \delta_{x_1 \text{mod} 2, 0} \delta_{r_1 \text{mod}2 , 1} \Theta(r-4t-1)\\
\cdot E(a_x^\alpha ,a_y^\beta ,t),
\end{multline}
where $\Theta$ denotes the Heaviside function. For further details on the graphical proof consult \cite{Piroli2020}. Using once more the results from \cite{Piroli2020}, for $x_1,y_1$ that give a non-zero $E$ we can utilise that $\mathcal{U}$ is unitary in both time and $x_1$-direction to cancel most of these operators. The MPU property \eqref{eq:MPU_cond} will in turn allow us to cancel some of the remaining local operators $\Lambda$ with their adjoint. To determine which operators and tensors remain after cancellation, we consider the points to which they are applied. According to \cite{Piroli2020} operators $\mathcal{U}$ will not cancel with their adjoint if they are applied to points in the set
\begin{equation}
K = \left\{ (z,\tau ) \, \Big| \, \left( z_1 = x_1 -2\tau \right) \lor \left( z_1 = y_1 + 2\tau \right) \right\}
\end{equation}
and the local tensors $\Lambda$ do not cancel if their physical legs represent the sites with coordinates in the set
\begin{equation}
\mathcal{T} = \left\{ z \, \Big| \, y_1 + 2t \leq z_1 \leq x_1 -2t \right\}.
\end{equation}
Now we return into the two-dimensional picture completely. The non-zero expectation values consist of two tilted walls of unitary operators connected at the base by rings of local tensors $\Lambda^{L_2}$. An example can be seen in Fig.~\ref{fig:tilted_opwall}.
We can see that in the second dimension we also have a light-cone structure around the operators $a_x^\alpha$ and $a_y^\beta$. Thus even more operators can be cancelled via the conventional unitary condition. Only those operators that are applied in a light-cone, cf.~\eqref{eq:LCy}, of either single-site operator do not cancel. Thus the remaining operators are those applied to events in
\begin{equation}
\mathcal{C}_{\text{opr}} = K \cap \left( \mathcal{LC} \left( a^\alpha_x \right) \cup \mathcal{LC} \left( a^\beta_y \right) \right) .
\end{equation}
An example is visualized in Fig.~\ref{fig:tilted_triangles}.
We now move our attention to the remaining local tensors. For later use define $\mathcal{C}_{\text{opr}}^x = K \cap \mathcal{LC} \left( a^\alpha_x \right)$ and $\mathcal{C}_{\text{opr}}^y$ analogously. These are the sets of points on which the two tilted disjoint triangles of ternary unitary operators are applied to. We further denote as $U^x_{\min}$ the operator that is applied to the event in $\mathcal{C}^x_{\text{opr}}$ with minimal $x_2$-coordinate. The corresponding site will be denoted as $x^{\min}$. We analogously define the operators $U_{\max}^x$, $U_{\min}^y$ and $U_{\max}^y$ and their coordinates, which we call extremal sites. On the other hand what remains of the local tensors is are $\ell_2$-many rings $\Lambda^{L_2}$-tensors with
\begin{equation}
\ell_2 = \frac{r_1-4t+1}{2} .
\end{equation}
Note that on the extremal sites the respective operators are connected to an $\ell_x$-long row of $\Lambda$-tensors. Let us assume $L_2> 8t + r_2 + \delta L_2$ with $\delta L_2 = 2 \left\lceil \frac{\ell_1}{2} \right\rceil$. This bound comes about, as we want to use the simple properties \eqref{eq:simple_prop_1} and \eqref{eq:simple_prop_2} to completely disconnect two neighbouring rows of local tensors. One can show graphically that they allow us to remove all local tensors $\Lambda$ that are applied outside of the set
\begin{multline}
\mathcal{C}_{\text{tens}} = \Big\{ z \in \mathcal{T} \, \Big| \, y^{\min}_2 - r^{\min}_1 -1 \leq\\
z_2 \leq x^{\min}_2 + r^{\min}_1 +1 \Big\}, 
\end{multline}
where $r^{\min}_1 = \min_{\tilde{z} = x^{\min}_1 y^{\min}_1} \abs{z_1 - \tilde{z}}$. The open legs are accordingly connected via the tensors $\lambda$ and $\gamma$.
To continue we may look at two different cases determined by the quantity $\Delta(x,y)$ in \eqref{eq:Delta}.
The simplest case occurs for $\Delta(x,y)>4t + \delta L_2$. Here the two sets $\mathcal{C}_{\text{opr}}^x$ and $\mathcal{C}_{\text{opr}}^y$ do not share any $x_2$ coordinate and are far enough apart, such that the application of the simple MPU properties allows us to separate two full rows of local tensors between the two sets. Therefore we can use the simple properties \eqref{eq:simple_prop_1} and \eqref{eq:simple_prop_2} to cancel all local tensors $\Lambda$. In turn the unitary condition of $U$ is now sufficient to cancel all unitary operators. Therefore
\begin{equation}
E\!\left( a^\alpha_x , a^\beta_y,t \right) \propto \Tr\!\left[ a^\alpha_x \right] = 0.
\end{equation}
In the other case, we cannot simplify any further.

\section{Details of the numerical simulations}

This section provides details of the numerical simulations that yield Fig.~\ref{fig:equal_time_corr_t1}. Most notably, we did not only assume that our local tensors $\Lambda$ to give rise to simple MPU, but furthermore we parametrised them specifically as
\begin{equation}
\begin{tikzpicture}[baseline=(current bounding box.center), scale=0.375]
\filldraw[fill=lightgray] (-1,-1) rectangle (1,1);
\node at (0,0){$\Lambda$};
\draw ( 0, 1) -- ( 0, 2) node[anchor=east]  {\small $(j,\mu_2)$};
\draw ( 0,-1) -- ( 0,-2) node[anchor=west]  {\small $(i,\mu_1)$};
\draw ( 1, 0) -- ( 2, 0) node[anchor=south] {\small $\eta_1$};
\draw (-1, 0) -- (-2, 0) node[anchor=north] {\small $\eta_2$};
\node[anchor=west] at (2.5,0){$=$};
\begin{scope}[shift = {(9,0)}]
% A-Block
\filldraw[fill=lightgray] (-2,2) rectangle (2,4);
\node at (0,3){$A$};
% B-Block
\filldraw[fill=lightgray] (-2,-2) rectangle (2,-4);
\node at (0,-3){$B$};
% Central legs
\draw (0,6)node[anchor=east]{\small $(j,\mu_2)$} -- (0,4);
\draw (0,2) -- (0,-2);
\node[anchor=west] at (0,0) {\small $c$};
\draw (0,-4) -- (0,-6)node[anchor=west]{\small $(i,\mu_1)$};
% Right leg A
\draw (2,2) to[in=180,  out=-80, looseness=1.5] ( 4,-0.25);
% Right leg B
\draw (2,-2) to[in=180, out=80,  looseness=1.5] ( 4, 0.25) node[anchor=south]{\small $\eta_1$};
% Left leg A
\draw (-2,2) to[in=0,   out=-100,looseness=1.5] (-4,-0.25) node[anchor=north]{\small $\eta_2$};
% Left leg B
\draw (-2,-2) to[in=0,  out=100, looseness=1.5] (-4, 0.25);
\end{scope}
\end{tikzpicture},
\end{equation}
where $A, B$ are unitary $(d \chi_1 )$-matrices and $\text{dim} (c) = \chi_c = d \chi_1 / \chi_2 \in \N$. One can easily check that this construction gives rise to a simple tensor.
Now we will explain the pattern of further pairs $(\Delta x_1, \Delta x_2)$ giving rise to $E=0$.

\begin{figure}[!ht]
\centering
\includegraphics[width=0.65\columnwidth]{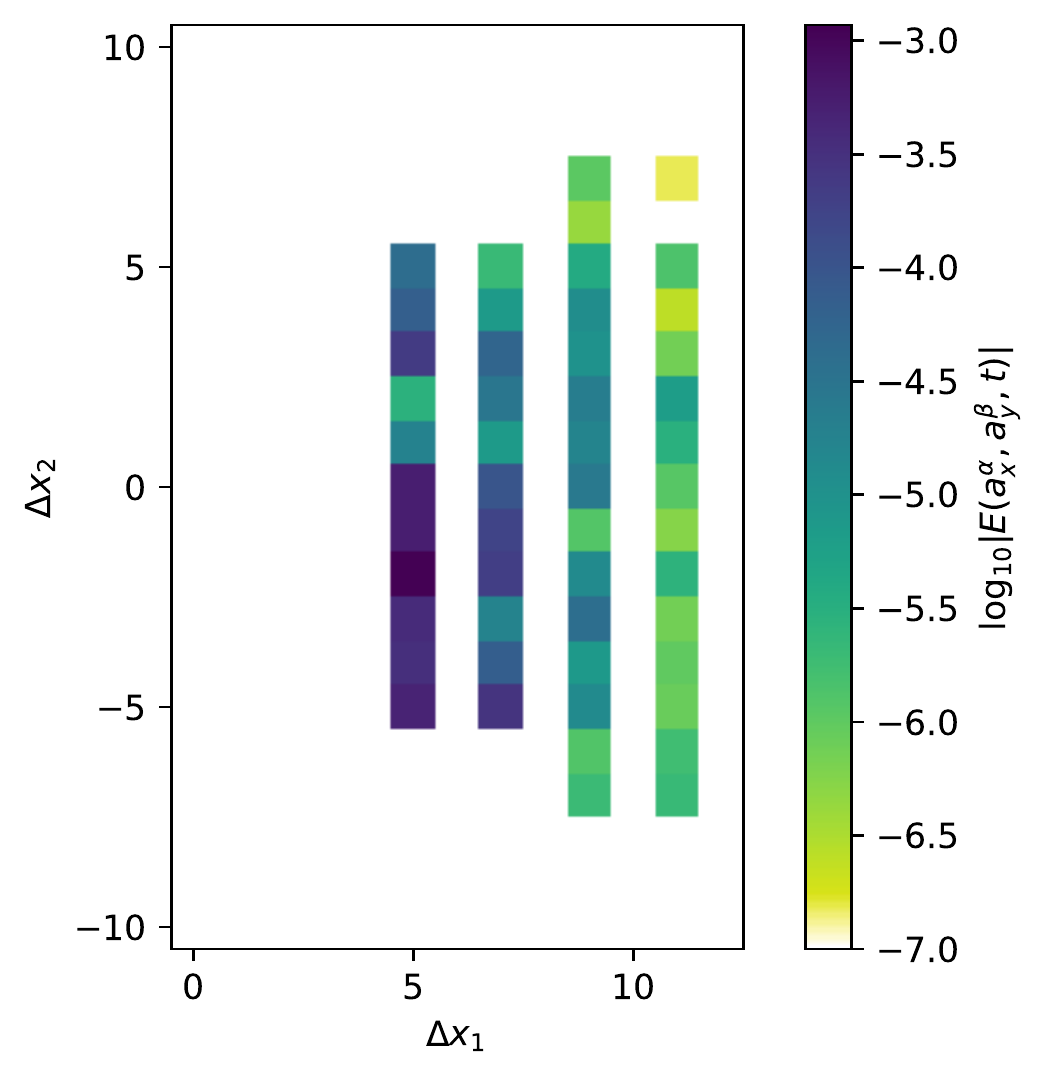}
\caption{Equal-time correlation functions for $t = 1$ and a random simple MPU tensor $\Lambda$ defining the sPEPS, an operator that is dual unitary with respect to the $x_1$-direction, and random traceless local operators $a_x^{\alpha}$ and $a_y^{\beta}$.}
\label{fig:equal_time_corr_dual_t1}
\end{figure}

Due to a finite $\Delta x_2$ the two triangles of operators are shifted against one another. Therefore some of the $\Lambda$ tensor on the outermost (with respect to the $x_1$-direction) rings $\Lambda^{L_2}$ are at one site connected to their adjoint via the identity. This allows us to cancel more of them via the simple condition. If $\abs{\Delta x_2}$ is sufficiently large, we can cancel the local tensor that describes one of the extremal sites. This in turn connects the one leg of the extremal operator applied to that site to its adjoint. Therefore, we can use the unitarity in $x_2$-direction an cancel that operator. This process can iteratively be continued, just as in the derivation of the dynamical correlation, until we reach the operator at the top of the triangle. Here it depends on the precice position $y$ relative to that operator, if we can cancel this last one. If we are able to, we find that $E \propto \Tr [ a_y^\beta ] = 0$. Therefore ternary unitary operators actually have an advantage over dual unitary operators for our computations. This can be seen more explicitely, if we compare the Fig.~\ref{fig:equal_time_corr_t1} to Fig.~\ref{fig:equal_time_corr_dual_t1}. For the first a ternary unitary was used, while for the latter we used an operator that is dual unitary with respect to the $x_1$-direction.

\begin{figure}[!ht]
\centering
\includegraphics[width=0.65\columnwidth]{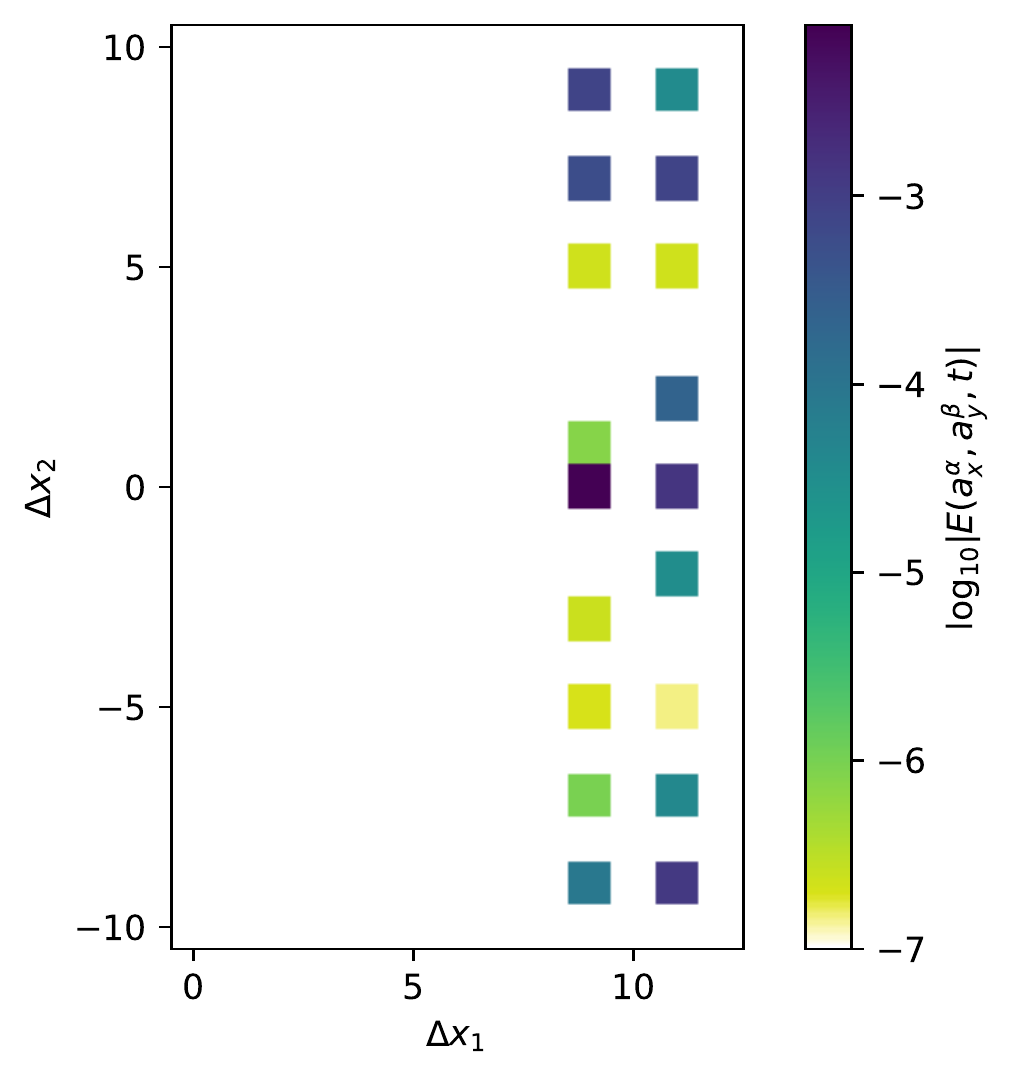}
\caption{Equal-time correlation functions for $t = 2$ and a random simple MPU tensor $\Lambda$ defining the sPEPS, a uniform ternary unitary gate of the form \eqref{eq:four_dual_construction}, and random traceless local operators $a_x^{\alpha}$ and $a_y^{\beta}$.}
\label{fig:equal_time_corr_t2}
\end{figure}

We also plotted the time evolution up to $t = 2$, see Fig.~\ref{fig:equal_time_corr_t2}. The distincitve features for large $\Delta x_2$ are not well visible. However, we can again see our theoretical consideration to be correct: For a longer times the correlation tends to decrease and a higher threshold value $\abs{\Delta x_2}_{\text{thresh}}$ is required for $E \equiv 0$ for $\abs{\Delta x_2}>\abs{\Delta x_2}_{\text{thresh}}$  compared to the shorter time $t=1$.

The code used for the numerical computations can be found at \url{https://github.com/cmendl/ternary_unitaries}.
\newpage

\end{document}